\documentclass{aa}  

\usepackage{graphicx}
\usepackage{txfonts}
\usepackage{gensymb}
\newcommand{\vsini}{$\mathrm{\upsilon} \sin i$}

\usepackage{color}

\begin{document} 

\title{From convective stellar dynamo simulations to Zeeman-Doppler images}

\author{T. Hackman
          \inst{1}
          \and O. Kochukhov\inst{2}
          \and M. Viviani\inst{3}
          \and J. Warnecke\inst{4}
          \and M.J. Korpi-Lagg
          \inst{5,4,6}%
          \and
          J.J. Lehtinen\inst{7,1}}
    
          \institute{Department of Physics, P.O. Box 64, FI-00014 University of Helsinki, Finland \\ \email{thomas.hackman@helsinki.fi}      
         \and
Department of Physics and Astronomy, Uppsala University, Box 516, S-75120 Uppsala, Sweden
\and
Wish s.r.l, Via Venezia 24, 87036 Rende (CS), Italy
\and
Max-Planck-Institut f\"ur Sonnensystemforschung, Justus-von-Liebig-Weg 3, D-37077 G\"ottingen, Germany
\and
Department of Computer Science, Aalto University, PO Box 15400, FI-00076
Espoo, Finland
\and
Nordita, KTH Royal Institute of Technology \& Stockholm University, Hannes Alfv\'ens v\"ag 12, Stockholm, SE-11419, Sweden
\and
Finnish Centre for Astronomy with ESO (FINCA), University of Turku, Vesilinnantie 5, FI-20014 University of Turku, Finland
}

   \date{Received ; accepted }

\abstract
{Zeeman-Doppler imaging (ZDI) is used to reconstruct the surface magnetic field of late-type stars from high-resolution spectropolarimetric observations. The results are usually described in terms of
characteristics of the field topology, such as poloidality versus toroidality and axisymmetry versus non-axisymmetry, in addition to the field strength.}
{In this study, we want to test how well these characteristics are preserved when applying the
ZDI method to simulated data. 
We are particularly interested in how accurately the field topology is preserved and to what extent stellar parameters, such as projected rotation velocity and rotation axis inclination, influence the reconstruction.}
{For these tests, we used 
published magnetic field 
vector data
from direct numerical magnetohydrodynamic simulations
taken near the surface of the simulation domain. These simulations 
have variable rotation rates and therefore represent
different levels of activity of an otherwise Sun-like
setup with a convective envelope of
solar thickness. Our ZDI reconstruction is based on spherical harmonics expansion.
By comparing the original values to those of the reconstructed images, we study the ability to reconstruct the surface magnetic field in terms of various characteristics of the field.}
{In general, the ZDI method works as expected. The main large-scale features are reasonably well recovered, but the strength of the recovered magnetic field is just a fraction of the original input. The quality of the reconstruction shows clear correlations with the data quality. Furthermore, there are some spurious dependencies between stellar parameters and the characteristics of the field.}
{Our study uncovers some limits of ZDI. Firstly, the recovered field strength will generally be lower than the `real' value, as smaller structures with opposite polarities will be blurred in the inversion. This is also seen in the relative distribution of magnetic energy in terms of the angular degree $\ell$. Secondly, the axisymmetry 
is
overestimated. The poloidality versus toroidality 
is
better recovered. The reconstruction works better for a stronger field and faster rotation velocity. Still, the ZDI method works surprisingly well even for a weaker field and slow rotation provided the data have a high signal-to-noise ratio and good rotation phase coverage.}


\maketitle

\section{Introduction}

Zeeman-Doppler imaging \citep[ZDI;][]{Brown1991,Kochukhov2016} is a powerful method to reconstruct surface magnetic field maps of late-type stars. In ZDI, high-resolution
spectropolarimetric observations (Stokes IVQU, Stokes IV, or just Stokes V) are
used as input data. The observations are normally limited to only Stokes IV, as the linear polarisation signal is usually too weak to measure within reasonable exposure times.
The solution is retrieved as magnetic field vector component
maps presenting the surface radial $B^r$, meridional $B^\theta$, and azimuthal $B^\phi$
components \citep[see e.g.][]{Hackman2016}.

In addition to estimating the strength of the magnetic field, its topology is an important characteristic.
Usually the field topology is described in terms of fractions of poloidal versus toroidal
and axisymmetric versus non-axisymmetric field energy. These are straightforward to
calculate when the field solution is derived in terms of a spherical harmonics expansion \citep[see e.g.][]{Donati2006}.
The above characteristics are used to illustrate the strengths and topologies of
stellar magnetic fields as functions of stellar parameters such as Rossby number, age, and spectral class \citep[see e.g.][]{Donati2008}. Apart from the obvious findings that the strength of the large-scale magnetic field is strongest for stars with low Rossby numbers, interesting conclusions drawn from these are for example that there are connections between the fractions of axisymmetric and toroidal field energies \citep{See2015}.
In the case of a series of ZDI maps from a particular star, topological characteristics can be used to illustrate the changes in the magnetic field topology possibly related to activity cycles \citep[see e.g.][]{Kochukhov2013}.

Testing the ZDI method is challenging, because the Sun is the only late-type star for which we can directly resolve the surface magnetic field vector. The study by \cite{Vidotto2016} shows what we could expect in terms of ZDI by observing the Sun as a star. 
A synoptic map of the solar surface magnetic vector field (Carrington Rotation CR2109) was used to reconstruct the surface field with spherical harmonics decomposition using different values for the highest angular degree ($\ell_{\max} =5$ and $\ell_{\max} = 150$). With $\ell_{\max} = 5$, which corresponds to the typical value used for ZDI inversions of slowly rotating active stars, most of the magnetic field remains undetected. Furthermore, the fractional strengths of the different field components are strongly biased towards $B^r$. The cumulative fraction of poloidal field energy is constantly over 0.9, showing some fluctuation at the lowest degrees ($\ell < 10$), reaching a local maximum of approximately $0.97$ around $\ell \approx 15$ and then smoothly settling towards 0.91 at $\ell =150$.

Another approach is to apply ZDI to stellar magnetic fields calculated from simulations. \cite{Lehmann2019} used 3D global magnetic field simulations  to study how well the ZDI method was able to reconstruct the magnetic field at
the surface. In these simulations \citep{Lehmann2018,Gibb2016}, 
a surface flux-transport model was adopted, where
only the surface evolution of the large-scale magnetic field was solved. Starspots emerged on the surface at a rate that is a free parameter of the model, as are also the rotation rate and the differential 
rotation
profile of the star. 
\cite{Lehmann2019} chose models with one, three, and five times the solar angular rotation velocity to mimic the Sun and more active stars.

In the present study, in contrast, we use direct numerical simulations of stellar dynamos \citep{Viviani2018,Viviani2019}. 
These models solve the full magnetohydrodynamic (MHD) equations in the stellar convection zone self-consistently without relying on a surface flux-transport approach, and therefore also produce strong magnetic fields on small scales.
In the full MHD approach, the differential rotation profile emerges as a solution to the Navier-Stokes equation, while the only input parameter is the rotation rate (or Coriolis number). However, the simulations we use do not completely realistically correspond to the surface magnetic fields of real stars. Firstly, the simulations only cover latitudes of between $-75 \degree$ and $+75 \degree$. Secondly, we use the magnetic field data just below the surface at a distance of $r=0.98 R_\star$ from the centre, with $R_\star$ being the stellar radius. The reasons for these discrepancies with a real case are described in Section \ref{prep}.

The rotation profiles of stars  other than the Sun are largely unknown. Therefore, one may argue that the direct numerical simulations, containing fewer free parameters and with differential rotation emerging as a natural solution to the equations, would give better representations of real stars.
Furthermore, the meridional circulation is also part of the solution and varies with the level of stellar activity, whereas it was kept constant in the 
simulations used by
\cite{Lehmann2019}.
In our convective dynamo simulations, the differential rotation together with the turbulent convection generates the magnetic field solution self-consistently and is not prescribed by a parameterized flux-emergence model.
Also, compared with the work of these latter authors, the simulations we use cover a larger parameter 
regime in terms of rotation and activity.
Moreover, because of the their setup, their solution will be dominated by the $m=0$ mode,
whereas our selected simulations contain cases with both $m=0$ and $m=1$ domination. 

Our main focus is to learn how well the magnetic field topology is preserved in the ZDI inversion using Stokes V spectra calculated from a simulated stellar magnetic field that is as realistic as possible. We mainly used three simulations in our study. For each simulation, we applied a set of different projected rotation velocities \vsini. For one simulation, we also tested the effect of different rotation axis inclinations $i$. These three simulations, although differing in magnetic field strength, all have quite  a low percentage of axisymmetric magnetic field energy. To explore the reconstruction of this parameter, we
introduced
a more axisymmetric case constructed from a wedge simulation.

We also want to roughly estimate what level of observation data quality (signal-to-noise ratio (S/N), and rotation phase coverage) is needed for ZDI in different cases. Contrary to \cite{Lehmann2019}, we do not test different descriptions of the magnetic field in terms of the relations between the spherical harmonics coefficients $\alpha_{\ell, m}$, $\beta_{\ell, m}$, and $\gamma_{\ell, m}$. Instead, we only use the description of divergence-free field
with maximum degrees of freedom \citep[case (iv);][]{Lehmann2019}.

\begin{figure}
      \centering
     \includegraphics[bb=240 230 390 790, width=2.4cm,clip,angle=90]{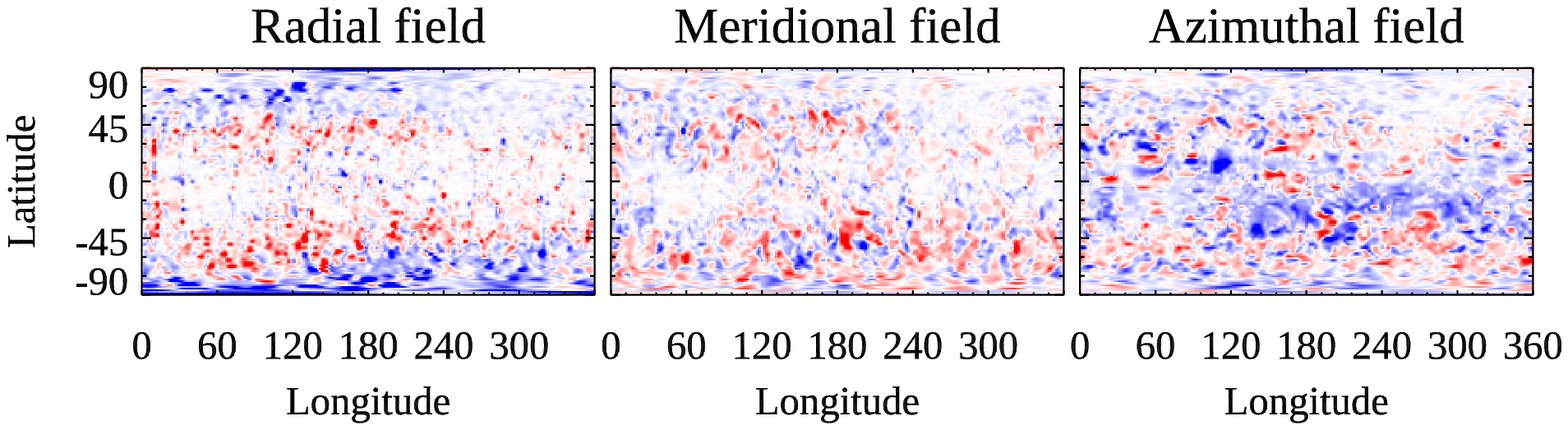}
     \includegraphics[bb=240 230 390 790, width=2.4cm,clip,angle=90]{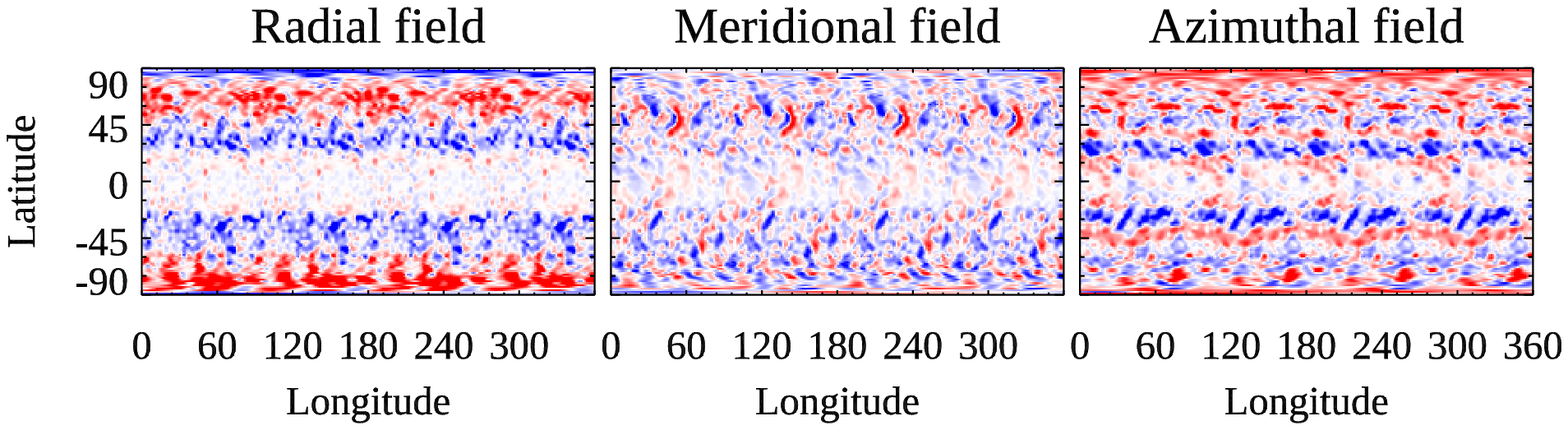}
    \includegraphics[bb=260 770 410 810, width=3cm]{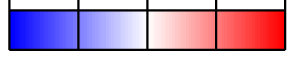}
\includegraphics[bb= 240 230 390 790, width=2.4cm,clip,angle=90]{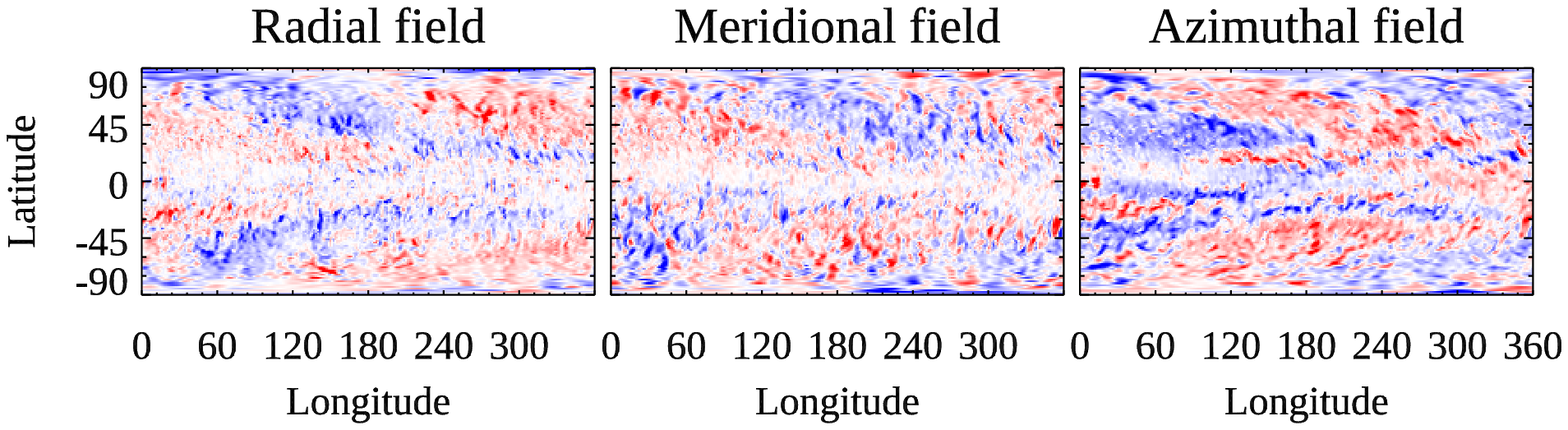}
\includegraphics[bb=240 230 390 790, width=2.4cm,clip,angle=90]{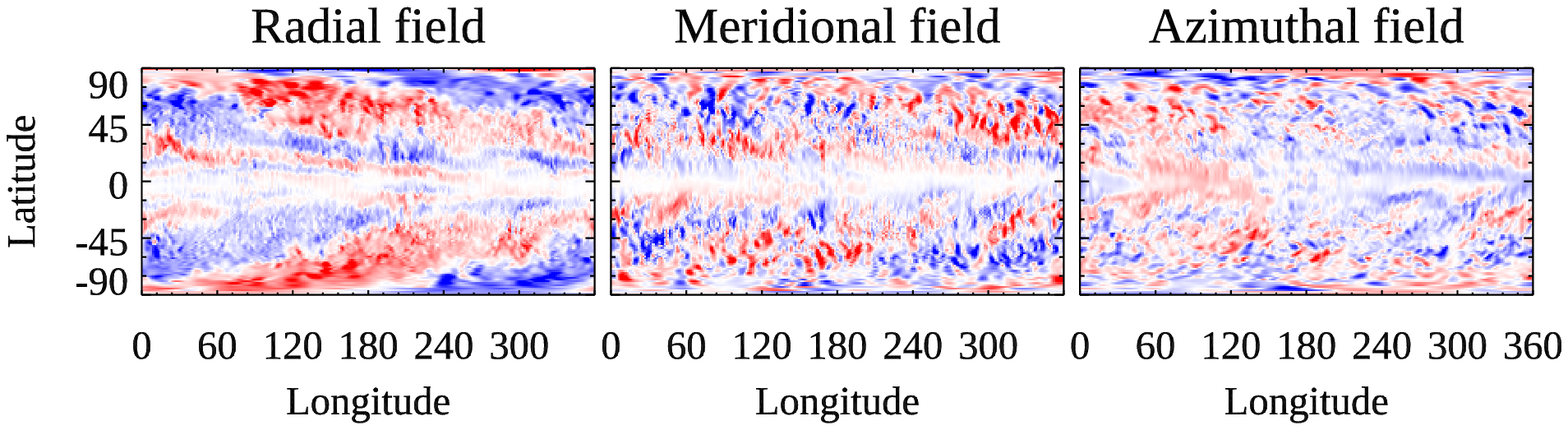}
\includegraphics[bb=260 770 410 810, width=3cm]{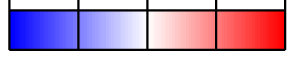}
\caption{Input magnetic field data for the ZDI calculations from top to bottom: C1, G$^W$,
H$^a$, and L$^a$. We highlight the different scales for C1/G$^W$ and L$^a$/H$^a$.}
   \label{origmap}
    \end{figure}

\begin{table*}
  \caption{Parameters and field characteristics for selected simulations.}
  \centering
\begin{tabular}{llrrrrrrrrrrr}
\hline \hline
Simulation & Original grid & $\Omega$ & Co & $B_\mathrm{RMS}$ & $B^r_\mathrm{RMS}$ & $B^\theta_\mathrm{RMS}$ & $B^\phi_\mathrm{RMS}$ & $B^{\ell\le 20}_\mathrm{RMS}$ & $p_\mathrm{pol}$ & $p_\mathrm{pol}^{\ell\le 20}$ &  $p_\mathrm{axi}$ & $p_\mathrm{axi}^{\ell\le 20}$\\
              &      & [$\Omega_\sun$] &  & [kG] & [kG] & [kg] & [kG] & [kG] & [\%] & [\%] & [\%] & [\%] \\
  \hline
C1 & $288 \times  576$  & 1.8  &  2.8 & 0.217 & 0.124 & 0.105 & 0.144 & 0.11 & 61.5 &  55.9 & 11.0 &  28.2 \\
  H$^a$ & $512 \times 1024$ & 7.8& 16.1 & 0.480 & 0.240 & 0.262 & 0.322 & 0.37 & 62.5 &  81.8 &  2.8 &   0.6 \\
L$^a$ & $512 \times 1024$ & 23.3& 83.4& 0.483 & 0.286 & 0.311 & 0.234 & 0.31 & 69.5 &  82.1 &  2.7 &   5.4 \\
G$^W$ &  $256 \times 128$  & 4.8& 8.3 & 0.305 & 0.191 & 0.140 & 0.193 & - & 63.3 & - & 31.8 & - \\ %
\hline
\end{tabular}
\tablefoot{$\Omega$ is the rotation rate of the simulations in terms of the solar rotation rate $\Omega_\sun$ and Co is the Coriolis number as defined in \cite{Viviani2018}. The characteristics of the magnetic
field are calculated from the preprocessed input data.}
\label{mhdpar}
\end{table*}

\begin{table*}
  \caption{Parameters and results of ZDI inversions.}
  \centering
\begin{tabular}{rlccrrcllllrr}
\hline \hline
ID & Data & $n$ & $n_\phi $ & $S/N$ & \vsini & $i$ & $B_\mathrm{RMS}$ & $B^r_\mathrm{RMS}$ & $B^\theta_\mathrm{RMS}$ & $B^\phi_\mathrm{RMS}$ & $p_\mathrm{pol}$ &  $p_\mathrm{axi}$ \\
    &    &     &           &       & [km s$^{-1}]$ & [$\degree$] & [kG] & [kG] &  [kG] & [kG]  & [\%]            & [\%] \\
\hline
1  & C1 & 2 & 20 &$10^5$ & 5  & 60 & 0.039 & 0.013 & 0.015 & 0.034 & 45.4   & 60.2 \\ 
2  & C1 & 2 & 20 &$10^5$ & 10 & 60 & 0.043 & 0.018 & 0.014 & 0.036 & 50.1   & 54.2\\  
3  & C1 & 2 & 20 &$10^5$ & 20 & 60 & 0.051 & 0.027 & 0.016 & 0.040 & 58.3   & 50.5\\  
4  & C1 & 2 & 20 &$10^5$ & 40 & 60 & 0.059 & 0.033 & 0.019 & 0.045 & 61.2   &  39.1\\ 
5  & H$^a$ & 2 & 10 & $10^4$ & 10 & 60 & 0.129 & 0.062 & 0.059 & 0.096 & 68.9   &  7.5 \\  
6  & H$^a$ & 2 & 10 & $10^4$ & 20 & 60 & 0.124 & 0.061 & 0.053 & 0.094 & 75.4 &  8.8 \\  
7  & H$^a$ & 2 & 10 & $10^4$ & 40 & 60 & 0.125 & 0.063 & 0.053 & 0.094 & 72.8  & 11.6 \\ 
8  & H$^a$ & 2 & 10 & $10^4$ & 60 & 60 & 0.148 & 0.076 & 0.060 & 0.112 & 69.7  & 13.7 \\ 
9 & L$^a$ & 2 & 10 & $10^4$ & 10 & 60 & 0.115 & 0.069 & 0.064 & 0.066 & 71.5  &  8.2 \\  
10 & L$^a$ & 2 & 10 & $10^4$ & 20 & 60 & 0.135 & 0.080 & 0.080 & 0.074 & 73.6  & 11.1 \\  
11 & L$^a$ & 2 & 10 & $10^4$ & 40 & 60 & 0.110 & 0.068 & 0.055 & 0.066 & 73.0  & 12.9 \\ 
12 & L$^a$ & 2 & 10 & $10^4$ & 60 & 60 & 0.141 & 0.087 & 0.083 & 0.073 & 69.8  & 13.5 \\ 
13 & L$^a$ & 2 & 10 & $5 \cdot 10^4$ & 40 & 10 & 0.203 & 0.162 & 0.105 & 0.060 &  82.6 & 6.6\\  
14 & L$^a$ & 2 & 10 & $5 \cdot 10^4$ & 40 & 20 & 0.201 & 0.159 & 0.104 & 0.065 &  82.1 & 6.2\\  
15 & L$^a$ & 2 & 10 & $5 \cdot 10^4$ & 40 & 30 & 0.191 & 0.150 & 0.098 & 0.065 &  81.3 & 6.0\\ 
16 & L$^a$ & 2 & 10 & $5 \cdot 10^4$ & 40 & 40 & 0.178 & 0.135 & 0.093 & 0.071 &  79.9 & 5.9\\  
17 & L$^a$ & 2 & 10 & $5 \cdot 10^4$ & 40 & 50 & 0.156 & 0.111 & 0.085 & 0.070 &  76.3 & 6.4\\ 
18 & L$^a$ & 2 & 10 & $5 \cdot 10^4$ & 40 & 60 & 0.140 & 0.090 & 0.074 & 0.078 &  73.4 & 7.6\\  
19 & L$^a$ & 2 & 10 & $5 \cdot 10^4$ & 40 & 70 & 0.121 & 0.068 & 0.058 & 0.081 &  75.3 & 10.0 \\ 
20 & L$^a$ & 2 & 10 & $5 \cdot 10^4$ & 40 & 80 & 0.125 & 0.067 & 0.061 & 0.086 &  77.7 & 10.3 \\ 
21 & L$^a$ & 2 & 10 & $5 \cdot 10^4$ & 40 & 85 & 0.140 & 0.076 & 0.075 & 0.090 & 76.0 &  7.4 \\ 
22 & L$^a$ & 2 & 10 & $5 \cdot 10^4$ & 40 & 90 & 0.148 & 0.082 & 0.081 & 0.093 & 75.4 &  5.2 \\ 
23 & L$^a$ & 1 & 10 & $5 \cdot 10^4$ & 40 & 10 & 0.175 & 0.145 & 0.086 & 0.046 & 85.6 &  7.3 \\ 
24 & L$^a$ & 1 & 10 & $5 \cdot 10^4$ & 40 & 20 & 0.176 & 0.145 & 0.083 & 0.053 & 85.0 & 6.8 \\ 
25 & L$^a$ & 1 & 10 & $5 \cdot 10^4$ & 40 & 40 & 0.160 & 0.129 & 0.072 & 0.062 & 83.8 & 6.1 \\ 
26 & L$^a$ & 1 & 10 & $5 \cdot 10^4$ & 40 & 60 & 0.127 & 0.087 & 0.059 & 0.072 & 77.4 & 8.0 \\ 
27 & L$^a$ & 1 & 10 & $5 \cdot 10^4$ & 40 & 70 & 0.110 & 0.063 & 0.047 & 0.076 & 75.2 & 10.3 \\ 
28 & L$^a$ & 1 & 10 & $5 \cdot 10^4$ & 40 & 80 & 0.111 & 0.060 & 0.048 & 0.080 & 74.2 & 9.1 \\ 
29 & L$^a$ & 1 & 10 & $5 \cdot 10^4$ & 40 & 85 & 0.119 & 0.067 & 0.055 & 0.081 & 74.7 & 7.2 \\ 
30 & L$^a$ & 1 & 10 & $5 \cdot 10^4$ & 40 & 90 & 0.123 & 0.072 & 0.060 & 0.081 & 74.8 & 6.4 \\ 
31 & G$^W$ & 2 & 20 &         $10^5$ & 40 & 60 & 0.126 & 0.079 & 0.049 & 0.085 & 59.9 & 66.6 \\ 
32 & G$^W$ & 2 & 10 &         $10^4$ & 40 & 60 & 0.074 & 0.044 & 0.034 & 0.049 & 59.4 & 61.3 \\ 

\hline
\end{tabular}
\tablefoot{Test ID, MHD simulation, exponent of regularisation function, number of rotational phases, S/N, projected
    rotation velocity at the equator, inclination of rotation axis, RMS values of the magnetic field,  percentages
    of poloidal and axisymmetric field energies.}
\label{zdires}
\end{table*}

\begin{figure}
   \centering
 \includegraphics[bb=240 230 390 790, width=2.4cm,clip,angle=90]{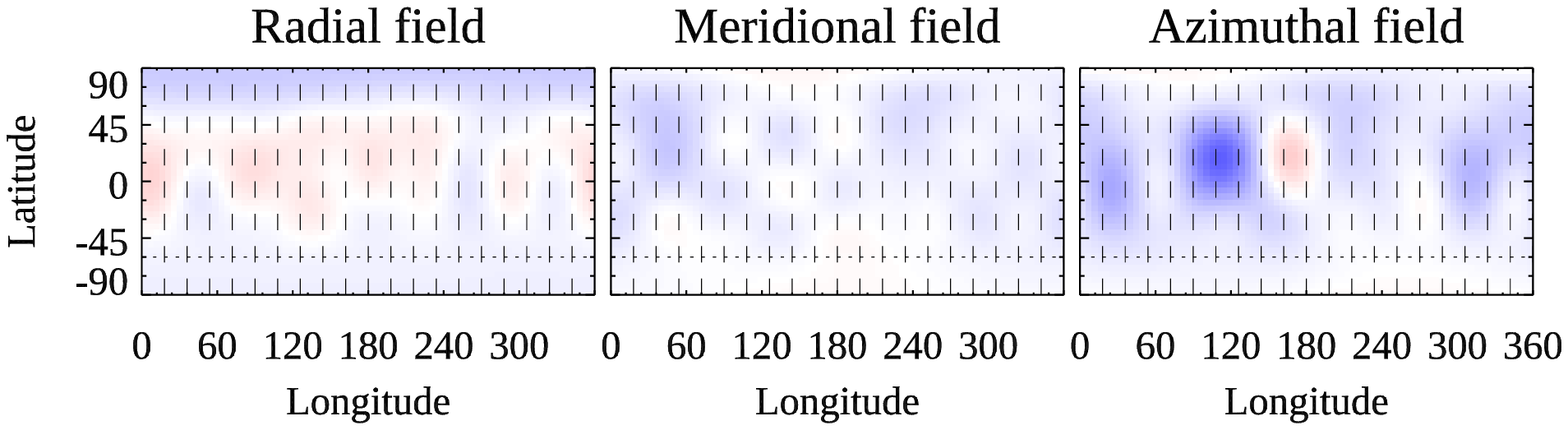}
 \includegraphics[bb=240 230 390 790, width=2.4cm,clip,angle=90]{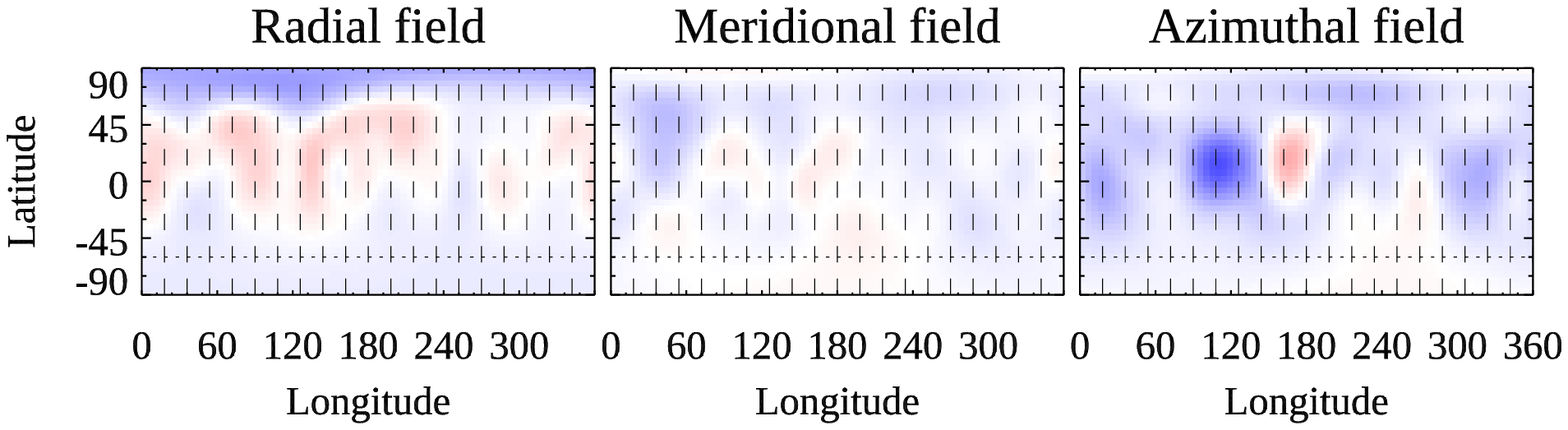}
 \includegraphics[bb=240 230 390 790, width=2.4cm,clip,angle=90]{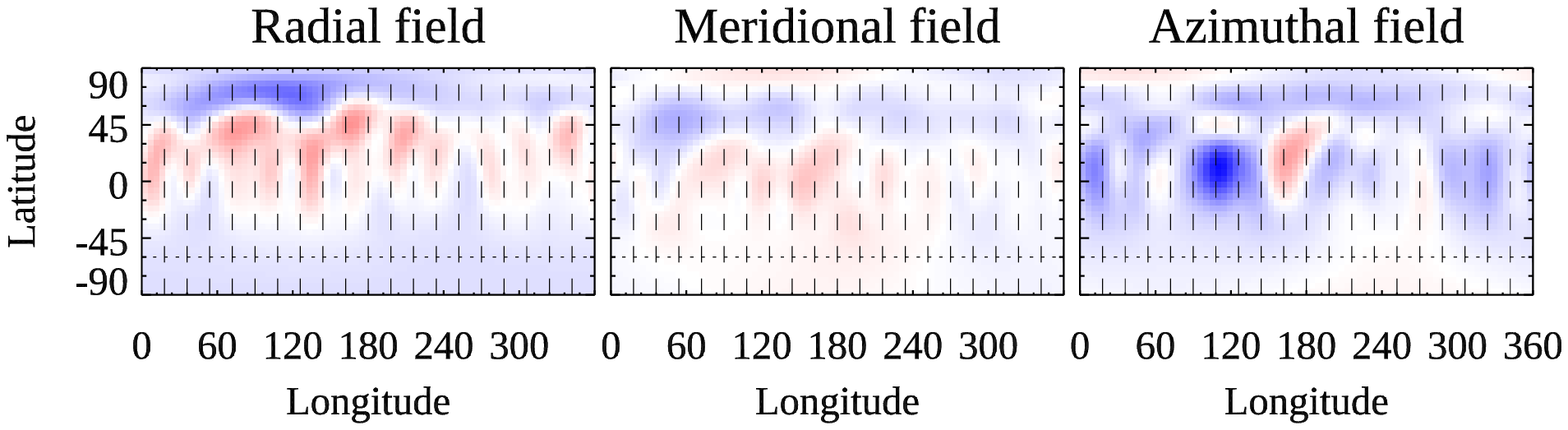}
 \includegraphics[bb=240 230 390 790, width=2.4cm,clip,angle=90]{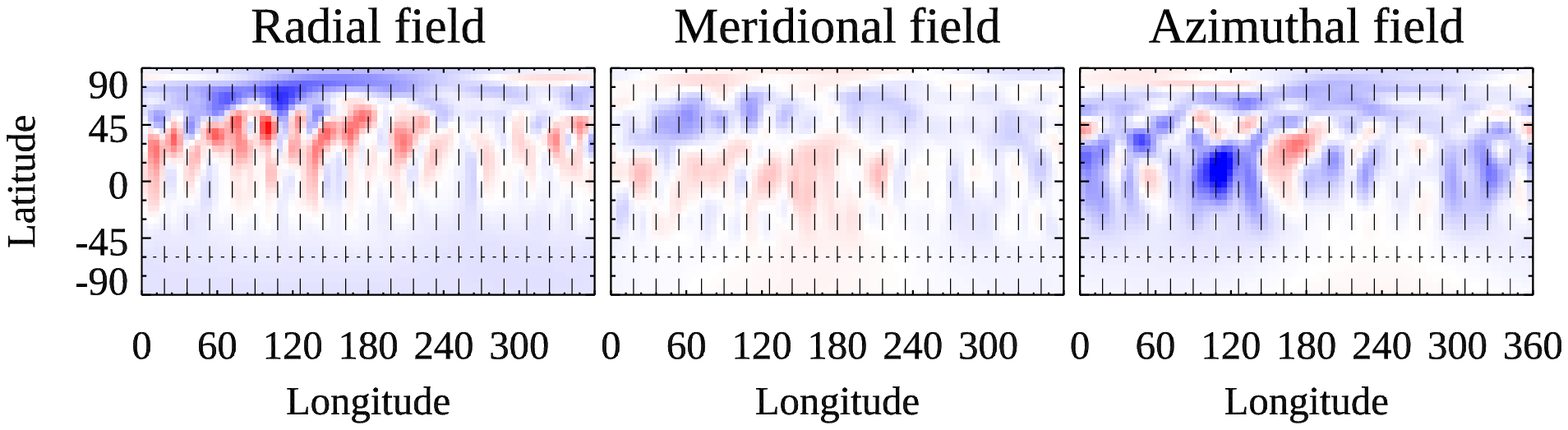}
\includegraphics[bb=260 770 410 810, width=3cm]{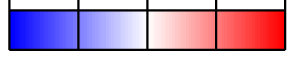}
 \caption{Resulting ZDI map for Tests 1--4 using simulation C1. Top to bottom 
 row: Increase in \vsini\, (5, 10, 20, and 40 km s$^{-1}$).
 The vertical lines mark the rotational phases of the simulated spectra. The horizontal line marks the visibility limit due to the rotational axis inclination $i$.}
              \label{resmapC1}
    \end{figure}

 \begin{figure}
   \centering
 \includegraphics[bb=240 230 390 790, width=2.4cm,clip,angle=90]{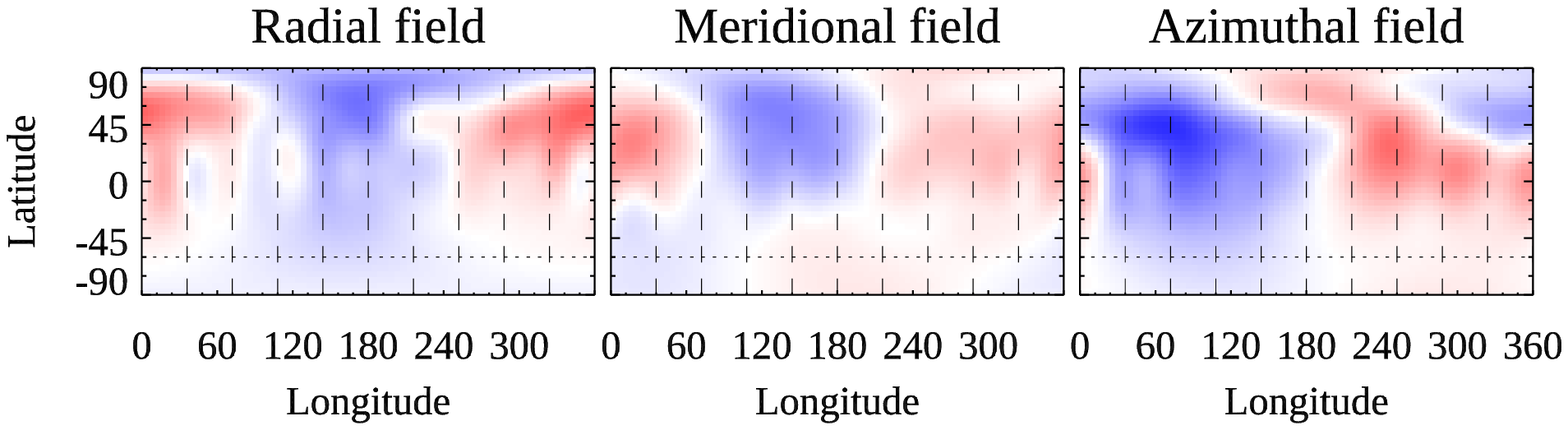}
 \includegraphics[bb=240 230 390 790, width=2.4cm,clip,angle=90]{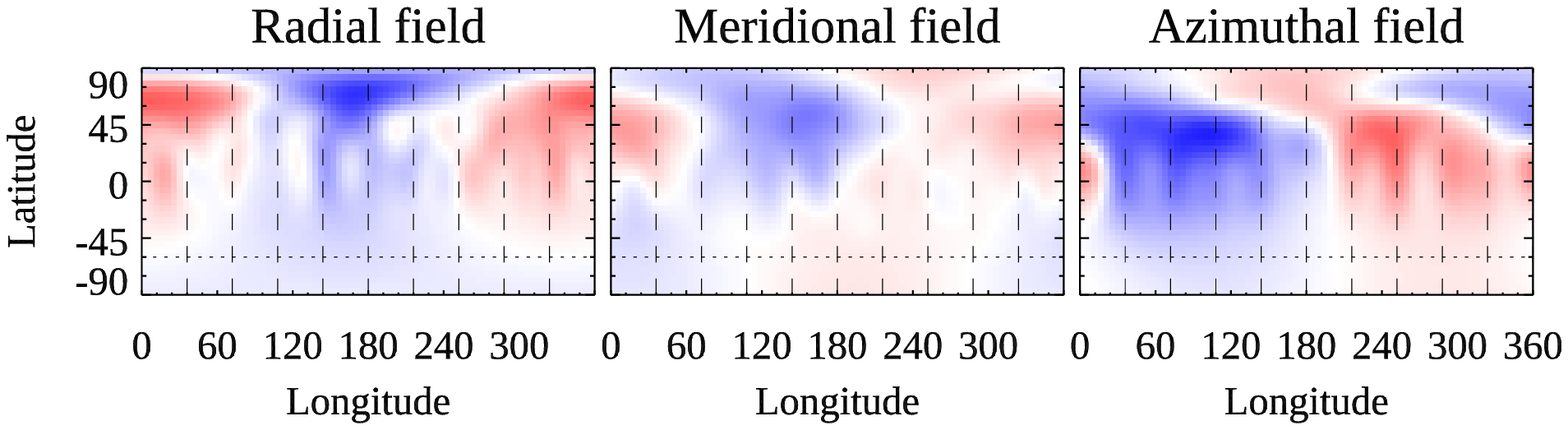}
 \includegraphics[bb=240 230 390 790, width=2.4cm,clip,angle=90]{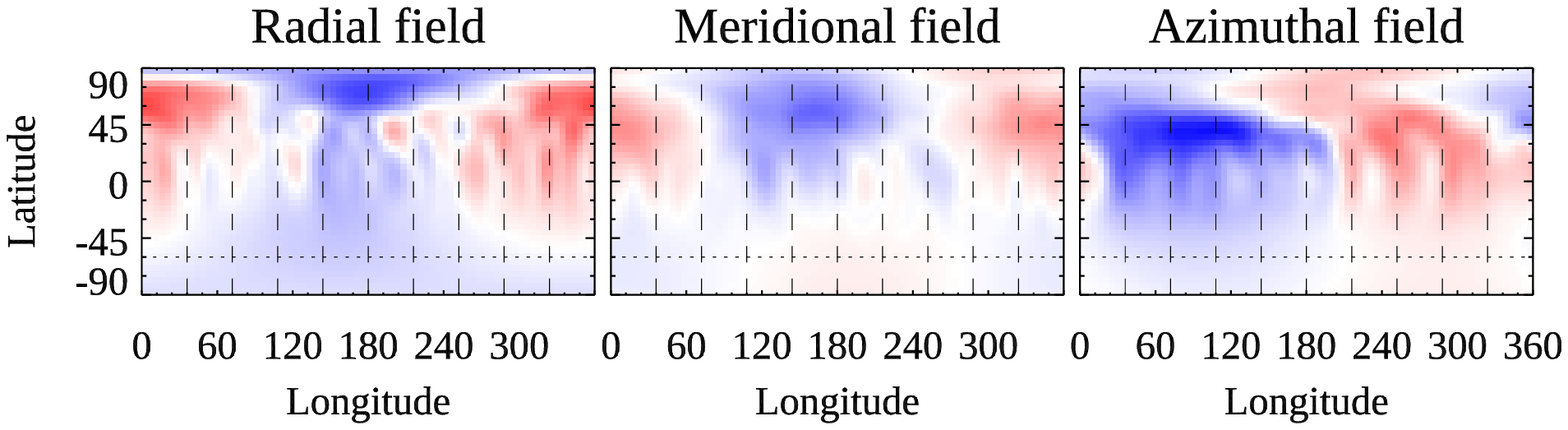}
 \includegraphics[bb=240 230 390 790, width=2.4cm,clip,angle=90]{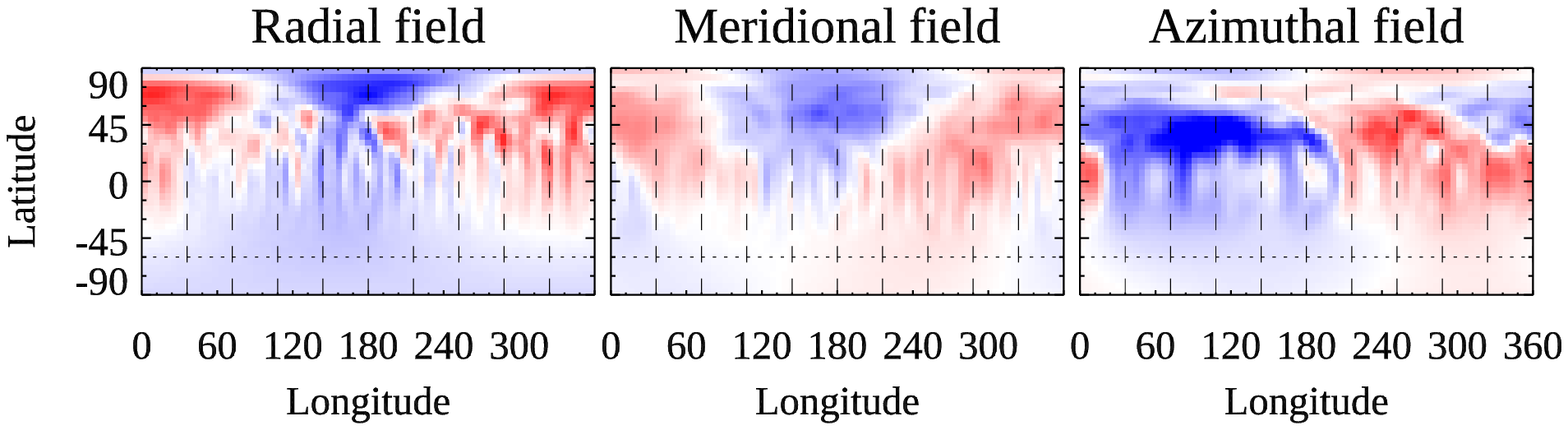}
 \includegraphics[bb=260 770 410 810, width=3cm]{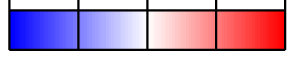}
 \caption{Resulting ZDI map for Tests 5--8 using simulation H$^a$ and \vsini~values 10, 20, 40, and 60 km s$^{-1}$. Otherwise same as in Fig. \ref{resmapC1}.}
              \label{resmapHa}
    \end{figure}

\begin{figure}
   \centering
 \includegraphics[bb=240 230 390 790, width=2.4cm,clip,angle=90]{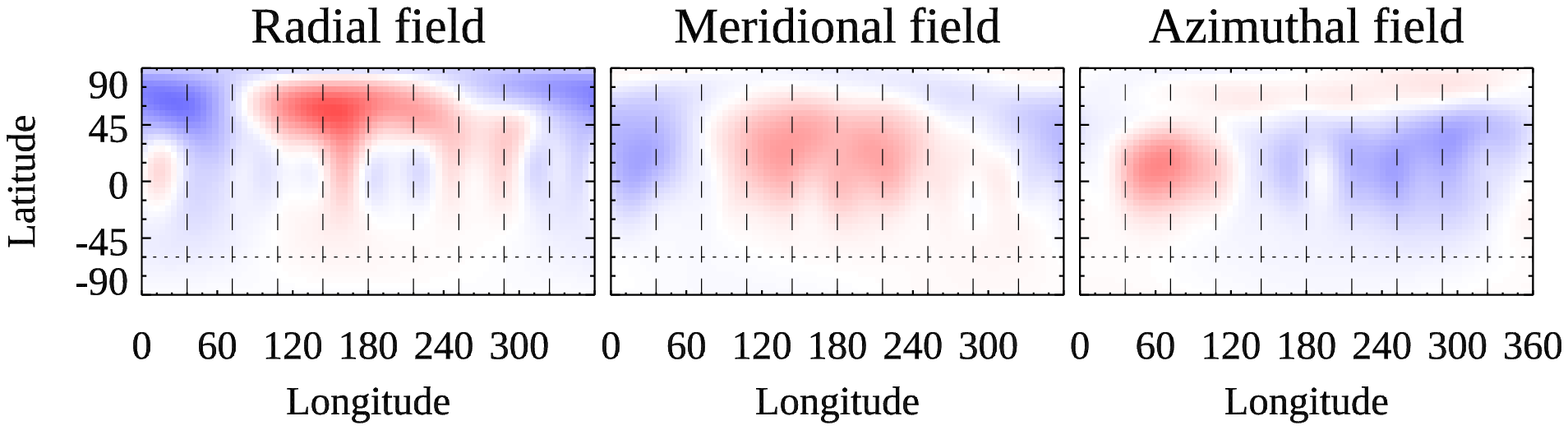}
 \includegraphics[bb=240 230 390 790, width=2.4cm,clip,angle=90]{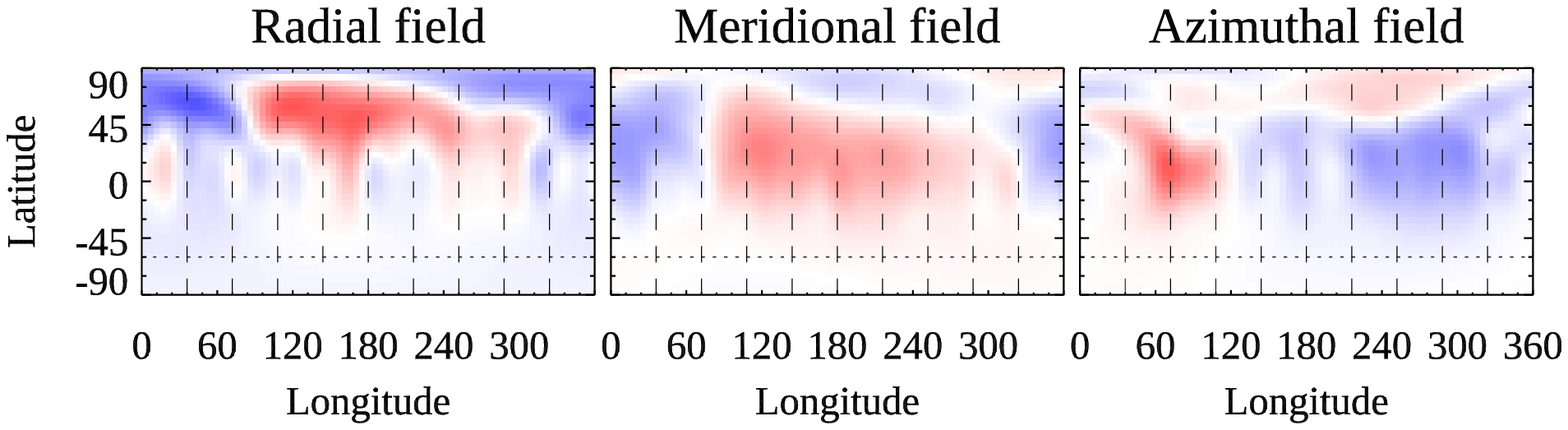}
 \includegraphics[bb=240 230 390 790, width=2.4cm,clip,angle=90]{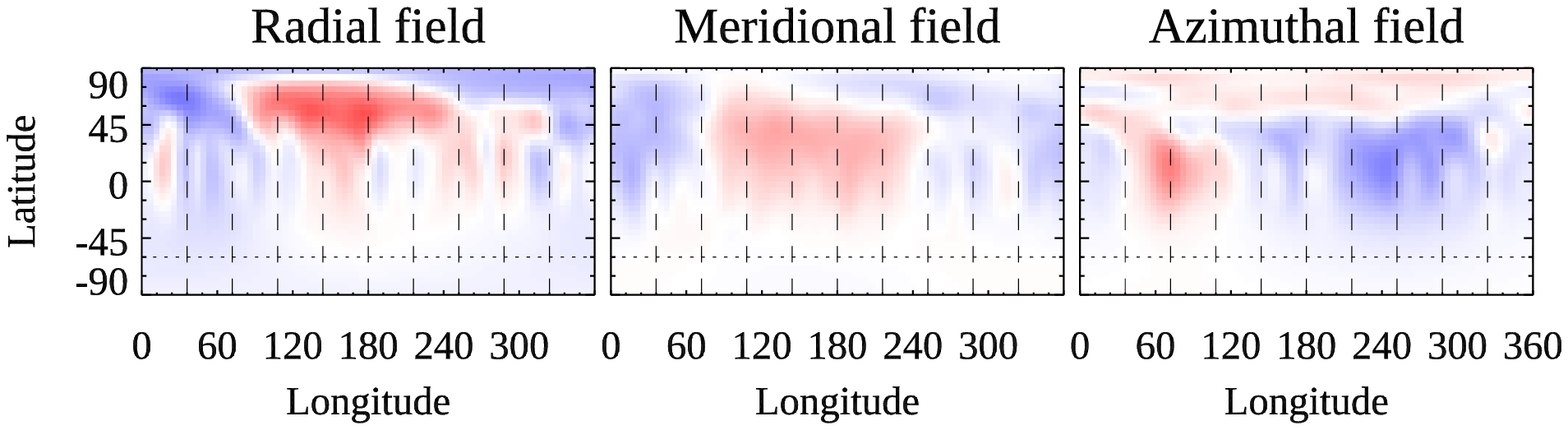}
 \includegraphics[bb=240 230 390 790, width=2.4cm,clip,angle=90]{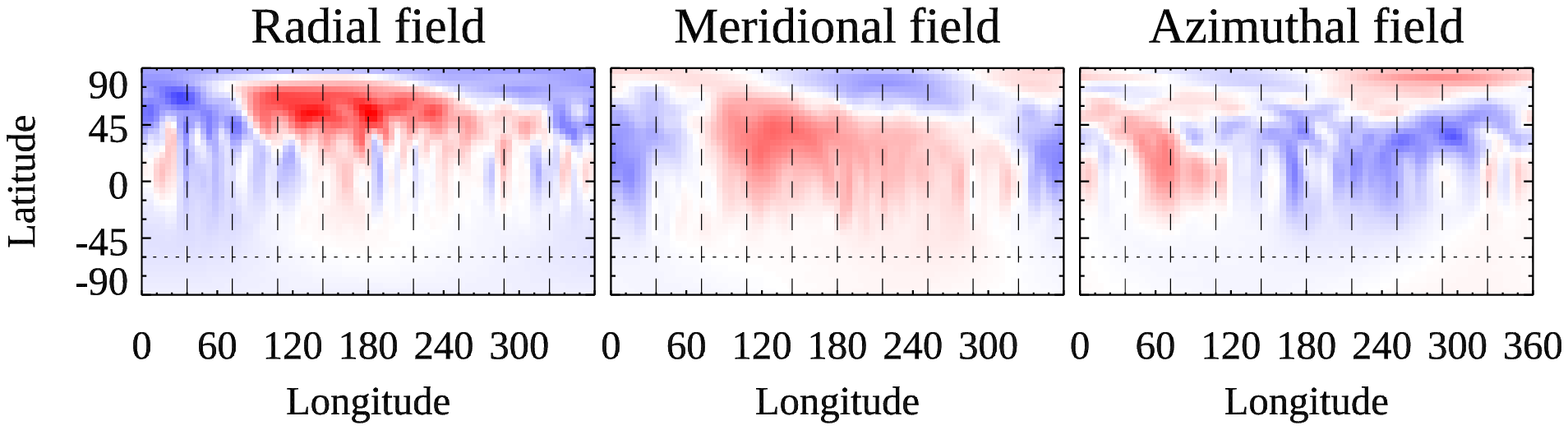}
\includegraphics[bb=260 770 410 810, width=3cm]{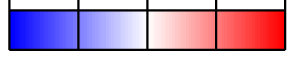}
\caption{Resulting ZDI map for Tests 9--12 using simulation L$^a$. Otherwise same as in
Fig. \ref{resmapHa}.}
              \label{resmapLa}
    \end{figure}

\begin{figure}
   \centering
\includegraphics[bb=240 230 390 790, width=2.4cm,clip,angle=90]{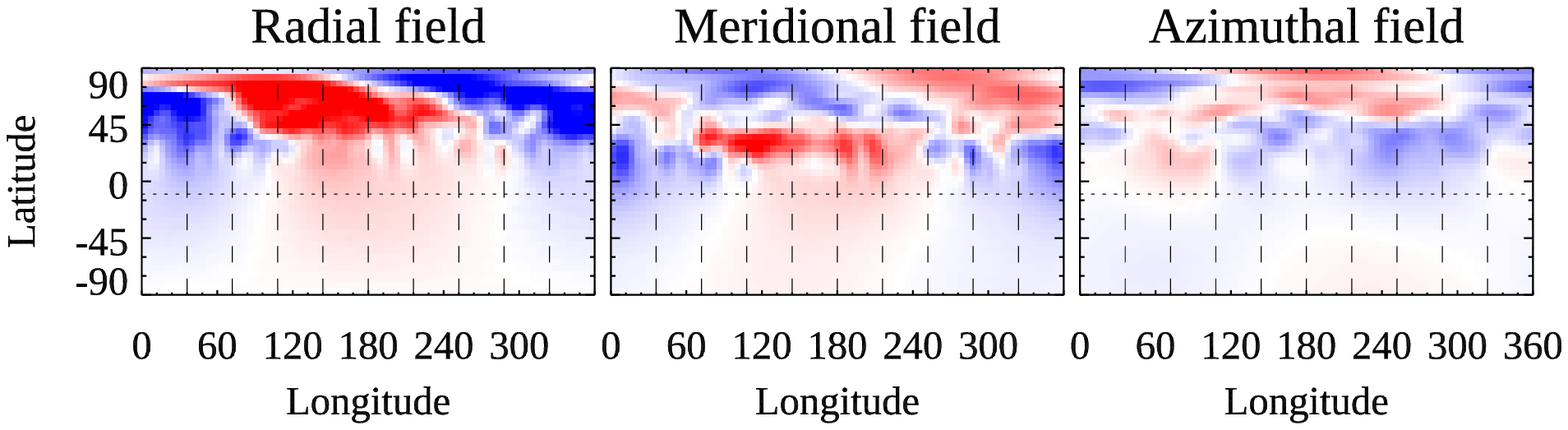}
\includegraphics[bb=240 230 370 790, width=2.05cm,clip,angle=90]{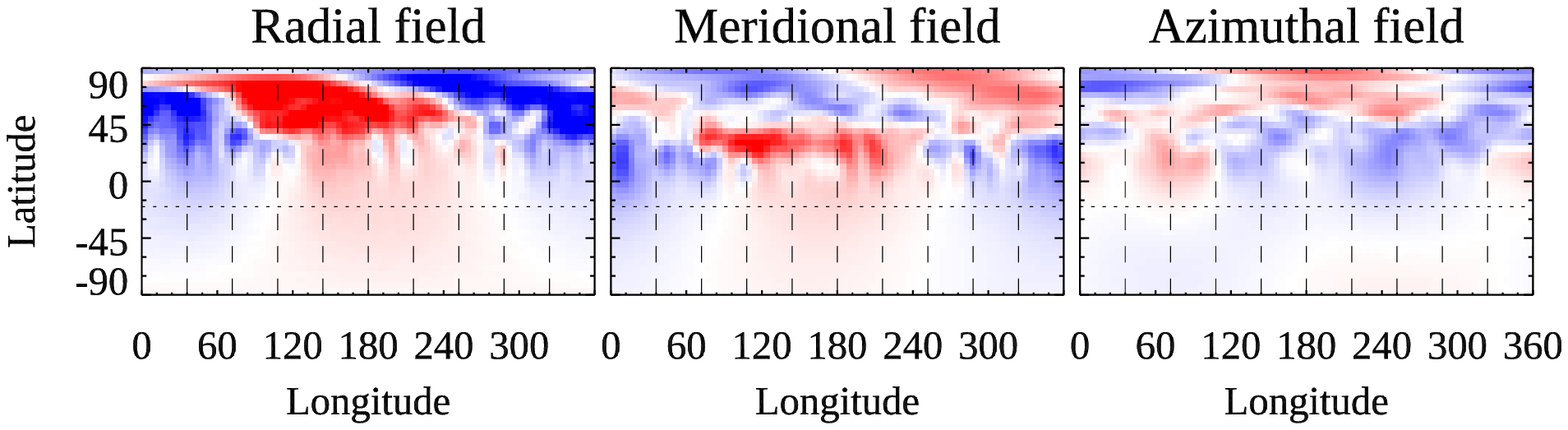}
\includegraphics[bb=240 230 370 790, width=2.05cm,clip,angle=90]{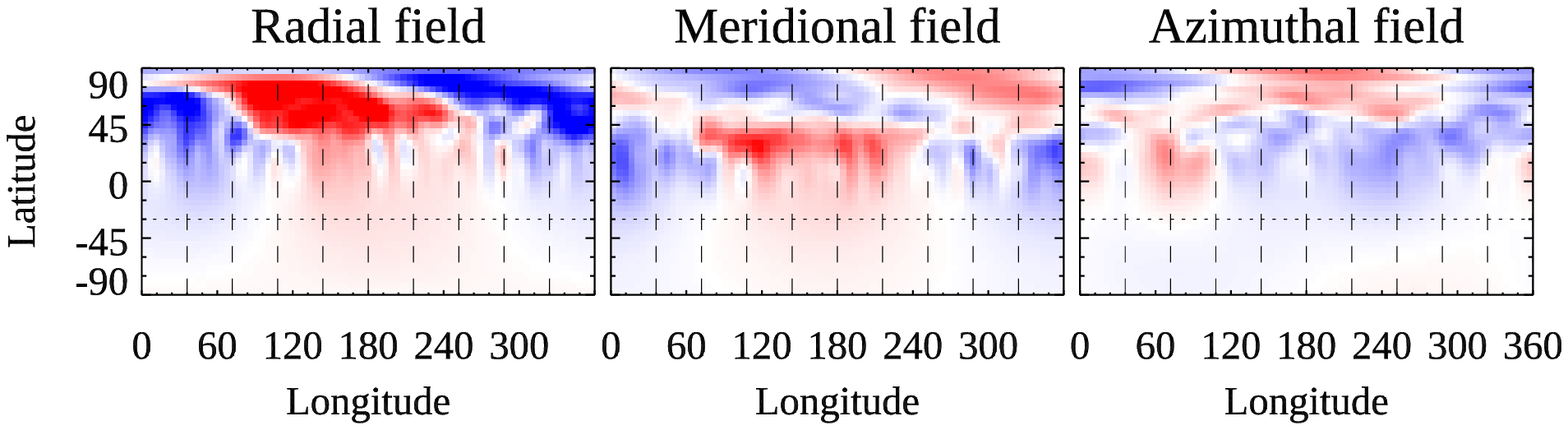}
\includegraphics[bb=240 230 370 790, width=2.05cm,clip,angle=90]{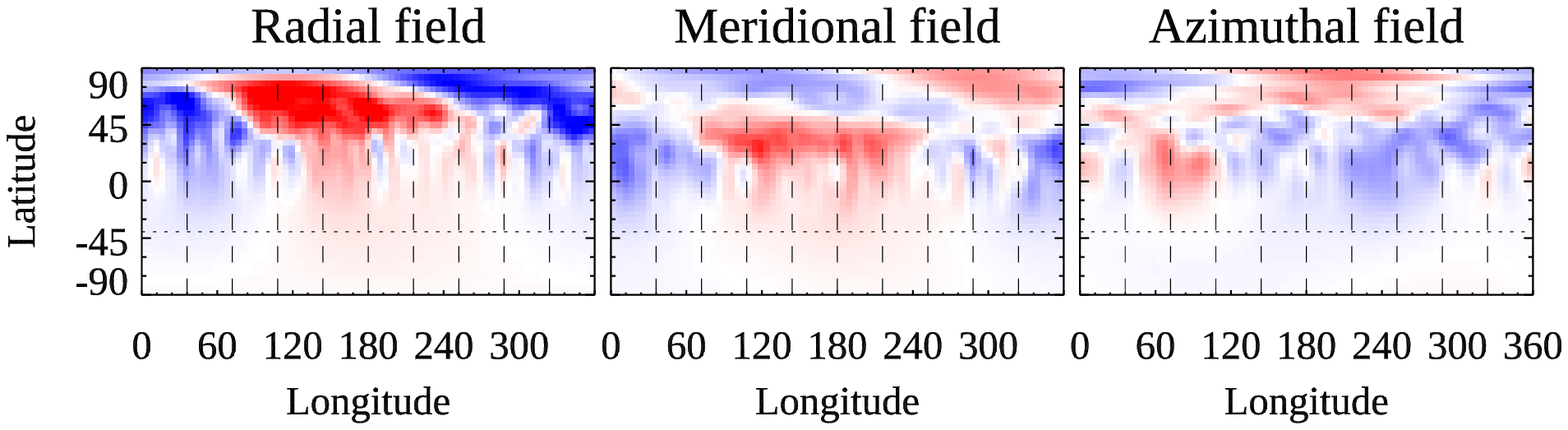}
\includegraphics[bb=240 230 370 790, width=2.05cm,clip,angle=90]{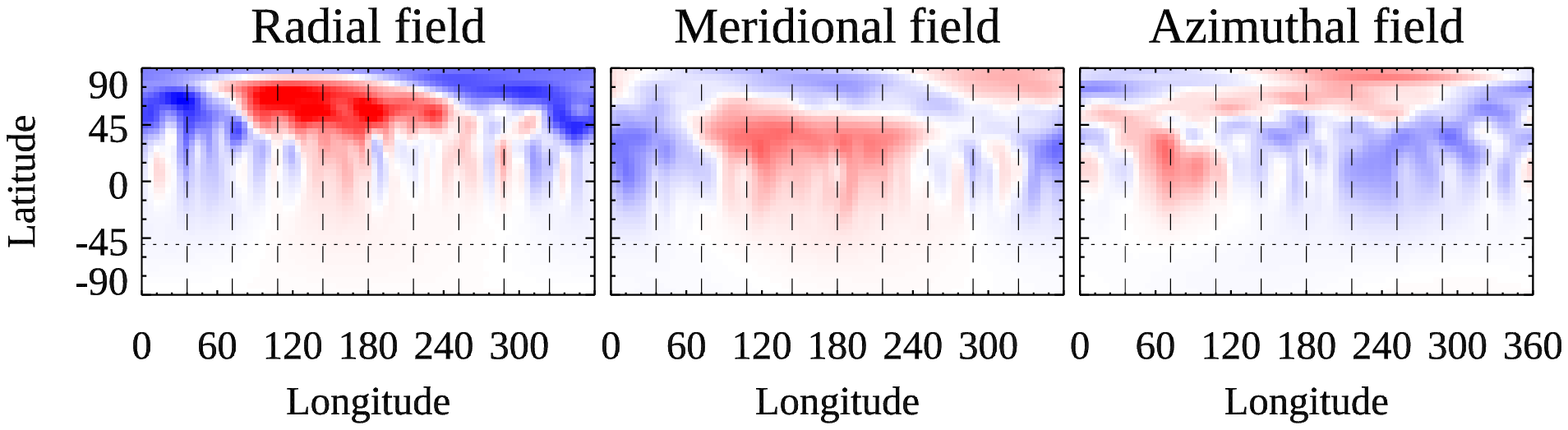}
\includegraphics[bb=240 230 370 790, width=2.05cm,clip,angle=90]{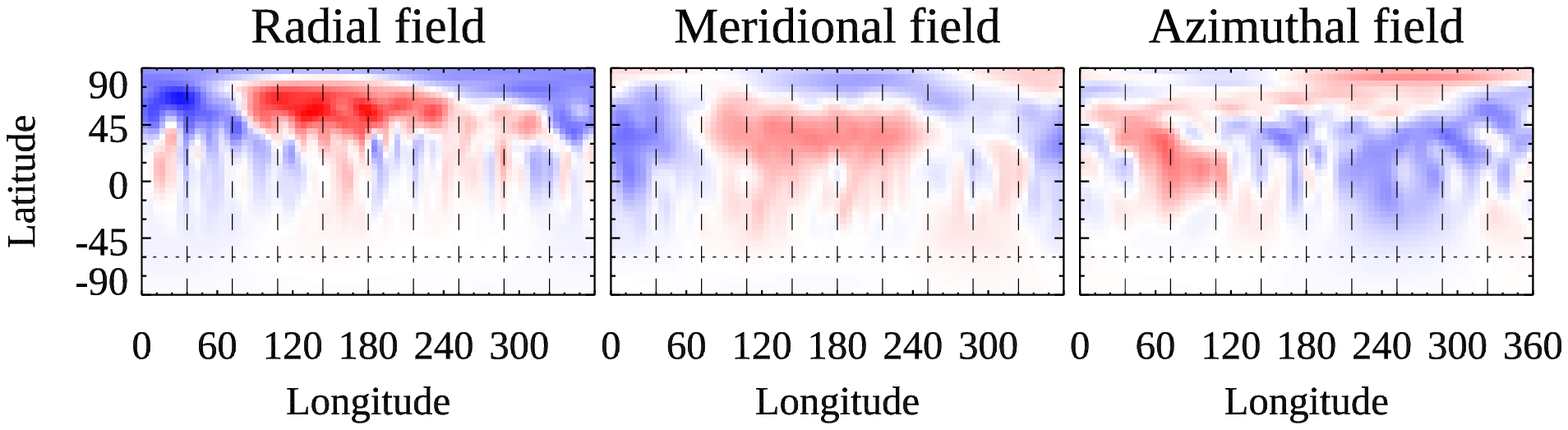}
\includegraphics[bb=240 230 370 790, width=2.05cm,clip,angle=90]{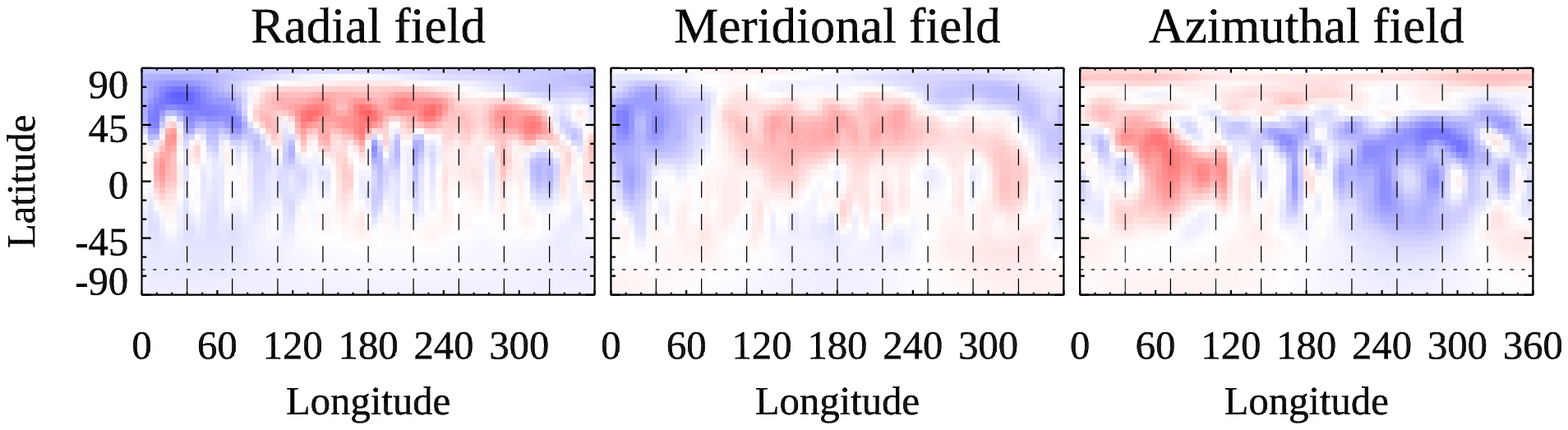}
\includegraphics[bb=240 230 370 790, width=2.05cm,clip,angle=90]{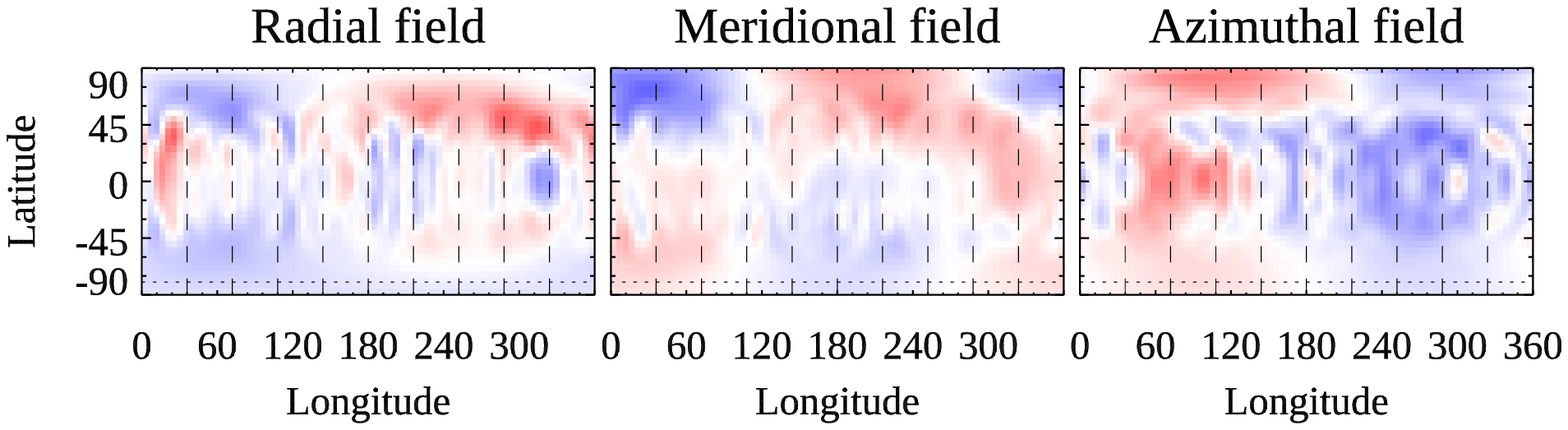}
\includegraphics[bb=240 230 370 790, width=2.05cm,clip,angle=90]{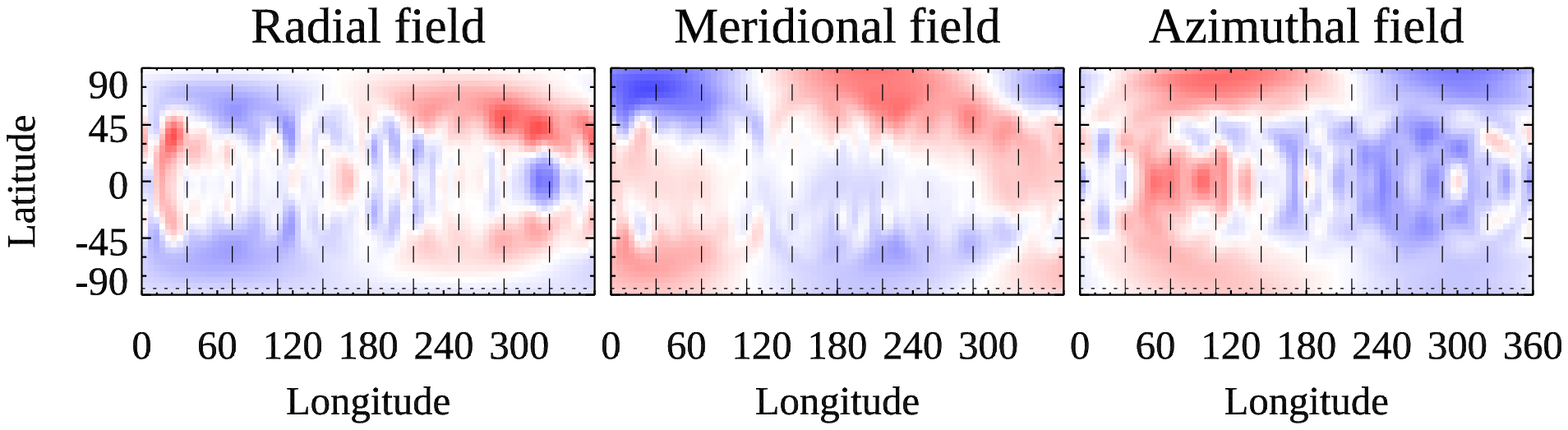}
\includegraphics[bb=240 230 370 790, width=2.05cm,clip,angle=90]{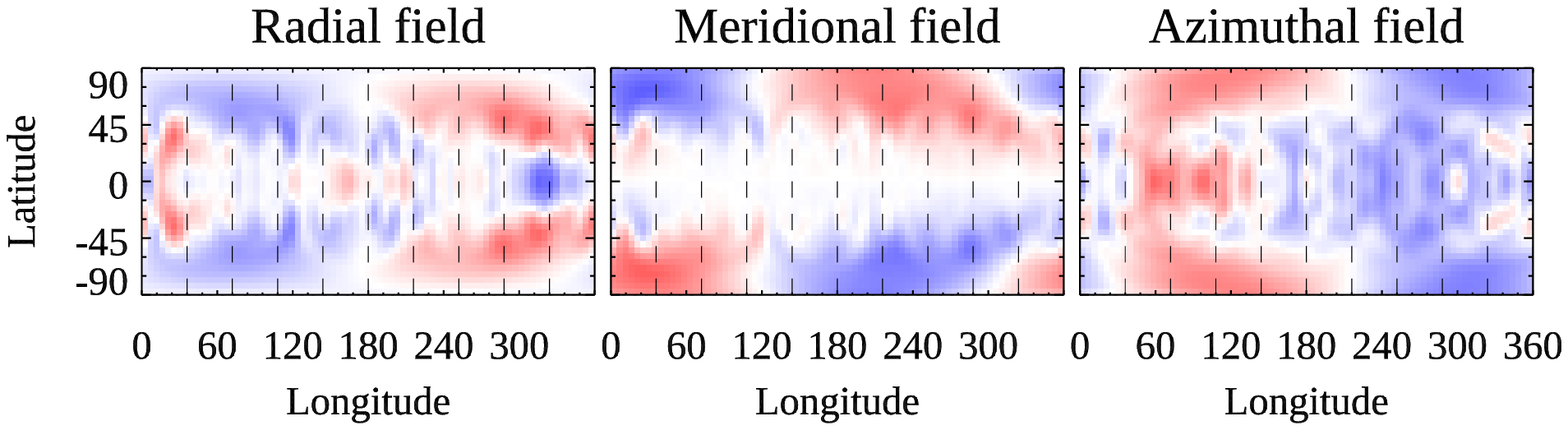}
\includegraphics[bb=260 770 410 810, width=3cm]{./scale_400.ps}
 \caption{Resulting ZDI map for Tests 13--22 using simulation L$^a$ at variable inclination with constant \vsini=40 km s$^{-1}$. The $i$ values are, from top to bottom, 10, 20, 30, 40, 50, 60, 70, 80, 85, and 90$\degree$.}
              \label{resmapLai}
    \end{figure}

\begin{figure}
   \centering
 \includegraphics[bb=240 230 390 790, width=2.4cm,clip,angle=90]{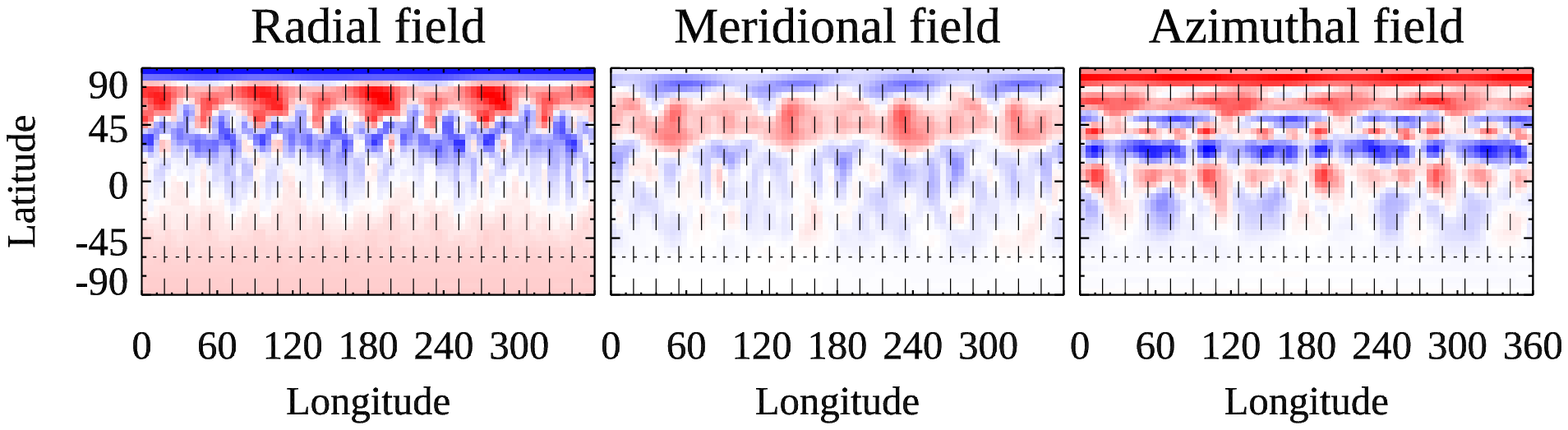}
 \includegraphics[bb=240 230 390 790, width=2.4cm,clip,angle=90]{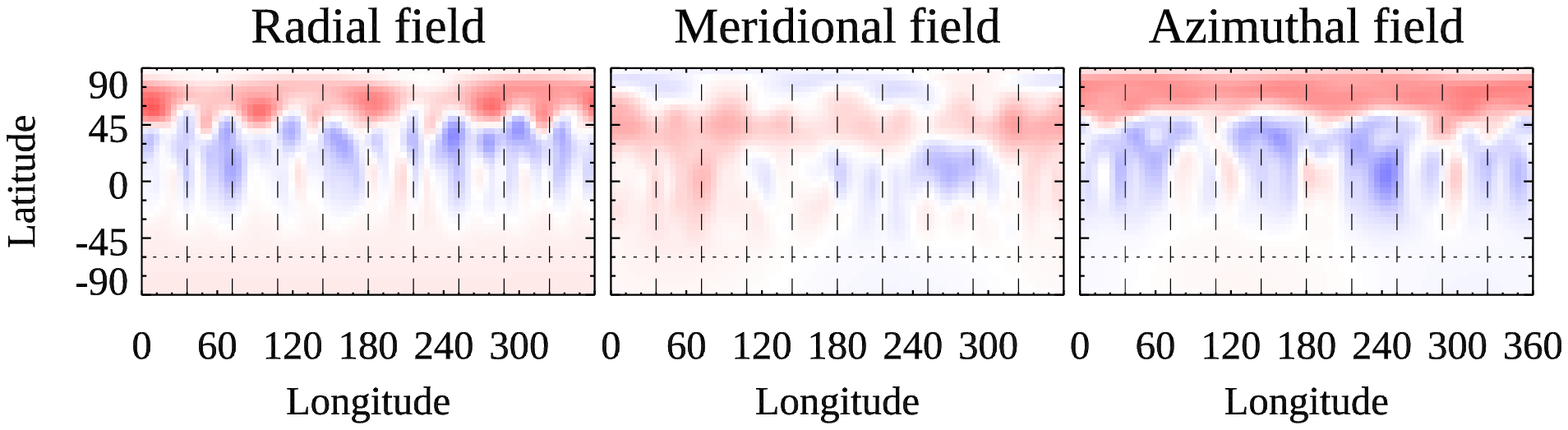}
 \includegraphics[bb=260 770 410 810, width=3cm]{./scale_300.ps}
\caption{Resulting ZDI for Tests 31 and 32 using simulation G$^W$. The difference in input is the number of `observed' phases 
($n_\phi=20$, $n_\phi=10$) and S/N ($S/N = 10^5$, $S/N=10^4$).}
              \label{resmapGW}
    \end{figure}

\section{Zeeman-Doppler imaging method}

The ZDI inversion code used in this study, {\sc InversLSD}, was introduced by
\cite{Kochukhov2014}. This code has been successfully applied to a number of
active late-type stars \citep{Kochukhov2015,Hackman2016,Rosen2016,Willamo2022}. 
The code is designed to model least-squares deconvolved \citep[see e.g.][LSD]{Donati1997} Stokes IV or IQUV profiles of photospheric absorption lines. Disk-integrated theoretical Stokes parameter spectra are calculated based on pretabulated grids of local line profiles computed for a set of continuum brightness values (normalised to 1), magnetic field strengths and orientations, and limb angles. Local line profiles can be computed using full polarised radiative transfer calculations \citep[e.g.][]{Kochukhov2014,Rosen2015} or using the analytical Unno-Rachkovsky solution \citep{Polarization2004} obtained under the assumption of Milne-Eddington atmospheres. Here, we use the latter approach, which does not rely on the assumption of a weak magnetic field. 

A detailed discussion of the method can be found in for example \citet{Rosen2016}. Of particular relevance for this study is the parameterisation of the stellar surface magnetic field. Both pixel-based specification of the three magnetic field vector components (radial, meridional, and azimuthal) and a description of these components in terms of a general harmonic expansion in poloidal and toroidal components are accepted as input formats in {\sc InversLSD} \citep[see][for details]{Kochukhov2014}. Here, we use the former (pixel-based) approach in the context of the forward calculations to simulate observational data while the latter approach (harmonic parameterisation) is employed for the inversions.

We used the same model to calculate synthetic spectra and retrieve the inverse solution. Therefore, systematic errors caused by the approximations in the line calculations become irrelevant. 
However, there is another source of systematic error; the ZDI solution is reconstructed in terms of a spherical harmonics expansion, and therefore a maximum angular degree $\ell_{\max}$ has to be set for the inversion. This is discussed further in section \ref{zdisetup}.

\section{Selected simulations and preprocessing}
\label{prep}

We selected three simulations from the study by \cite{Viviani2018} ---namely C1,
H$^a$, and L$^a$---
to investigate the impact of different stellar parameters. 
These three models represent a large range of rotation rate and activity and harbour three different and particularly interesting magnetic field solutions. 

The Run C1 has the highest rotation rate (1.8 solar), which still produces anti-solar-like differential rotation (i.e. slow equator, fast poles); it has a unique solution of a cyclic axisymmetric field in the anti-solar differential regime, which was studied in detail by \cite{Viviani2019}. For this run, the $m=0$ (axisymmetric) magnetic field mode is the strongest mode, with a cycle period of around 3.3 yr. Run H$^a$ has a higher rotation rate (7.8 solar) and exhibits a non-axisymmetric large-scale magnetic field, where the $m=1$ mode dominates. This mode is generated by a dynamo wave travelling in the retrograde azimuthal direction. Run L$^a$ has an even higher rotation rate (23.3 solar) and exhibits a strong $m=1$ mode, which is two orders of magnitude larger than the $m=0$ mode. This run shows the feature of a standing azimuthal dynamo wave.
From these runs, we chose a 
time point of the time series where the magnetic field growth has saturated and the properties of the dynamo and the magnetic field configuration no longer change significantly.
These three simulations were selected to represent three different levels of magnetic activity. The C1 data representing the lowest level of activity would in principle most closely match those for the Sun. However, these data cannot be seen as a model for the Sun, because for example the differential rotation is opposite to that of the Sun.

The simulations of \cite{Viviani2018} are based on the setup of \cite{Kapyla2013}, which uses the {\sc Pencil Code} \citep{PencilCode2021}.
The simulation scheme uses finite differences in spherical coordinates, due to which the grid cell dimensions decrease towards the poles. The time step, which is proportional to the grid-cell dimension, for integrating forward in time would correspondingly decrease, leading to unrealistically long computing times. Hence, the simulations leave out 
15$\degree$ of latitude around the 
poles. As this does not correspond to
the situation in real stars, we 
artificially stretched the magnetic field
data to cover these regions. 
Furthermore, the simulations do not describe the stellar surface properties correctly. 
The density stratification is much lower near the surface than in a 
Sun-like
star, and the magnetic field is set to be purely radial at the $r=R_{\star}$ stellar surface.
Therefore, we take only the magnetic field data from a slice just below the surface ($r=0.98 R_{\star}$), where the influence of the boundary conditions is not strong and all magnetic field components are present. The simulations do not provide an absolute scale of the magnetic field strength, as they deal with non-dimensional numbers. However, they do allow a unit system to be selected so that the simulated magnetic field strength can be matched with the observed ones. The simulations units are scaled to physical units by assuming that the simulations rotate with 1.8, 7.8, and 23.3 times the solar value, that the density at the bottom of the convection zone is the solar one, and that the radial extent of the domain is the size of the solar convection zone (0.3 $R_{\star}$); see  \cite{Viviani2018} for details. 
To compensate for the missing density stratification near the surface, we scale the surface magnetic field strength down to values comparable to 
observations.

The geometry of the simulated
magnetic field 
data
did not correspond to the input format for the ZDI code.
The 
grid dimensions of the C1 
simulated data were
$288 \times 576$ 
in latitude and longitude, respectively,
and $512 \times 1024$
for the H$^a$ and L$^a$. The 
data
were first down-sampled to half of the grid size, and were then converted
to a Mollweide-type format with 20862 surface elements of approximately equal area \citep[see][]{Piskunov2002} using bilinear interpolation.

The change in geometry slightly reduced the root-mean-square (RMS) values of the magnetic field components, as small structures were smoothed,  especially those at high
latitudes. Otherwise, the preprocessing did not significantly alter the field topology, except that it led to the regions around the rotation poles being filled.
    
The parameters for the preprocessed simulated magnetic field data are listed in Table~\ref{mhdpar}. The magnetic field is characterised by the RMS values of the total, radial, meridional, and azimuthal field ($B_\mathrm{RMS}$, $B^r_\mathrm{RMS}$, $B^\theta_\mathrm{RMS}$ and $B^\phi_\mathrm{RMS}$). 
The magnetic field topology is characterised by the percentages of the poloidal ($p_\mathrm{pol}$) and axisymmetric ($p_\mathrm{axi}$) components in terms of magnetic energy. 
In Table~\ref{mhdpar}, we also list the total RMS field, as well as the poloidal and axisymmetric magnetic energy fractions for spherical harmonics decompositions of the C1, H$^a$, and L$^a$ maps reconstructed to the same $\ell_{\max}$ = 20 as used in some of the ZDI inversions. These data were preprocessed in the same way as the original simulations, after which the magnetic field characteristics where calculated. These values are marked with the upper index `$\ell \le 20$' and can be compared with the results reported in Section \ref{results}.

As is evident from Table \ref{mhdpar}, all three simulations represented a relatively narrow range in terms of axisymmetry. To test the performance of the ZDI inversion in a more axisymmetric case, we added the wedge-simulation G$^W$ from \cite{Viviani2018} by replicating it to $360 \degree$ in longitudes. In other respects, these data were preprocessed in the same way as the C1, H$^a$, and L$^a$ data. For the G$^W$ case, we only tested the impact of the observational S/N and phase coverage.
The input magnetic field data are plotted in
Fig.~\ref{origmap}.

\section{Simulated observations and ZDI setup}
\label{zdisetup}

Synthetic Stokes IV spectropolarimetric observations were calculated for the four different MHD simulations using a set of different rotation velocities \vsini, inclinations $i,$ and $S/N$ values. For the C1, L$^a$, and H$^a$ data, we initially tested reconstruction with both $n_\phi=10$ and $n_\phi=20$ rotation phases evenly distributed between 0 and 1. We also tested the impact of noise by using $S/N = 10^4, 5 \cdot 10^4$, and $10^5$. For the C1 map, we noticed significant differences when using only ten phases and the lowest $S/N$ compared to the denser phase grid and highest $S/N$. Therefore, the final calculations were done with the higher values. For the tests with the H$^a$ and L$^a$, we used $n_\phi=10$ and $S/N=10^4$, except when testing different $i$ values. For the G$^W$ data, we fixed the inclination and rotation velocity to $i=60 \degree$ and \vsini = 40 km s$^{-1}$ and only tested two different cases for $S/N$
and $n_\phi$.

The tested \vsini~range was 5 -- 60 km s$^{-1}$. In most tests, the inclination was set to $i = 60 \degree$. This combination would correspond to rotation periods of 0.7 d < $P_\mathrm{rot}$ < 8.8 d.
This is within a typical range for stars studied with the ZDI method, such as AB Dor ($P_\mathrm{rot}\approx 0.5$ d, \vsini $\approx$ 70 km s$^{-1}$) and $\kappa^1$ Cet ($P_\mathrm{rot} \approx 9.2$ d, \vsini $\approx$ 5.2 km s$^{-1}$).
The \vsini~range for the C1 reconstruction differed slightly from the other reconstructions, as C1 represents a more slowly rotating case. The effect of $i$ was tested only for the L$^a$ simulation. This choice was based on the fact that L$^a$ represents the most rapidly rotating model, meaning that this would be the most optimal for ZDI. In order to reduce the influence of random errors, we used $S/N = 5 \cdot 10^4 $ for these cases. The tested range in inclination was $10 \degree - 90 \degree$, which combined with \vsini = 40 km s$^{-1}$ would correspond to $P_\mathrm{rot}$ values of between 0.2 d and 1.3 d.

The synthetic Stokes IV profiles were calculated using the same procedure as in \cite{Hackman2016}. The intrinsic spectral line width was set to 3 km s$^{-1}$. This is a combination of both the physical FWHM of the local line profile and the spectral resolution, and would correspond to a spectral resolution of $R \sim 100 000$. For the spectral parameters, such as the central line depth and effective Land\'e factor $g$$_\mathrm{eff}$, we used the same values as in \cite{Willamo2022b}; for example $g_\mathrm{eff}=1.215$.

Two different values for $\ell_{\max}$ were tested. For \vsini\,$\le 20$ {km s$^{-1}$} $\ell_{\max} = 20$ was found to be sufficient as the contribution from degrees $\ell > 20$ was negligible. For  \vsini\,$\ge 40$ {km s$^{-1}$,} this was not the case, and we adopted a higher value of $\ell_{\max} = 30$.

The regularisation of the ZDI inversion was done in the same way as described in \cite{Rosen2016}. 
Higher degrees of $\ell$ were penalised using the regularisation function 

\begin{equation}
    R = \Lambda \Sigma_{\ell,m} \ell^n(\alpha_{\ell,m}^2 + \beta_{\ell,m}^2 + \gamma_{\ell,m}^2),
\label{reg}
\end{equation}

\noindent
where $\alpha_{\ell,m}$, $\beta_{\ell,m}$, and  $\gamma_{\ell,m}$ are the harmonic expansion coefficients, $m$ is the azimuthal order, and $\Lambda$ is the regularisation parameter. For the exponent of $\ell$ in the regularisation, we mostly used the `standard' value of $n=2$ (Tests 1--22 and 31--32 in Table \ref{zdires}). For some of the runs, we also tried the exponent $n=1$ in order to also explore this parameter (Tests 23--30).

As no temperature maps were used in the process, we only used the Stokes V profiles for the reconstruction. Magnetic fields will clearly also alter the Stokes I profiles, but in our cases the magnetic fields were too small to induce any significant effect. This was verified by some initial trials for the C1, H$^a$, and L$^a$ data; inclusion of the Stokes I profiles did not have a notable effect on the result in any of these, other than increasing the computation time.

The inversions were started with an initial value of $\Lambda \sim 10^{-8} - 10^{-7}$ (Eq. \ref{reg}).
The value of $\Lambda$ was gradually reduced, until we reached a $\chi^2$ value of the Stokes V-profiles corresponding to the level of the induced noise in the synthetic data. With real observations, aiming to achieve a reduced $\chi^2 \sim 1$ is seldom a reasonable approach. Firstly, modelling of the Stokes I and V spectra will contain systematic errors. Secondly, the $\ell$-degrees are limited  in
the inversion. 

The value of $\ell_{\max}$ is equivalent to the surface resolution of the solution.
The resolution will be dependent on the projected rotation velocity of the star combined with the intrinsic line profile. The latter will be a combination of the instrumental spectral resolution and local line-broadening in the stellar photosphere.  However, there is no straightforward way to estimate a `correct' value for $\ell_{\max}$, as the rotational resolution varies with the latitude combined with the inclination $i$. Furthermore, the rotational phase coverage will also play a role, in addition to the $S/N$. 
The usual way to set $\ell_{\max}$ is to try out different values, and find the value at which the remaining magnetic field energy starts to become negligible.
Also,  there is a problem with synthetic data, as the forward modelling does not limit the
$\ell$-degrees. In practice, this only becomes an issue if the $\ell_{\max}$ value is set to an excessively low value compared to the theoretically achievable image resolution.
Therefore, our target for the reduced $\chi^2$ is defendable.
With real observations, one can use a pragmatic approach, and stop the inversions at a level where the decrease in $\chi^2$ reaches a flat stage as a function of decreasing $\Lambda$ \citep{Kochukhov2017}.

\section{Results and discussion}
\label{results}

The results are listed in Table \ref{zdires} and the ZDI maps are plotted in Figs. \ref{resmapC1} -- \ref{resmapGW}. In addition to the tests listed in the table, we experimented with different values of $S/N$, $\ell_{\max}$, $n_{\phi}$, and the exponent $n$ of the regularisation function (Eq. \ref{reg}). The results of these tests are not reported in this study, as they were rather used for choosing reasonable values for the different parameters.

\begin{figure}
\centering
 \includegraphics[width=3.4cm]{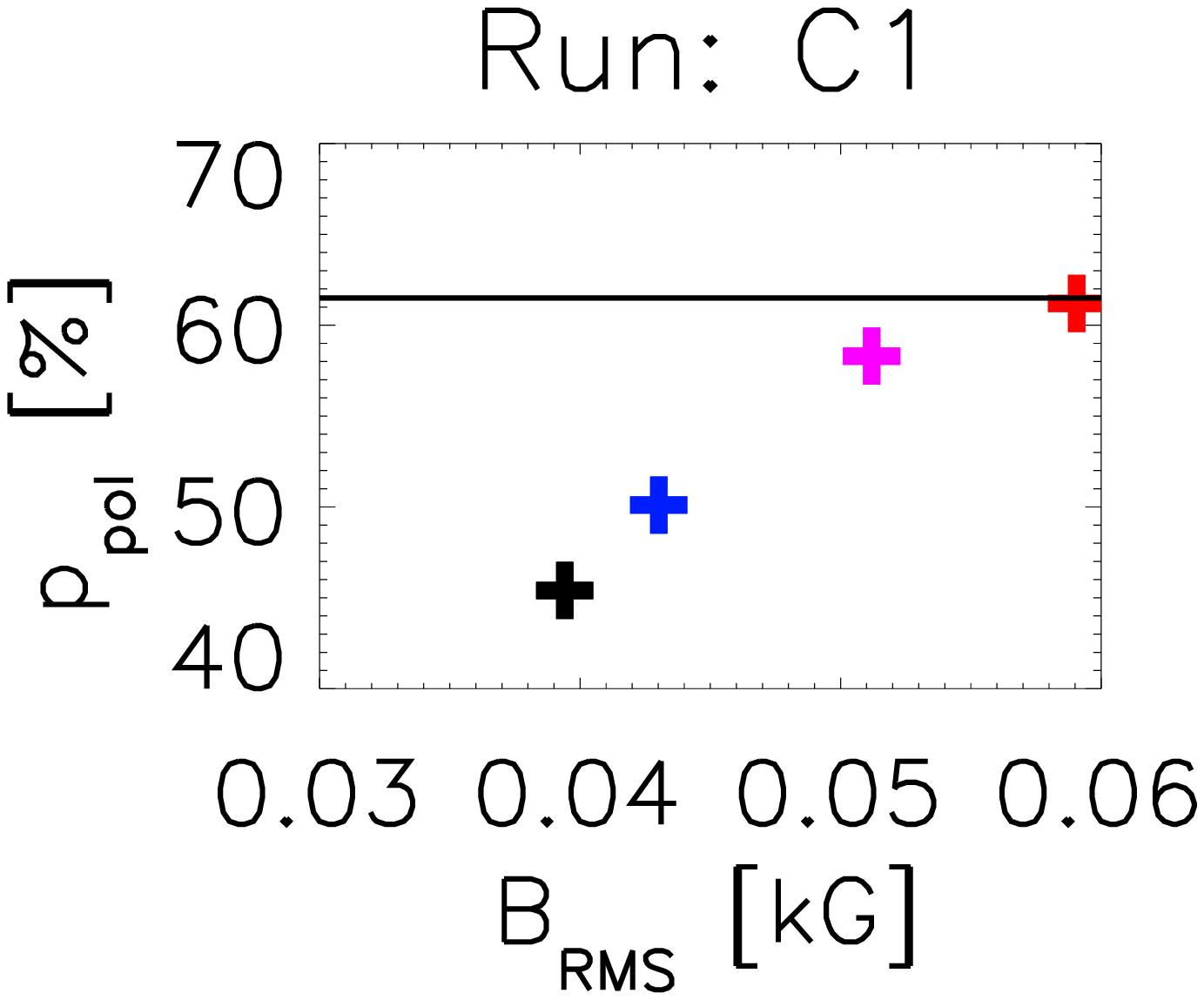} 
\hspace{-0.8cm}
 \includegraphics[width=3.4cm]{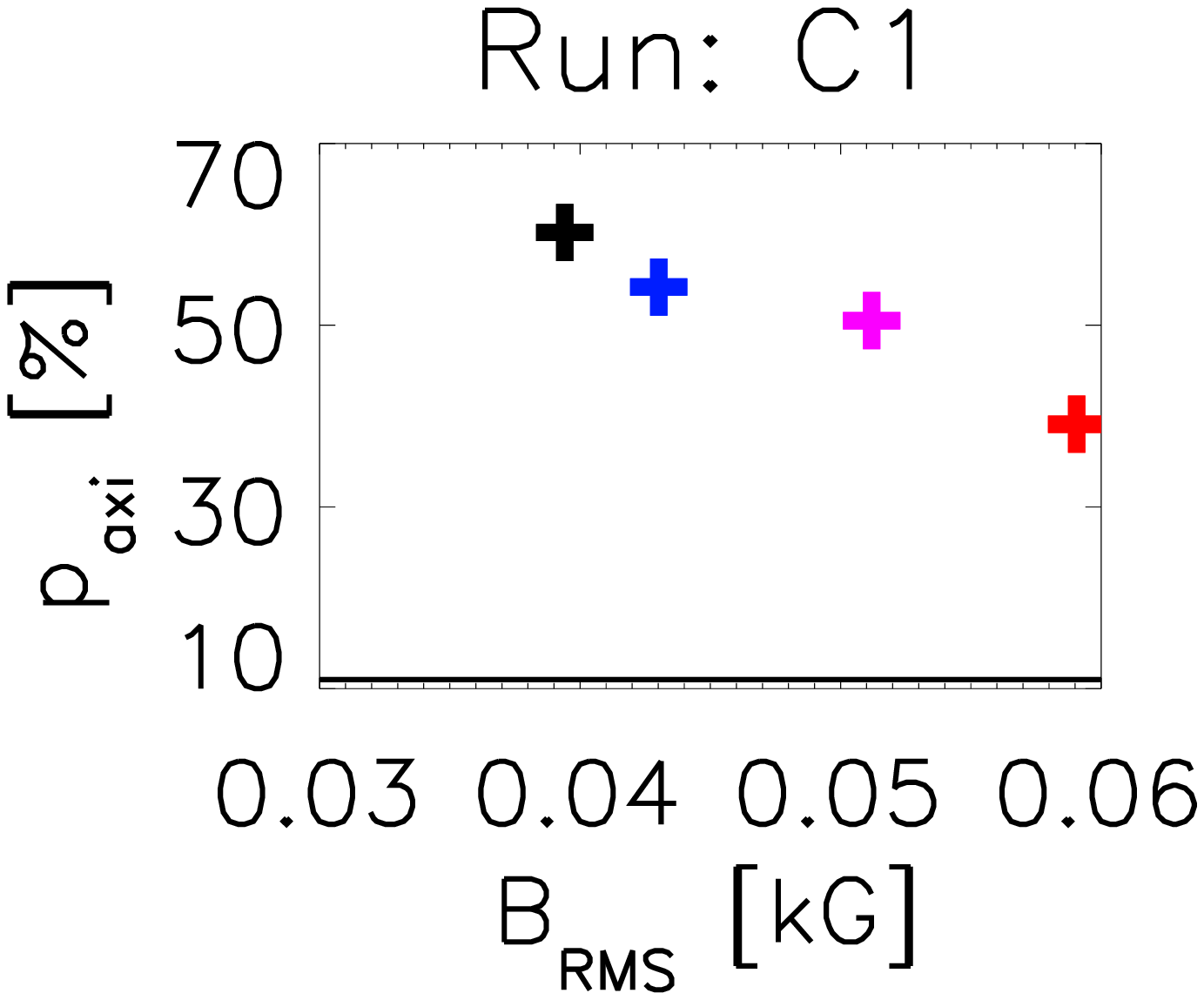} 
\hspace{-0.8cm}
 \includegraphics[width=3.4cm]{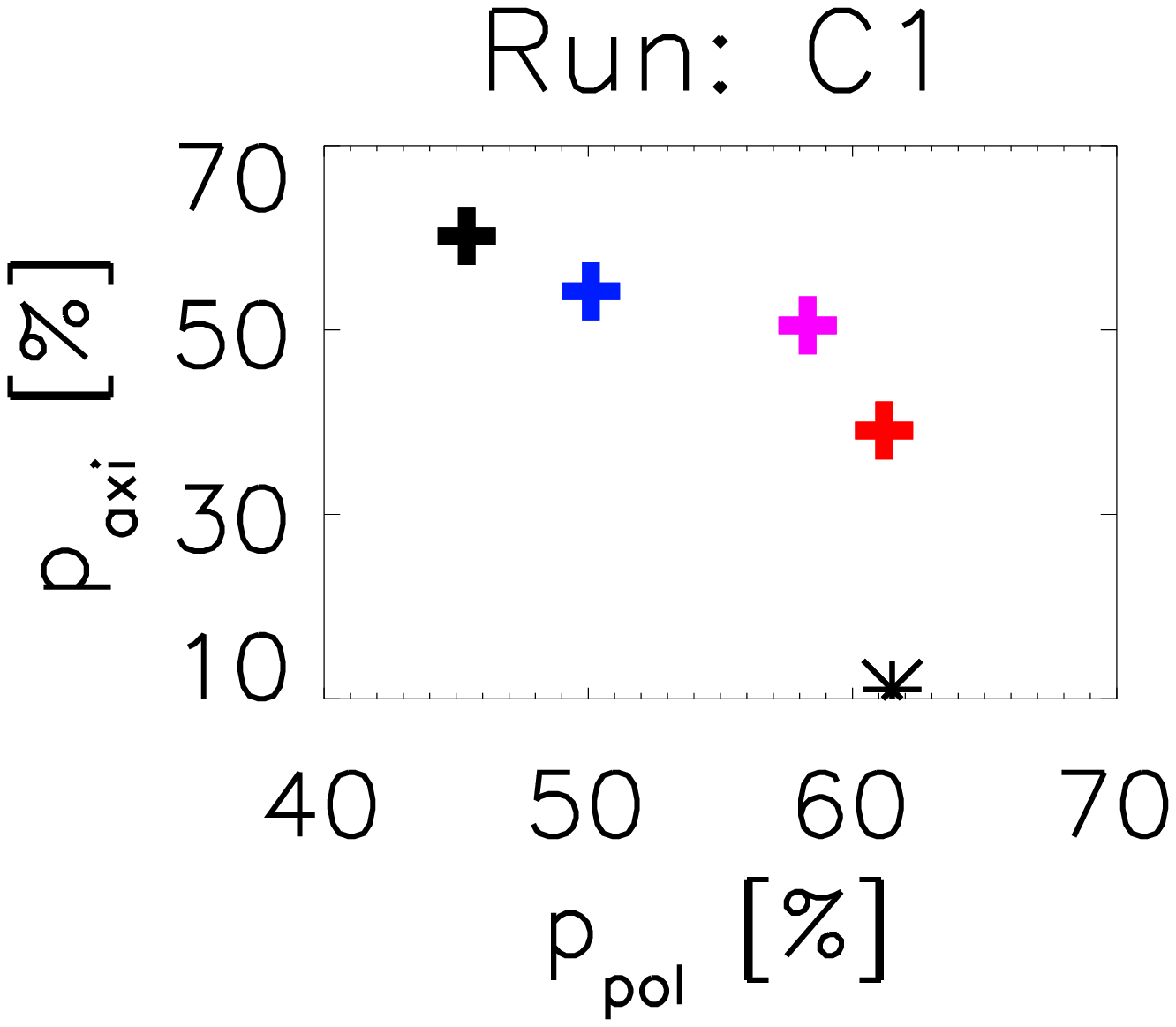}
\hspace{-0.65cm}
 \includegraphics[width=3.4cm]{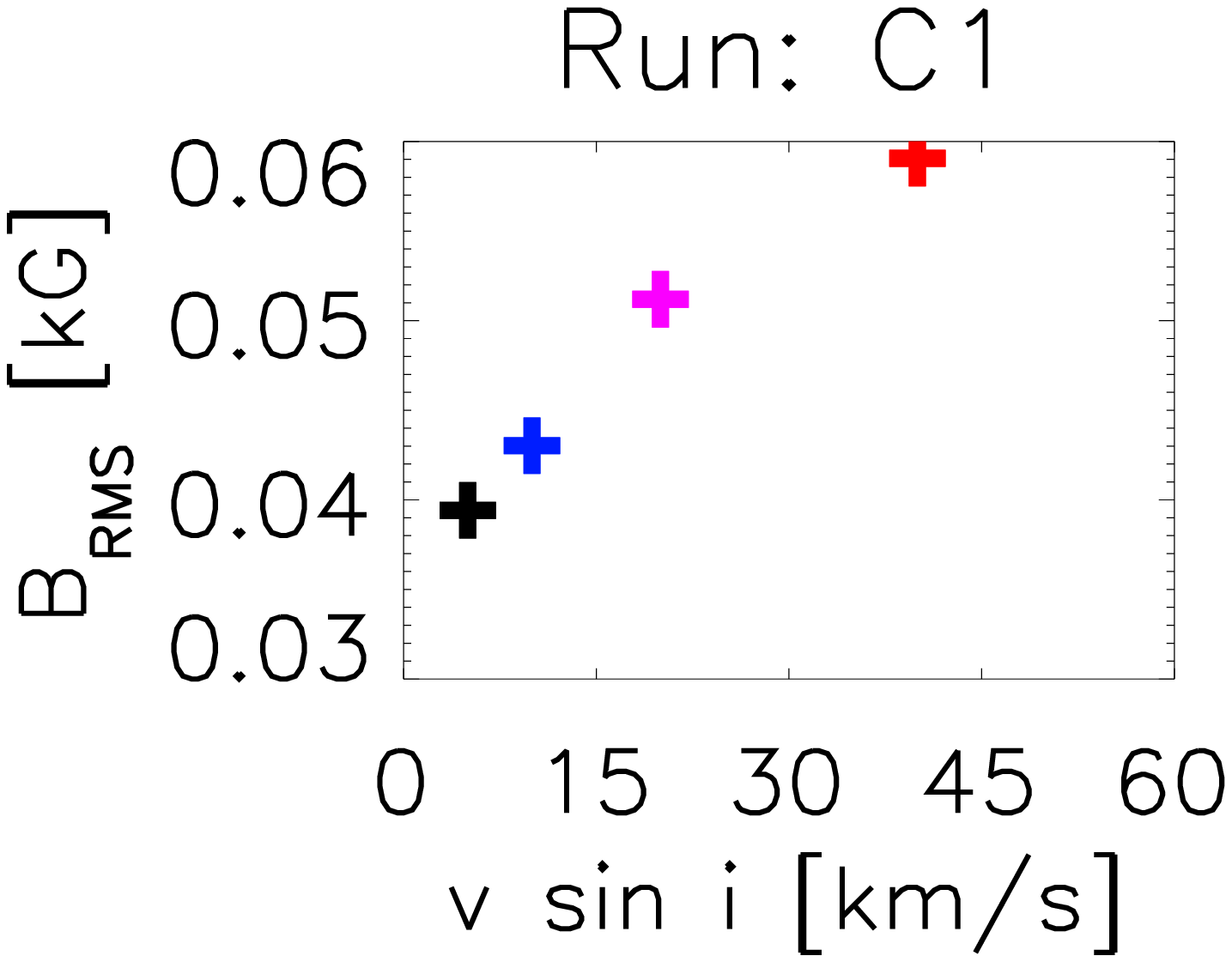}
\hspace{-0.8cm}
 \includegraphics[width=3.4cm]{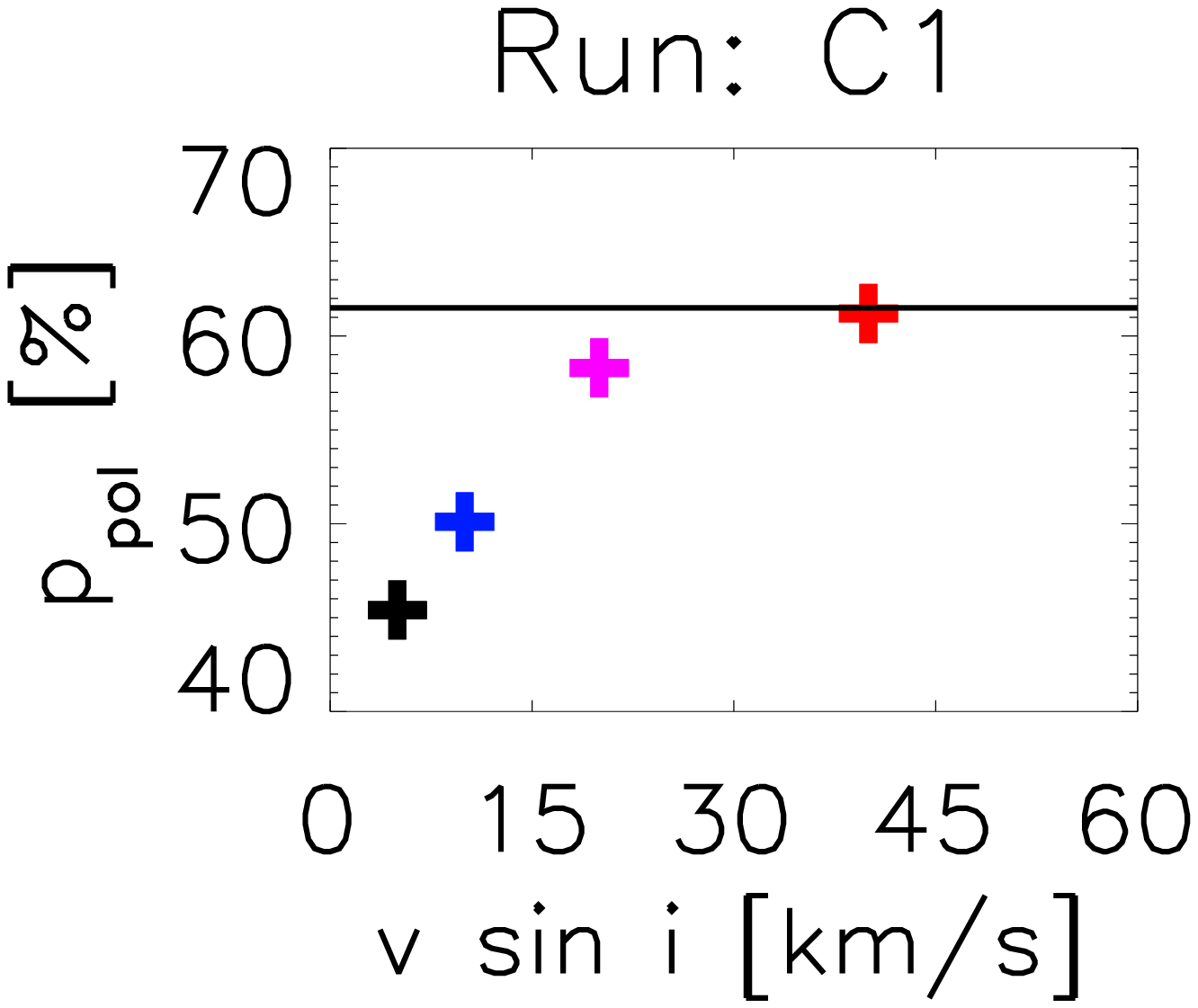}
\hspace{-0.8cm}
\includegraphics[width=3.4cm]{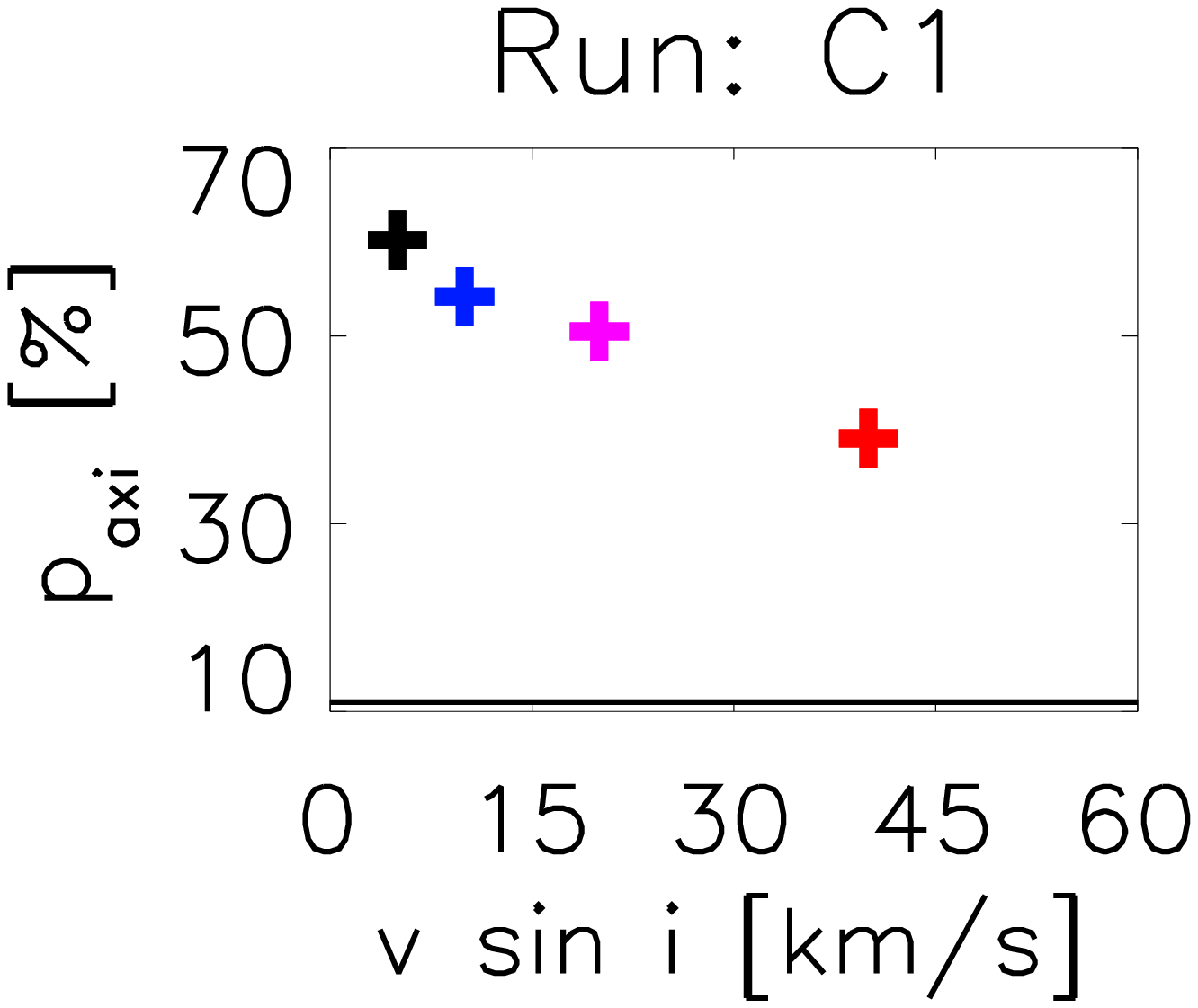}
 \caption{Results for Tests 1--4 using the simulation C1 showing dependencies of magnetic field characteristics and \vsini. The asterisk and horizontal lines mark the original values. The different colours are for different values of \vsini:~ in ascending order, black (5 km s$^{-1}$), blue (10 km s$^{-1}$), pink (20 km s$^{-1}$), and red (40 km s$^{-1}$).}        
 \label{C1_res}
    \end{figure}

\begin{figure}
   \centering
 \includegraphics[width=3.3cm]{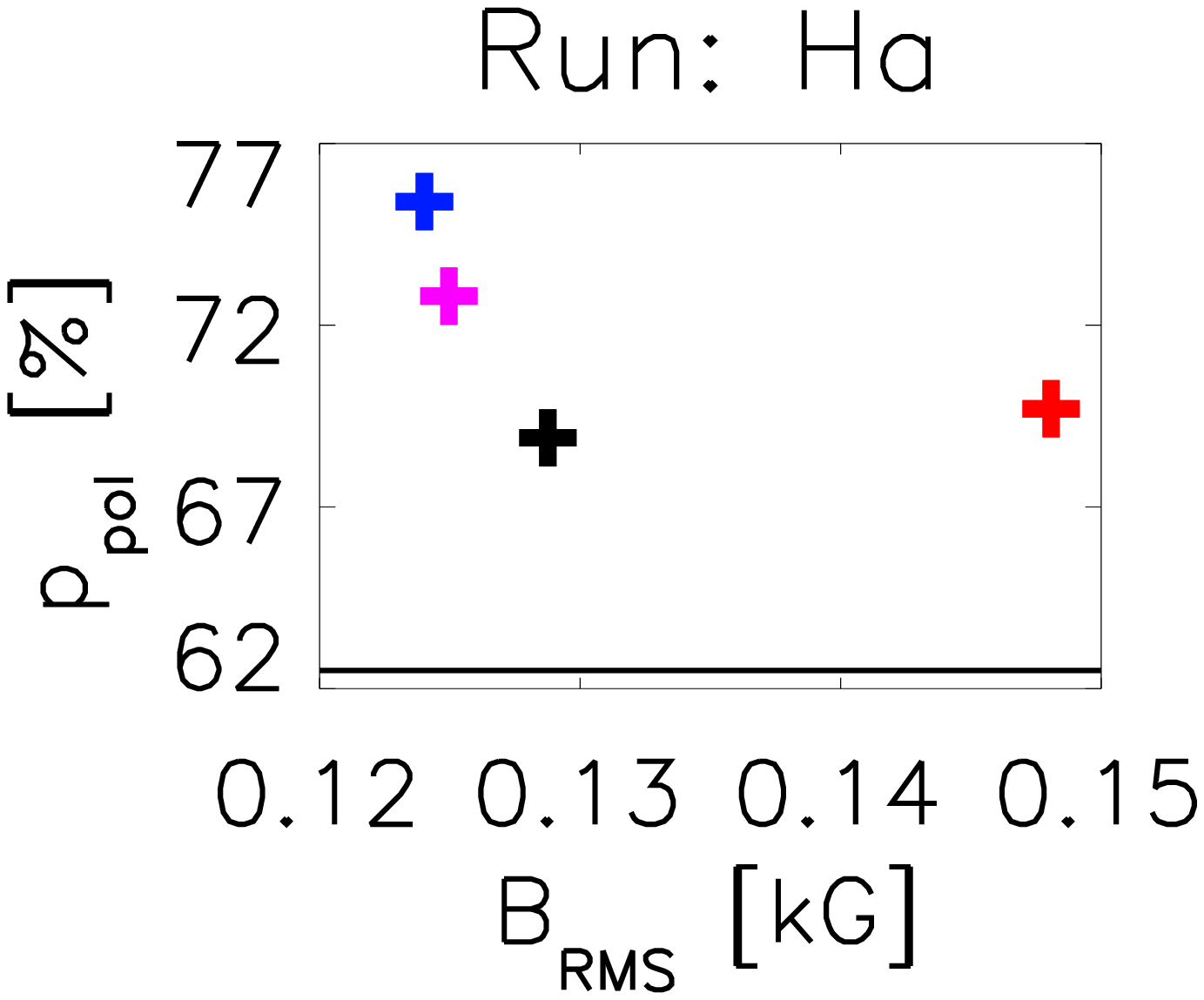} 
\hspace{-0.8cm}
 \includegraphics[width=3.3cm]{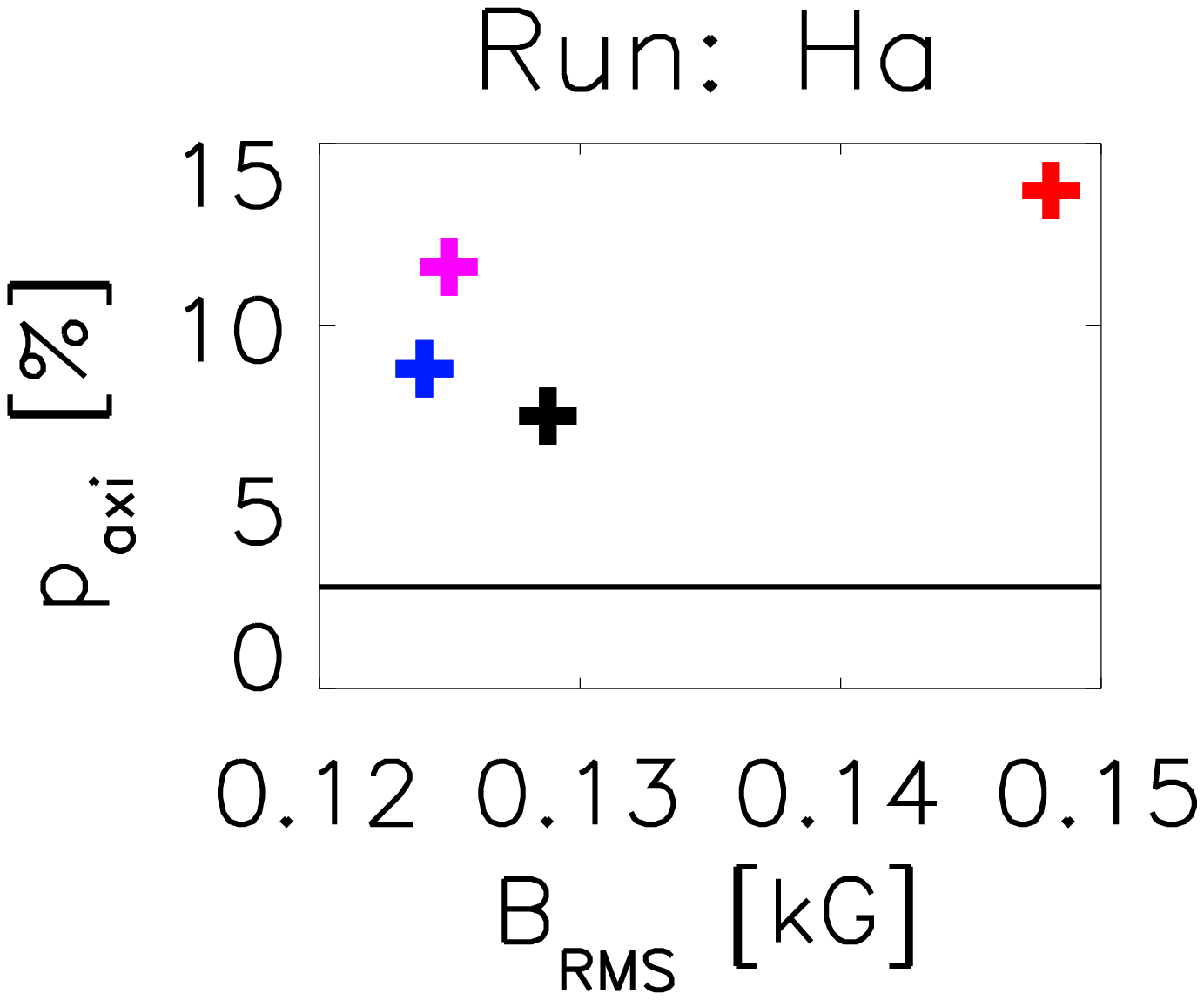}
\hspace{-0.8cm}
 \includegraphics[width=3.3cm]{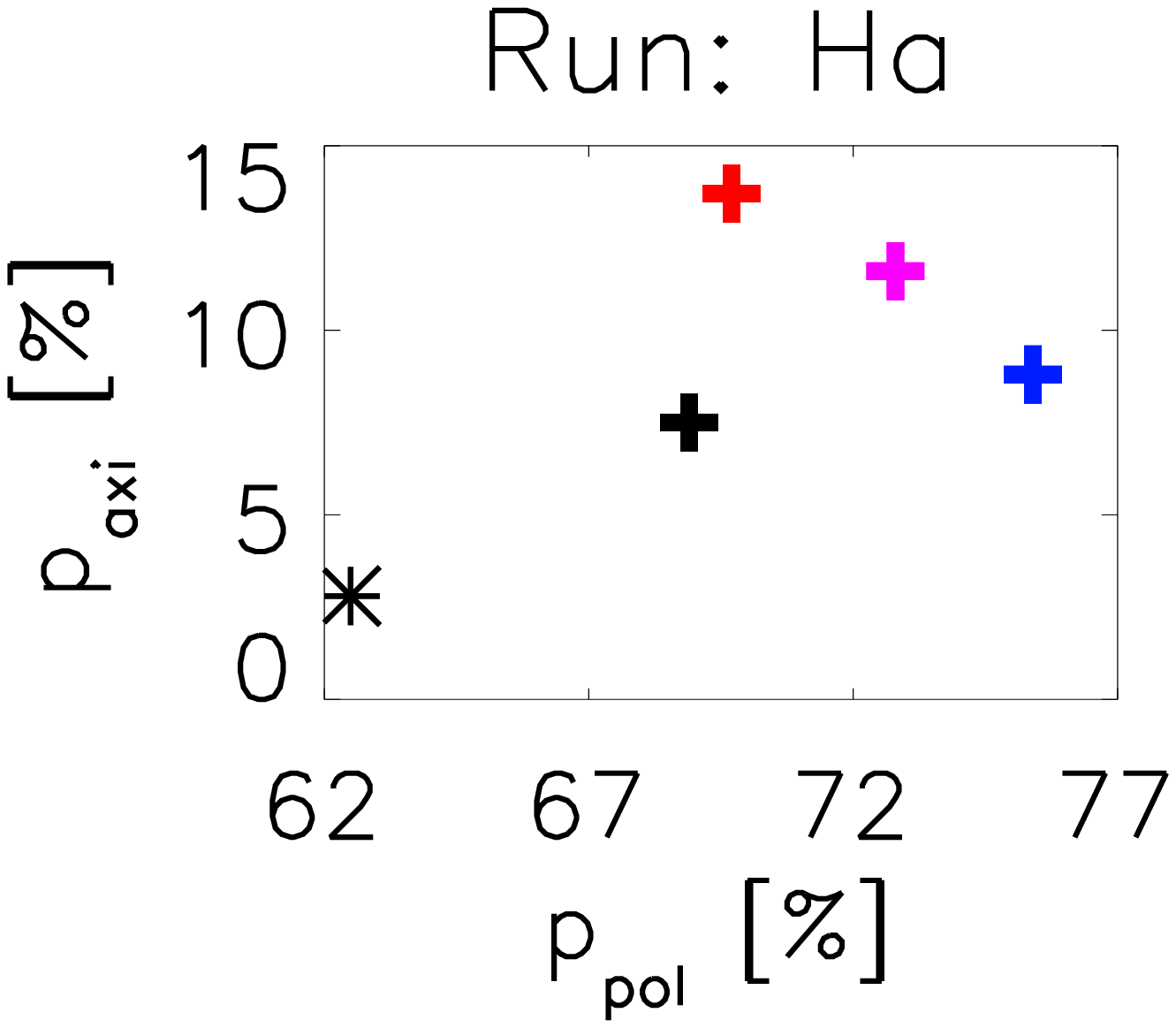}
\hspace{-0.8cm}
 \includegraphics[width=3.3cm]{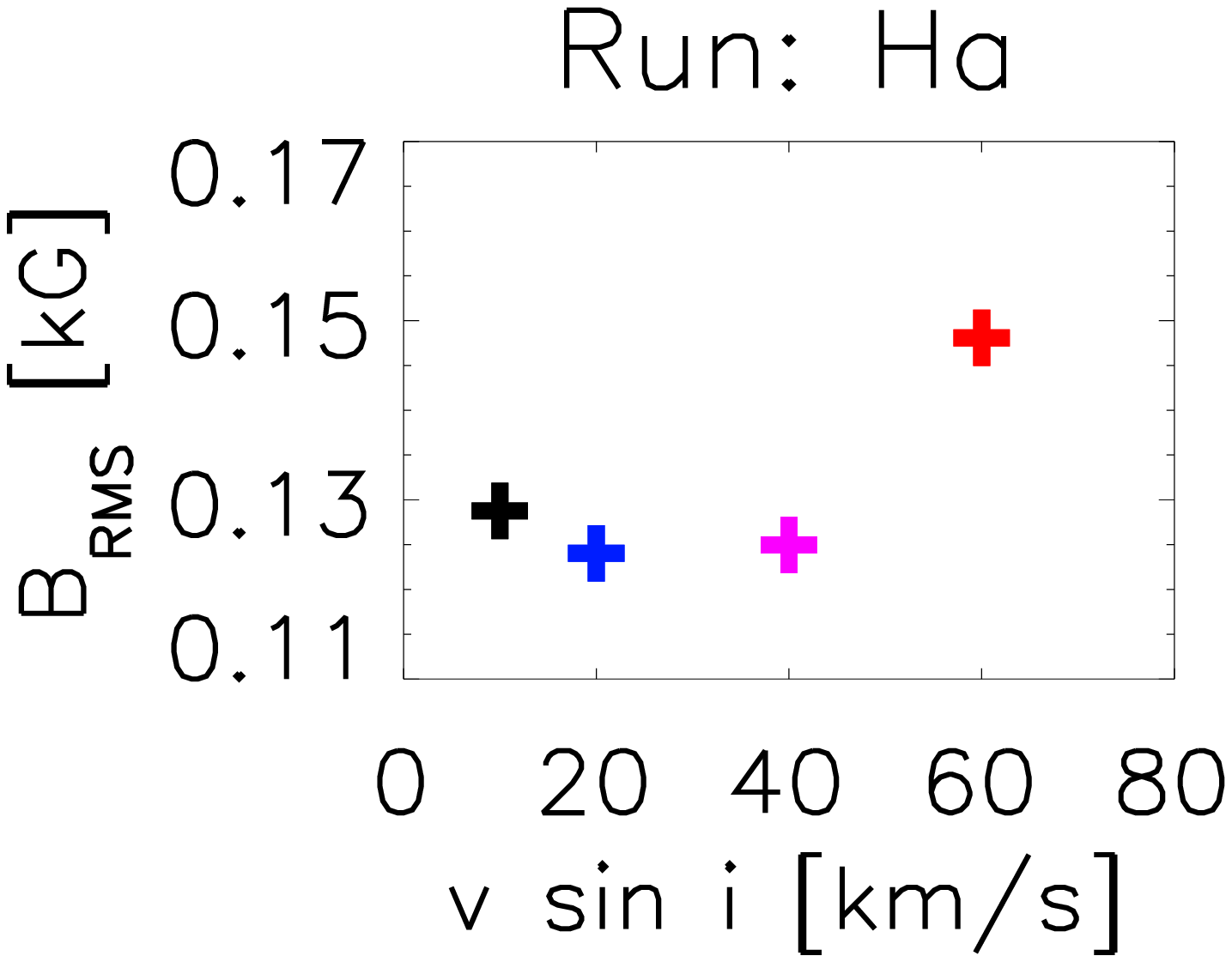}
\hspace{-0.8cm}
 \includegraphics[width=3.3cm]{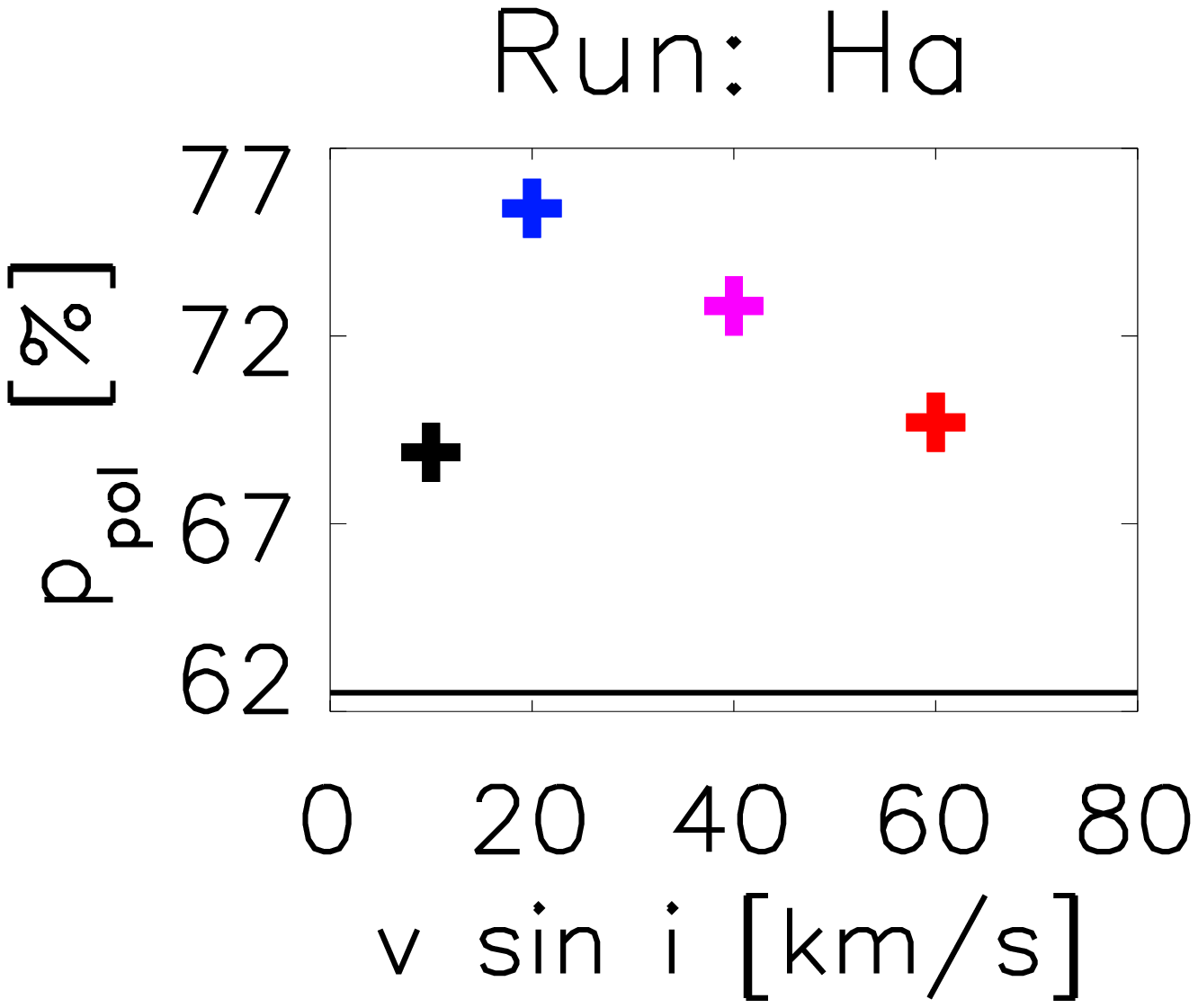}
\hspace{-0.8cm}
 \includegraphics[width=3.3cm]{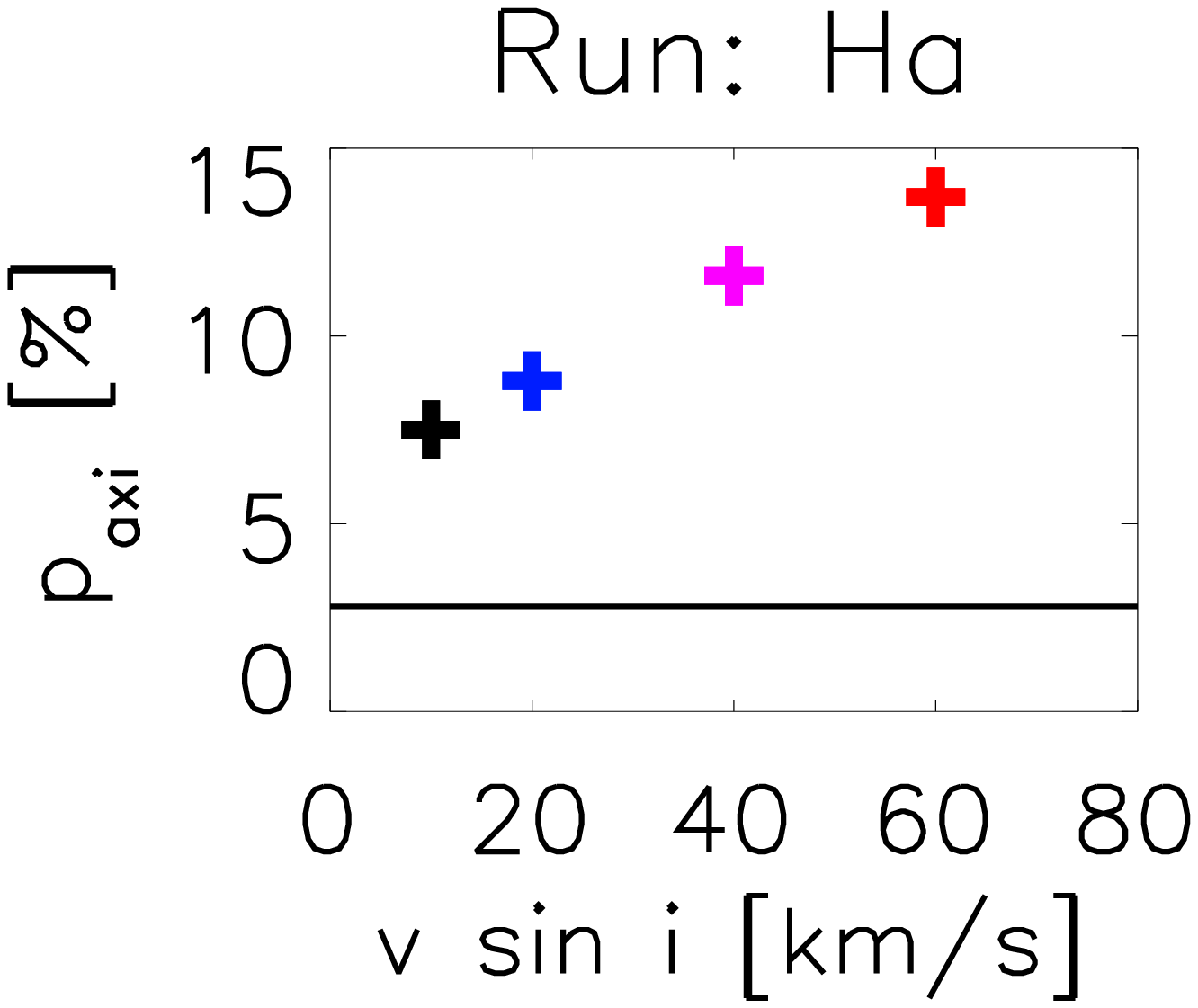}
 \caption{Results for Tests 5--8 using the simulation H$^a$. Markings are the same as in Fig. \ref{C1_res} except that the \vsini~ values are 10 (black), 20 (blue), 40 (pink), and 60  km s$^{-1}$ (red).}             
 \label{Ha_res}
    \end{figure}

\begin{figure}
  \centering
 \includegraphics[width=3.3cm]{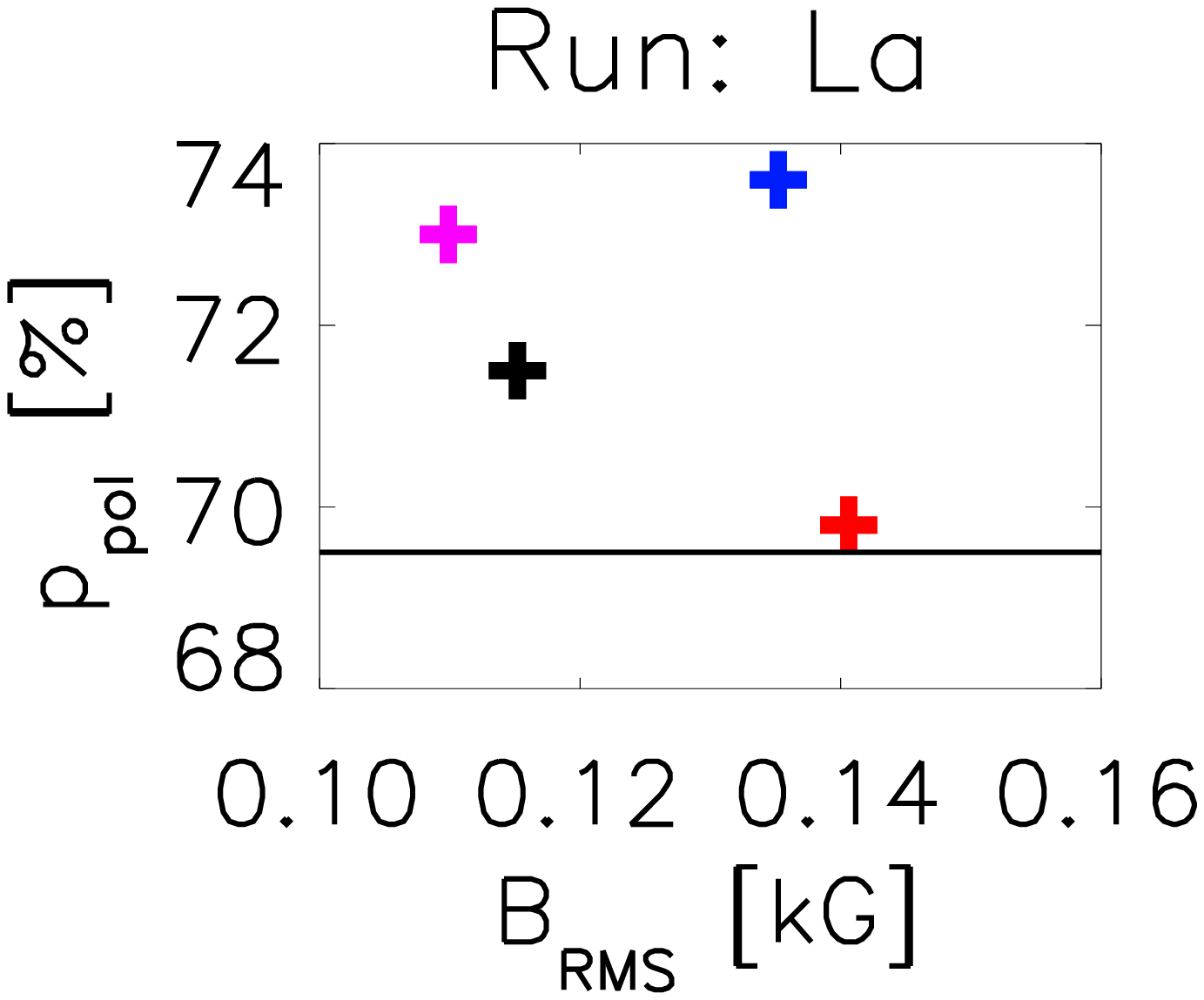} 
\hspace{-0.8cm}
 \includegraphics[width=3.3cm]{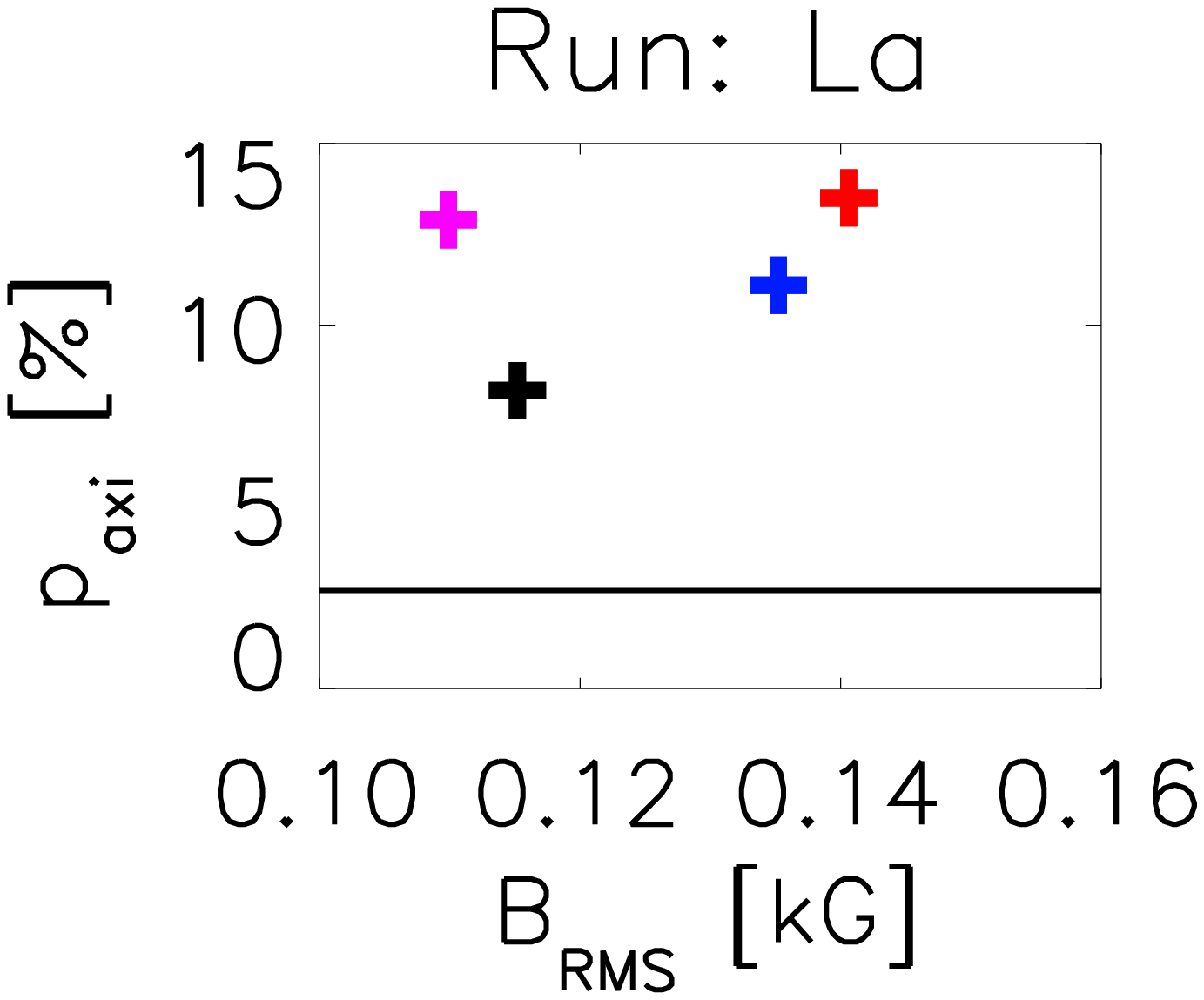}
\hspace{-0.8cm}
 \includegraphics[width=3.3cm]{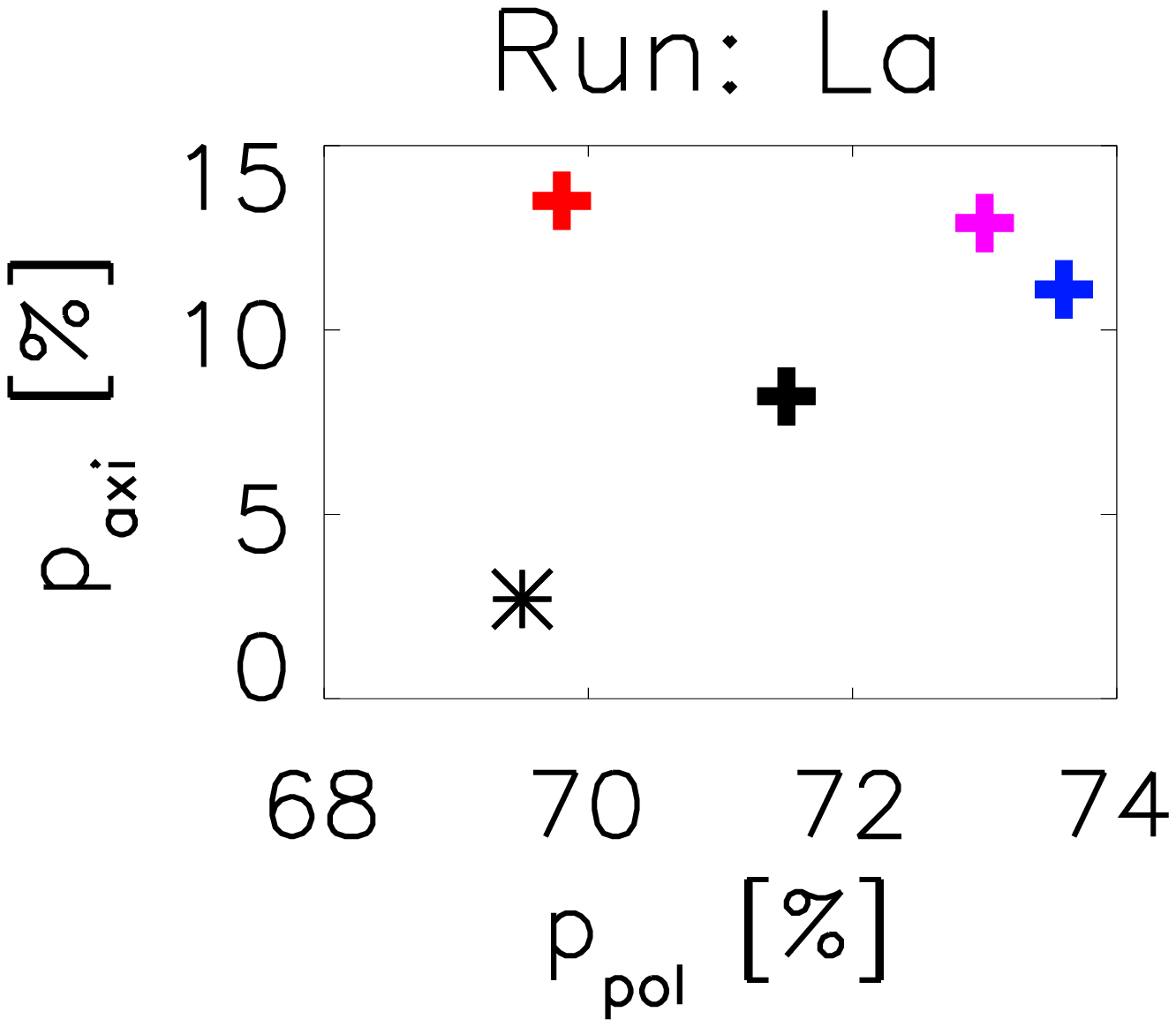}
\hspace{-0.65cm}
 \includegraphics[width=3.3cm]{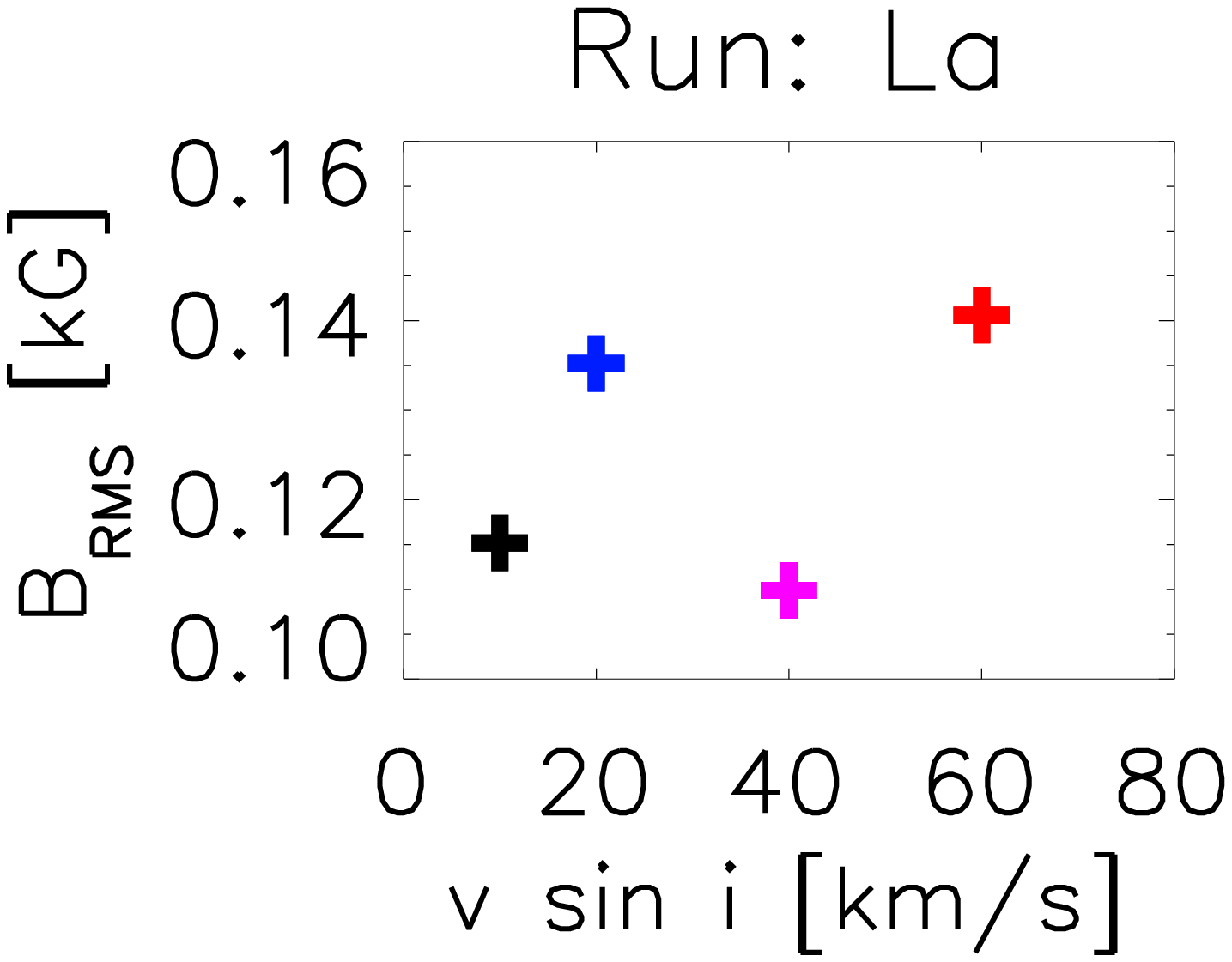}
\hspace{-0.8cm}
 \includegraphics[width=3.3cm]{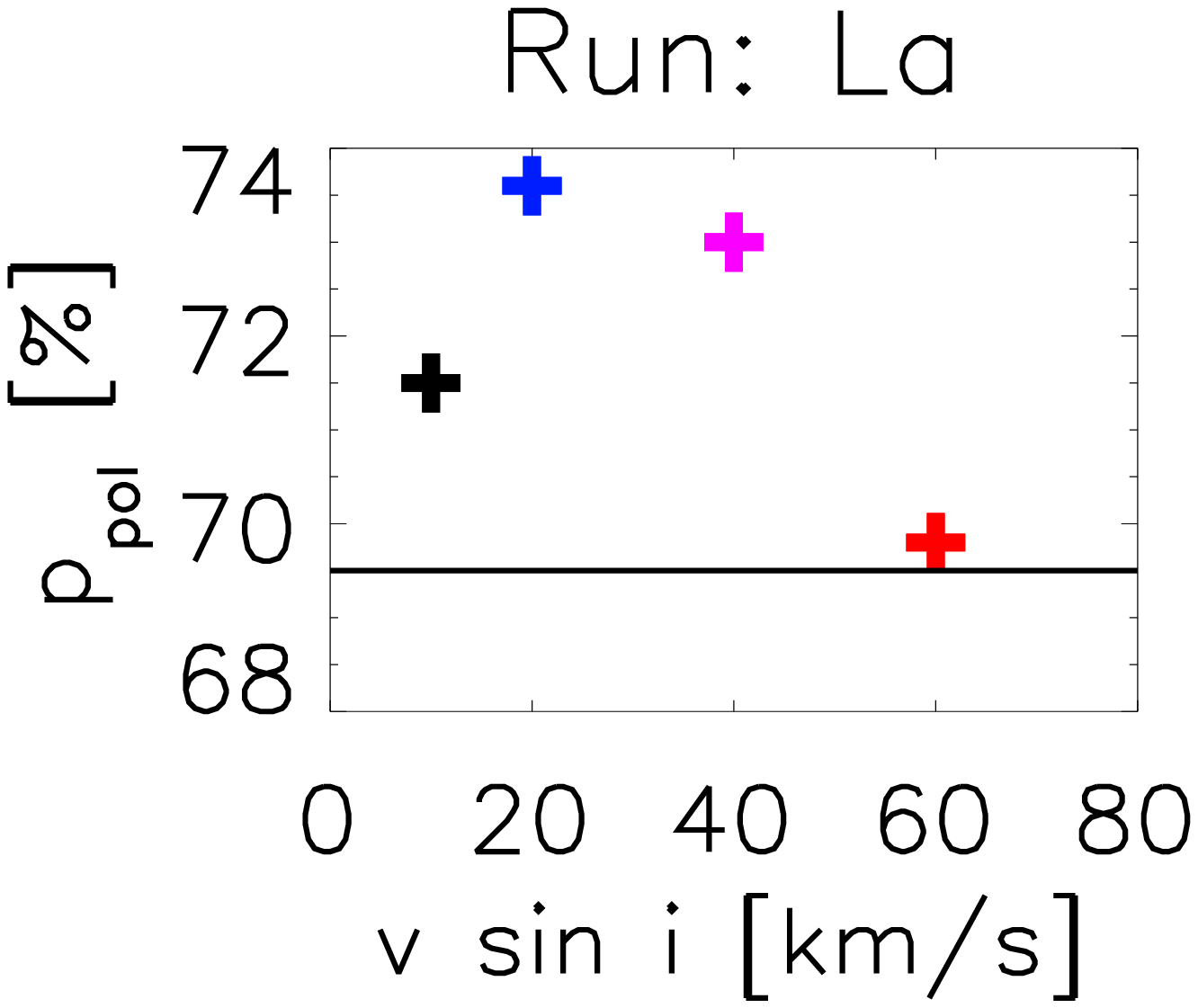}
\hspace{-0.8cm}
\includegraphics[width=3.3cm]{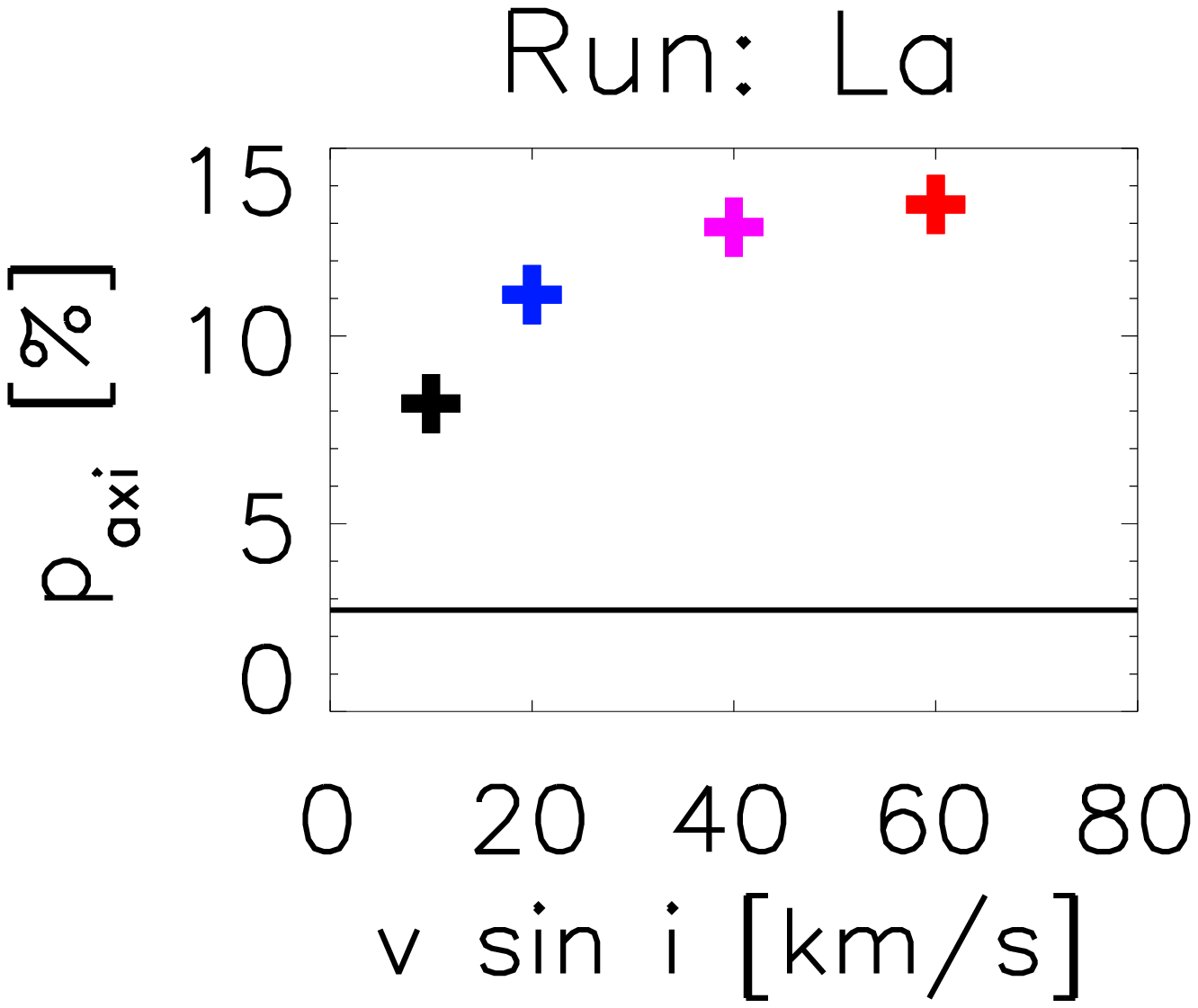}
  \caption{Results for Tests {\ 9--12} using the simulation L$^a$. Markings are the same as in Fig. \ref{Ha_res}.}             
  \label{La_res}
    \end{figure}

\begin{figure}
   \centering
 \includegraphics[width=4.7cm]{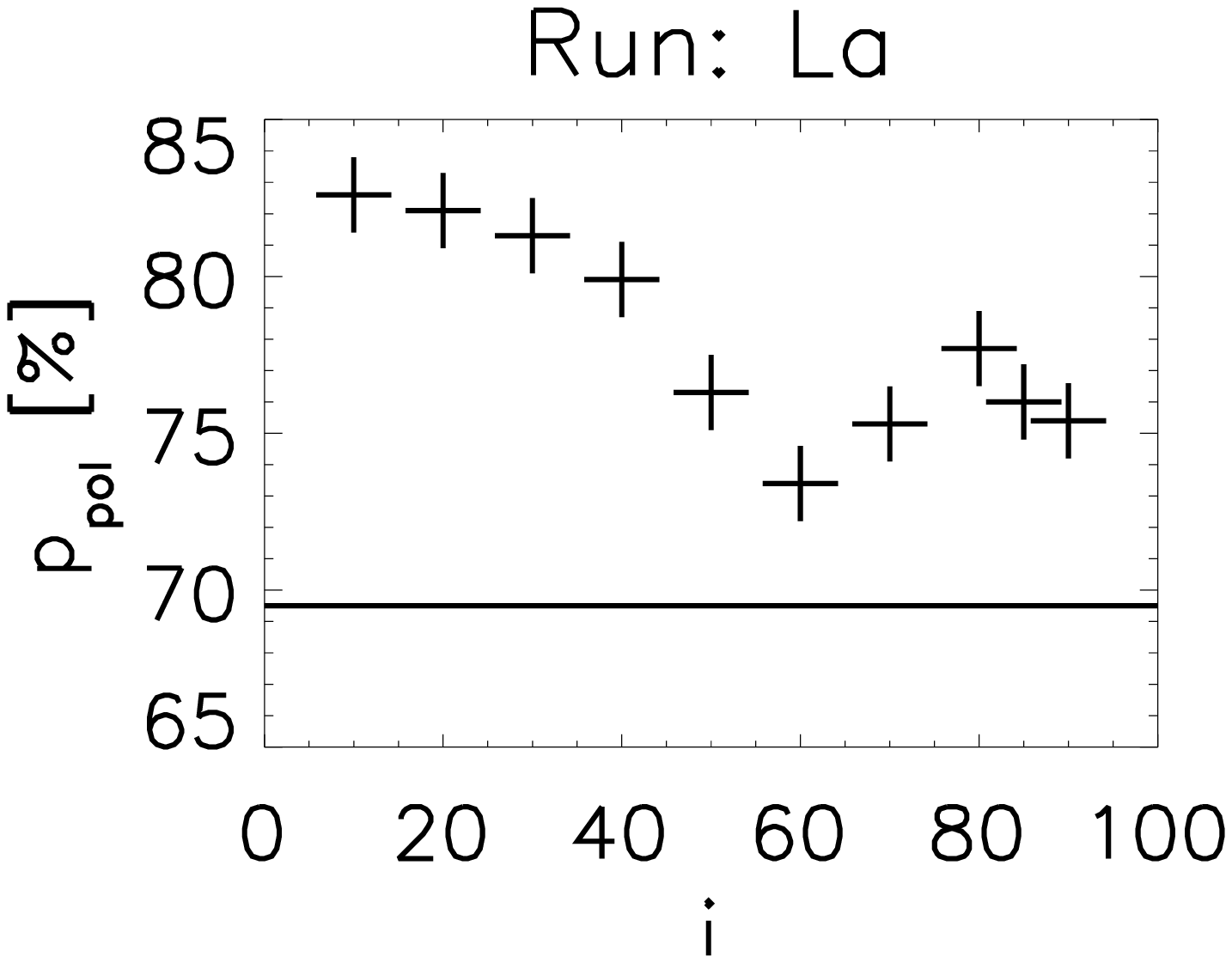}
 \hspace{-1cm}
 \includegraphics[width=4.7cm]{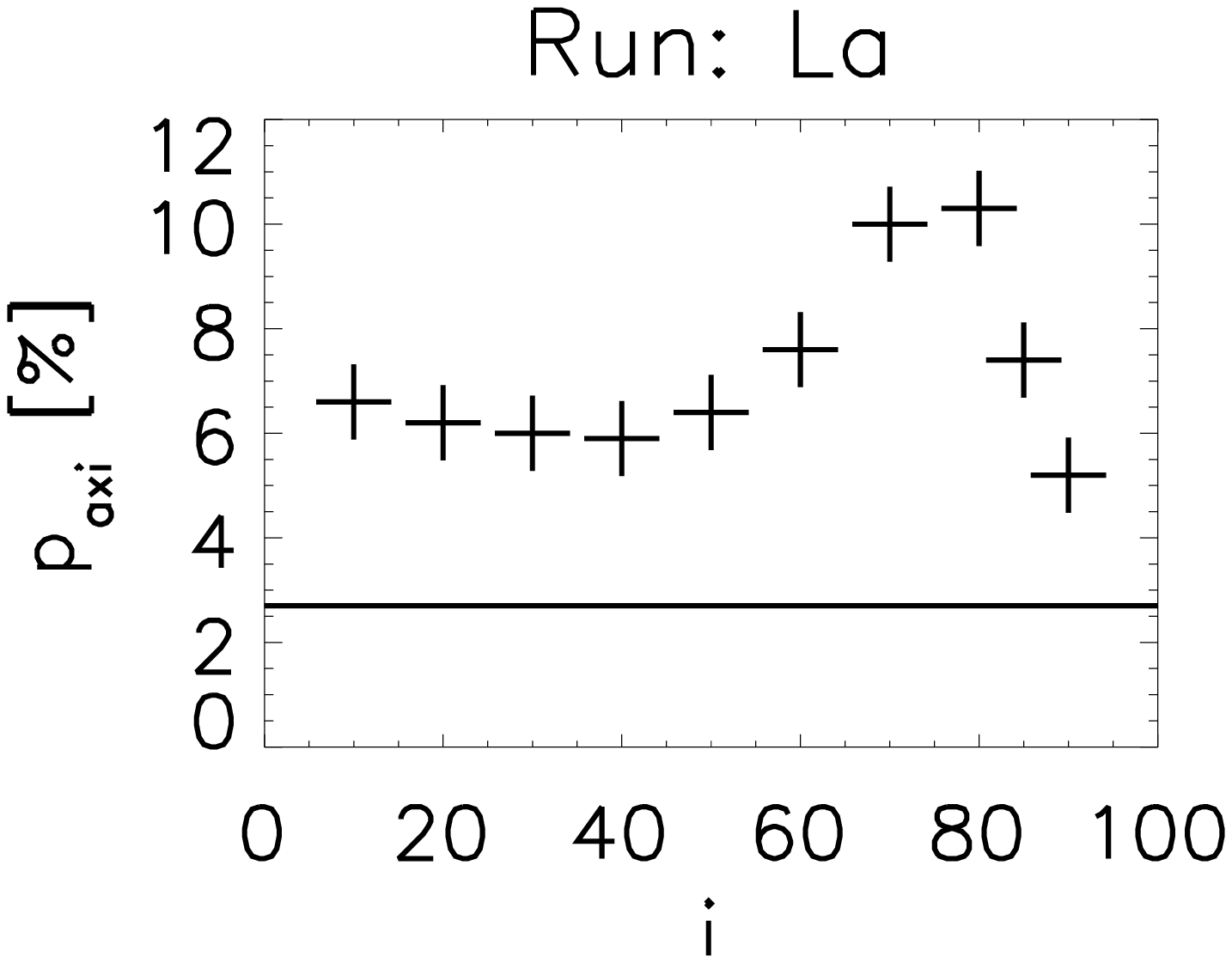}
\caption{Effect of inclination on the fractions of poloidal and axisymmetric field energies for Tests 13--22. The horizontal lines show the original value.}             
 \label{Lai_res}
    \end{figure}

\begin{figure}
   \centering
 \includegraphics[width=8cm]{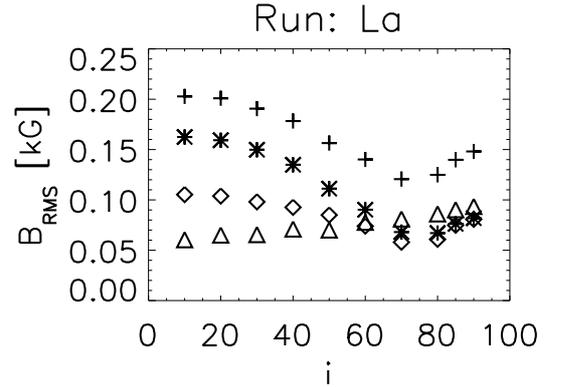}
\caption{Effect of inclination on the RMS values of the magnetic field (Tests 13--22). The total field is plotted with crosses, the radial field with asterisks, the meridional field with diamonds, and the azimuthal field with triangles.}             
 \label{Laib_res}
    \end{figure}

\begin{figure*}
   \centering
  \includegraphics[height=4cm]{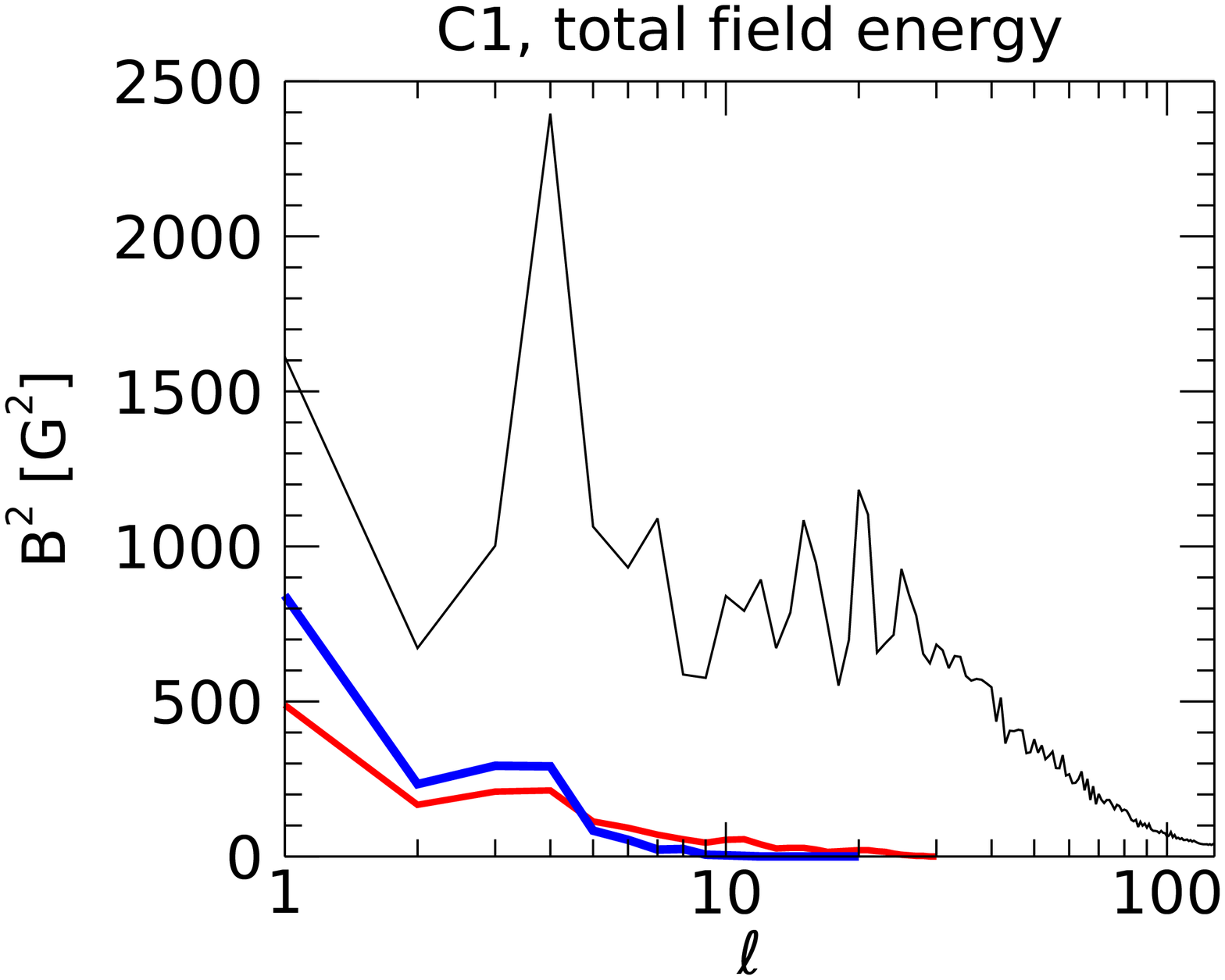}  
  \includegraphics[height=4cm]{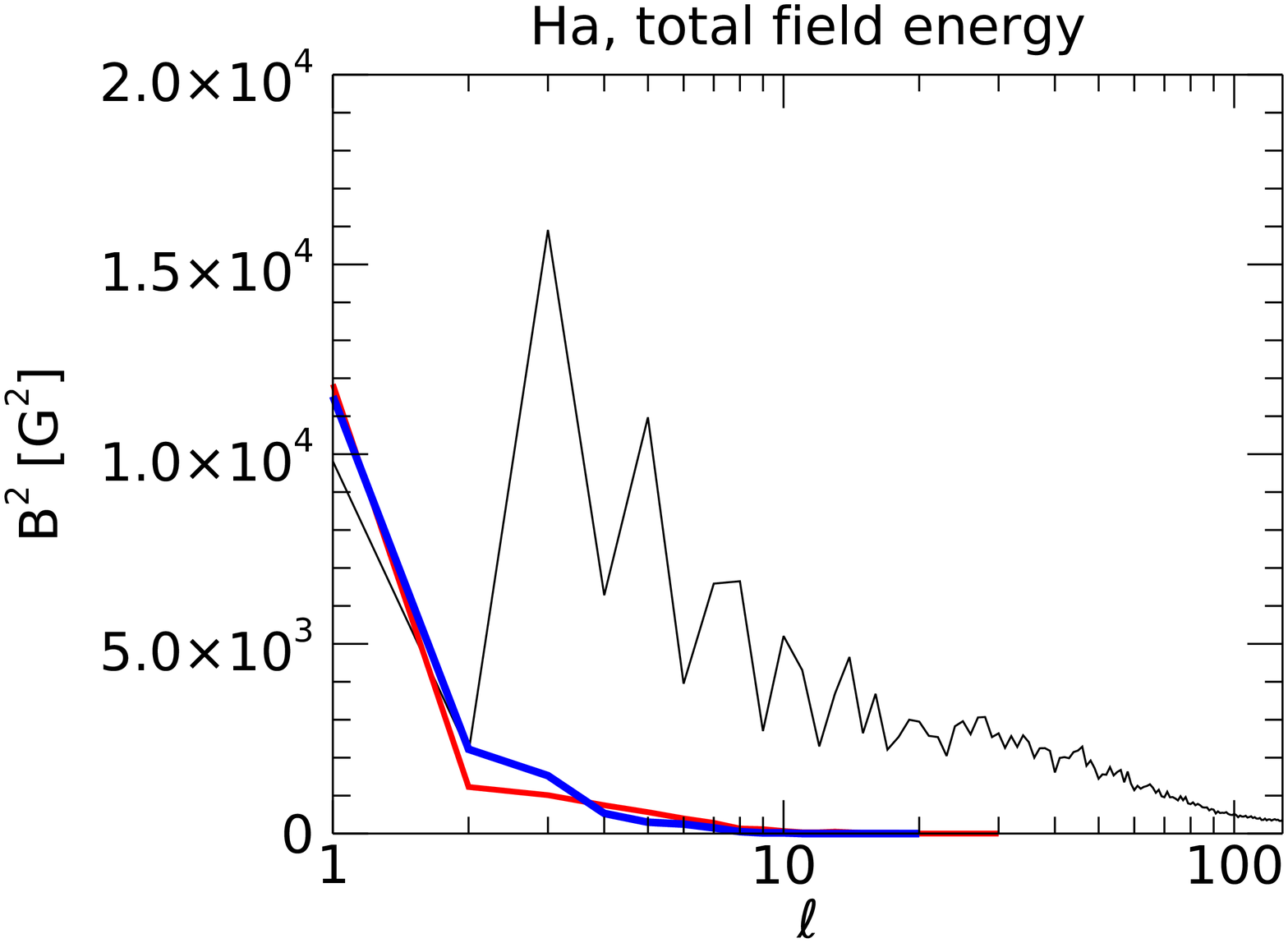} 
  \includegraphics[height=4cm]{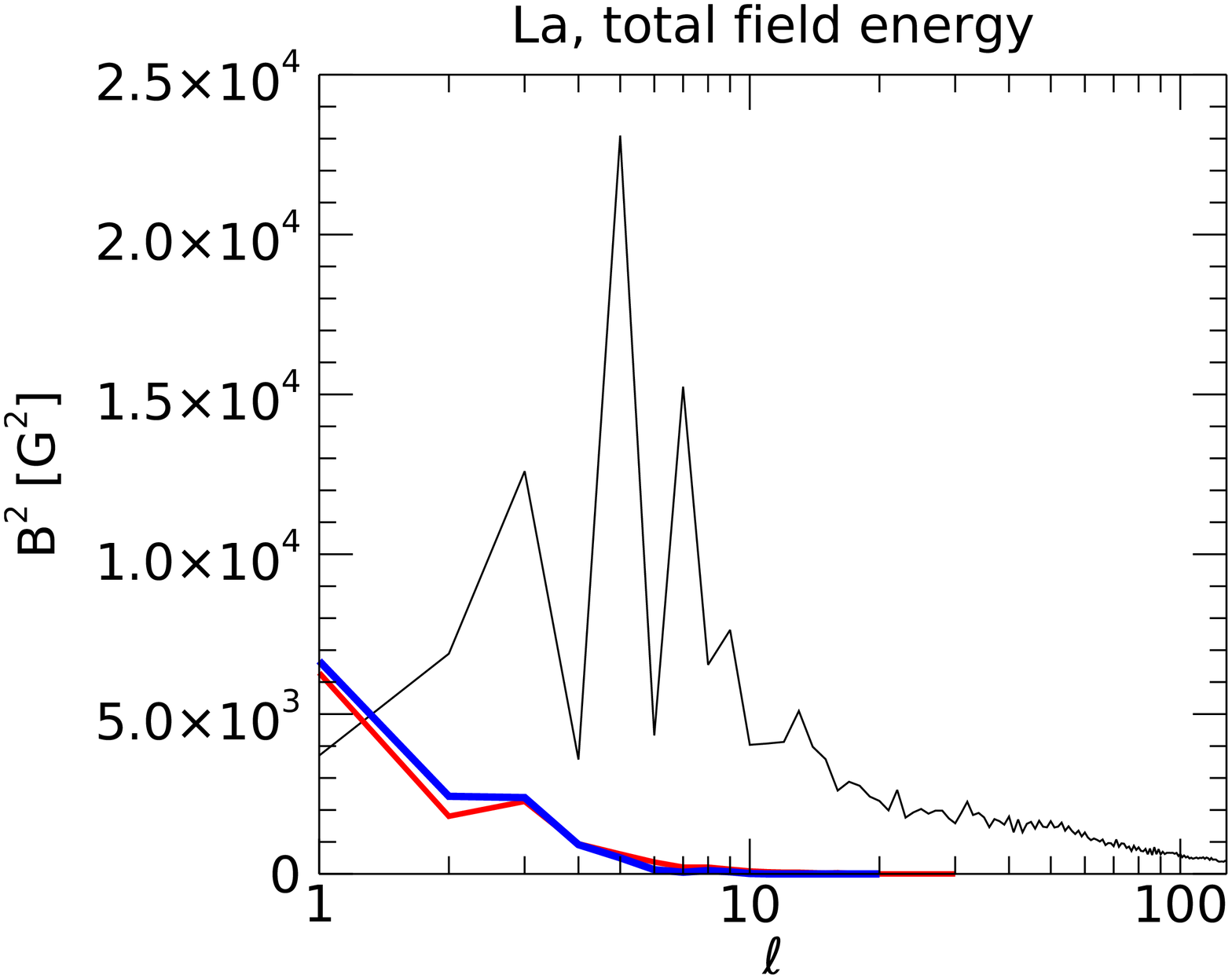} 
\caption{Contribution to the mean magnetic energy by $\ell$-degrees: Original images (black line). Tests 2, 5, and 9 (\vsini = 10 km s$^{-1}$, blue line); Tests 4, 7, and 11 (\vsini = 40 km s$^{-1}$, red line).}             
 \label{B2_lcomp}
    \end{figure*}

\begin{figure*}
   \centering
\includegraphics[height=4cm]{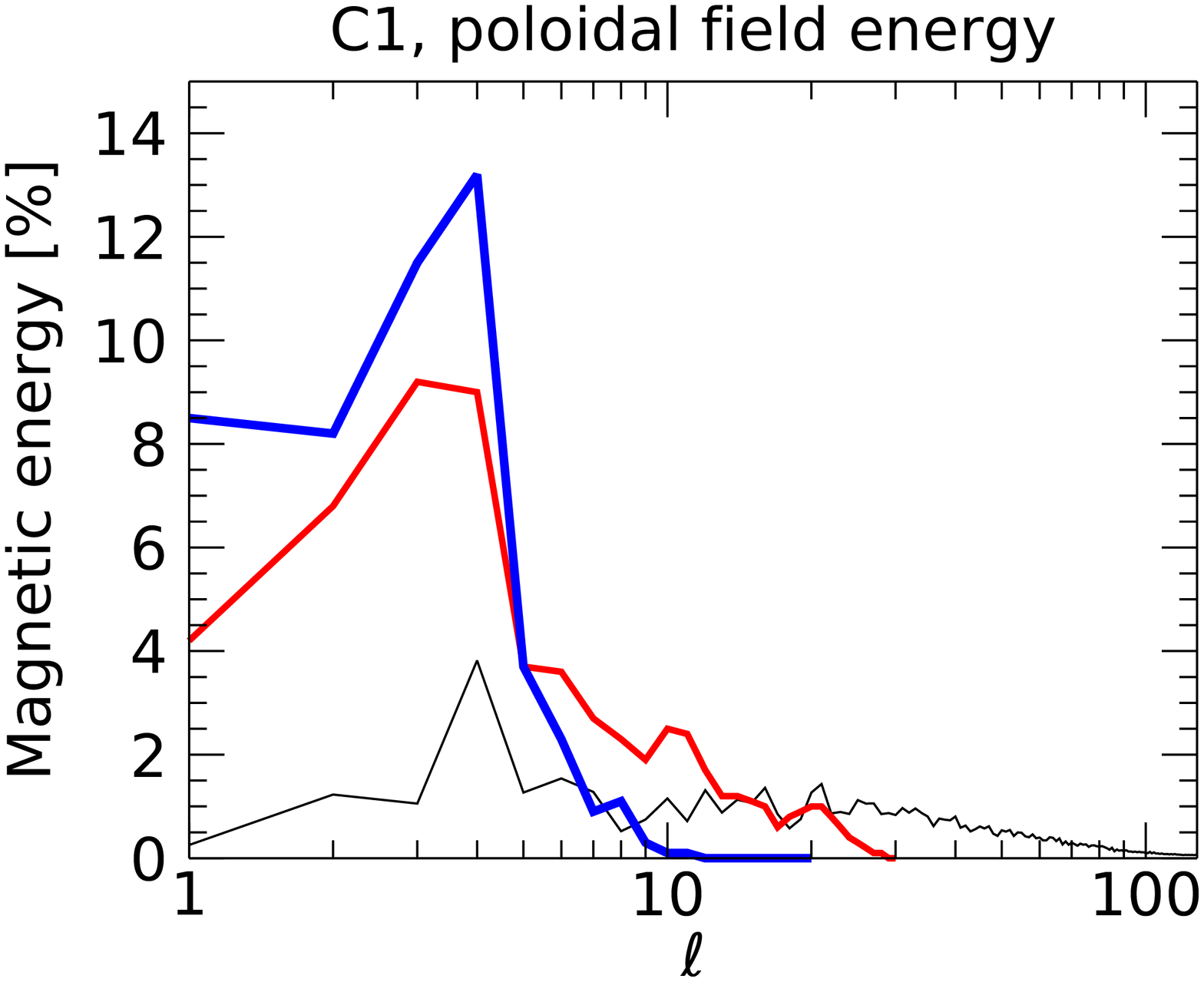}
\includegraphics[height=4cm]{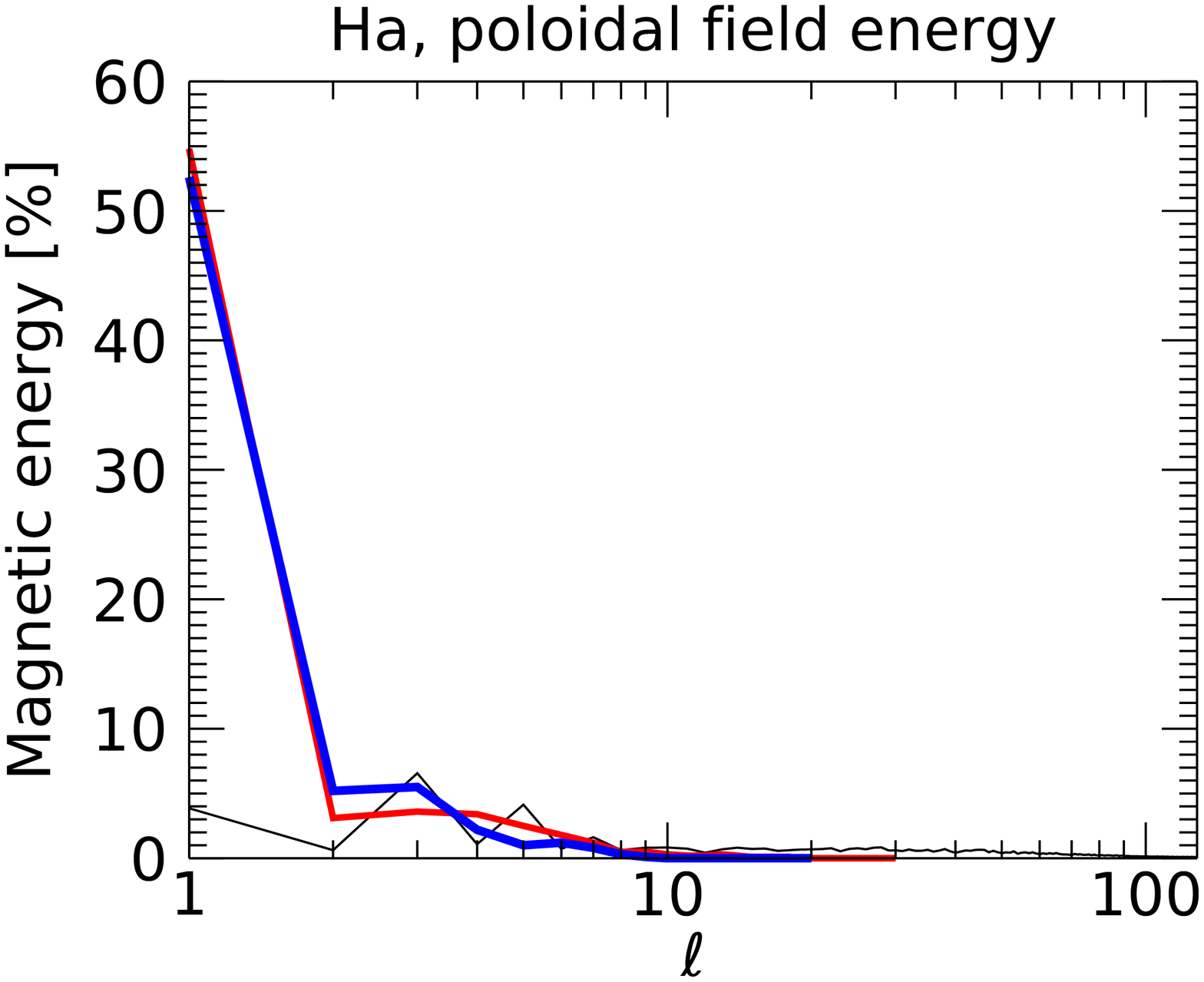}
\includegraphics[height=4cm]{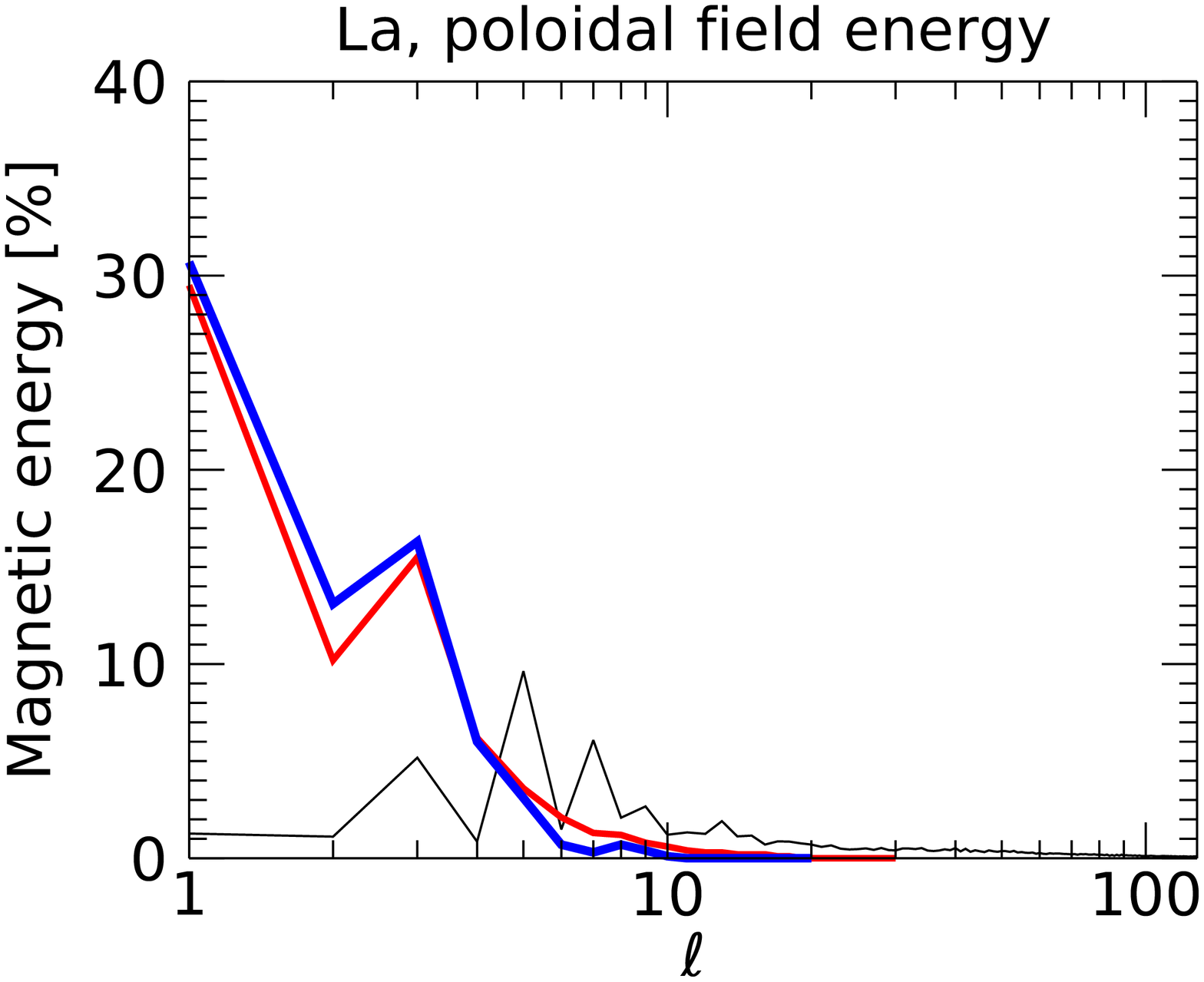}
\caption{Poloidal fraction of the total magnetic energy by $\ell$-degrees. Colour markings are the same as in Fig. \ref{B2_lcomp}.}             
 \label{pol_lcomp}
    \end{figure*}

\begin{figure*}
   \centering
\includegraphics[height=4cm]{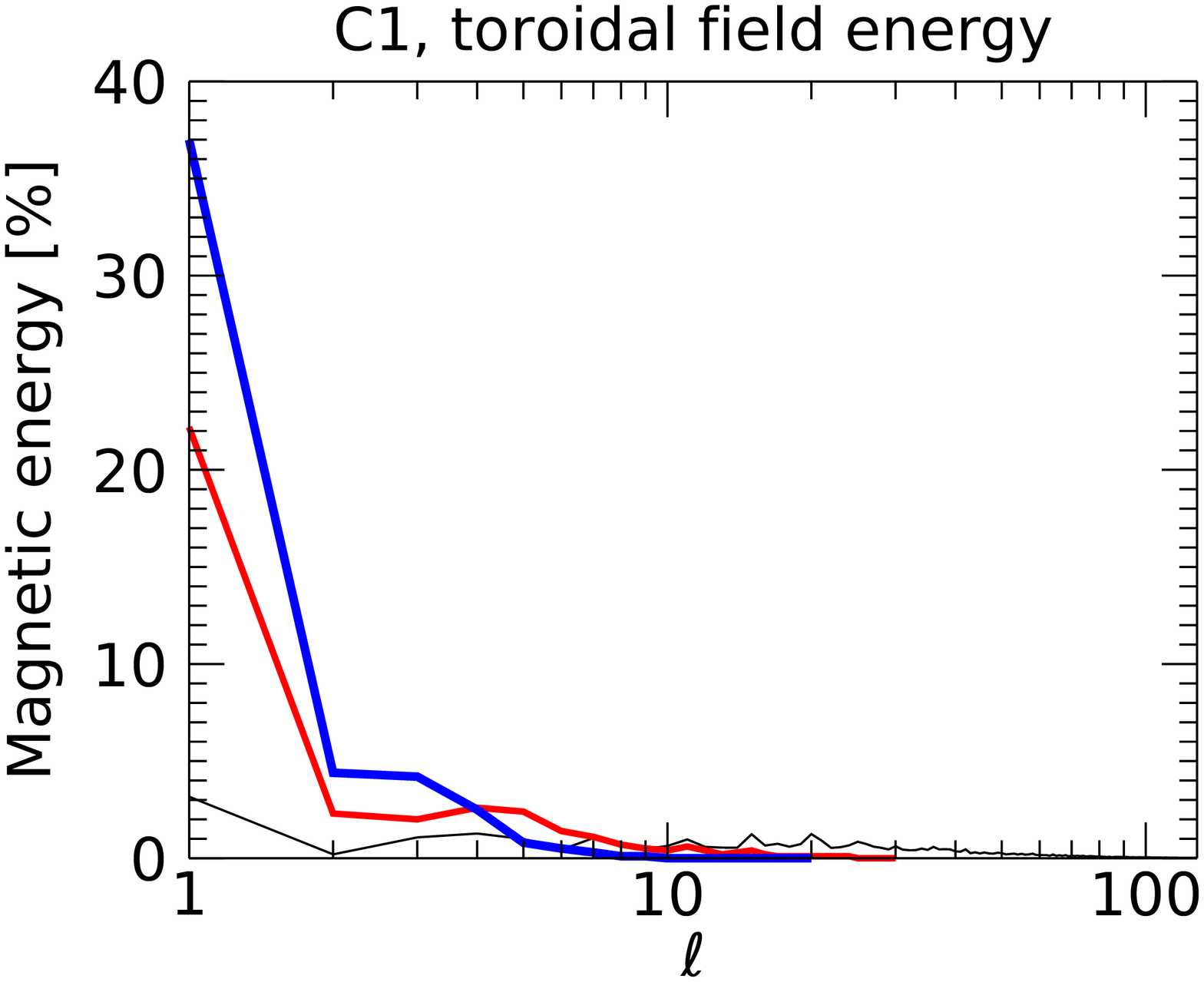}
\includegraphics[height=4cm]{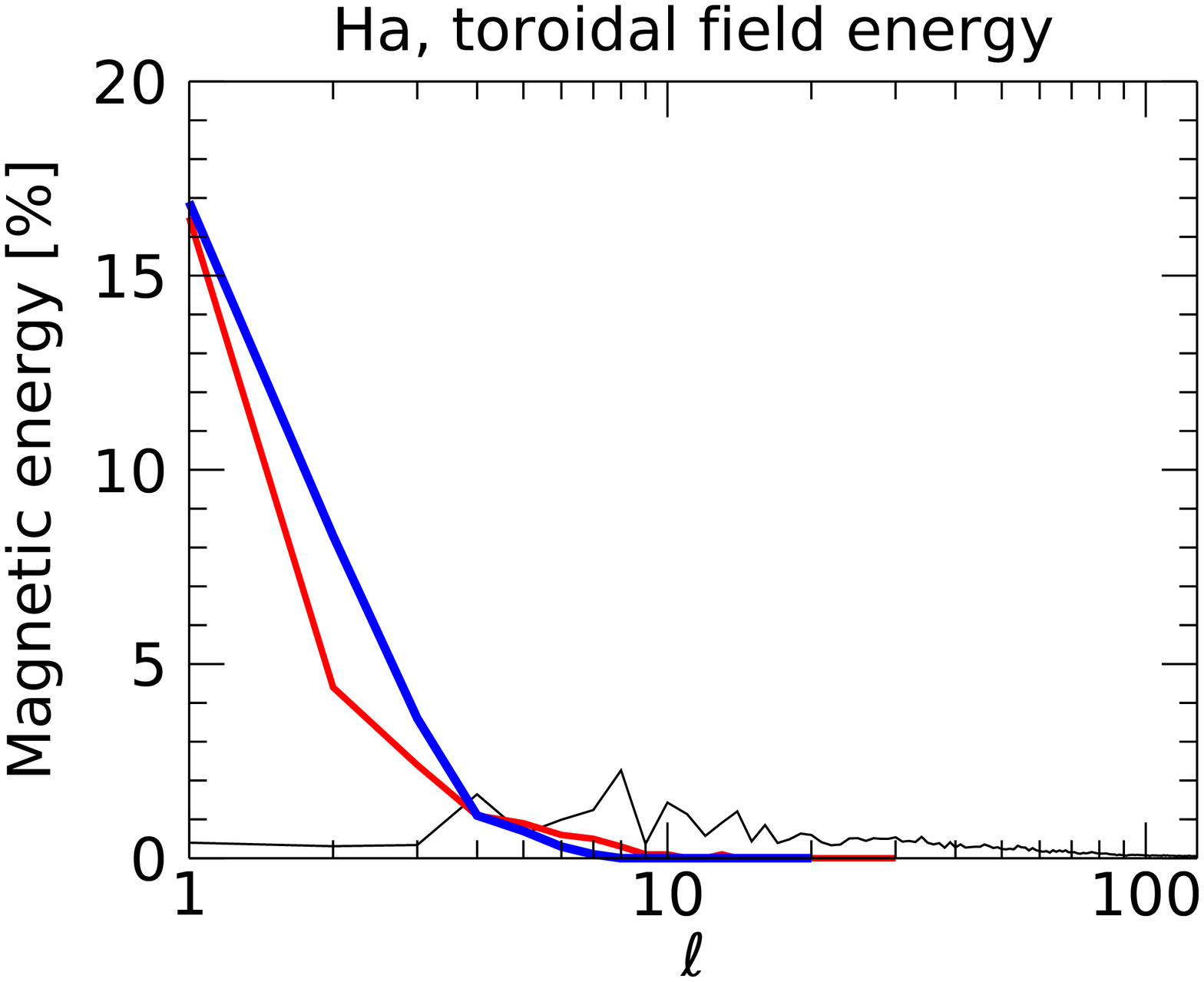}
\includegraphics[height=4cm]{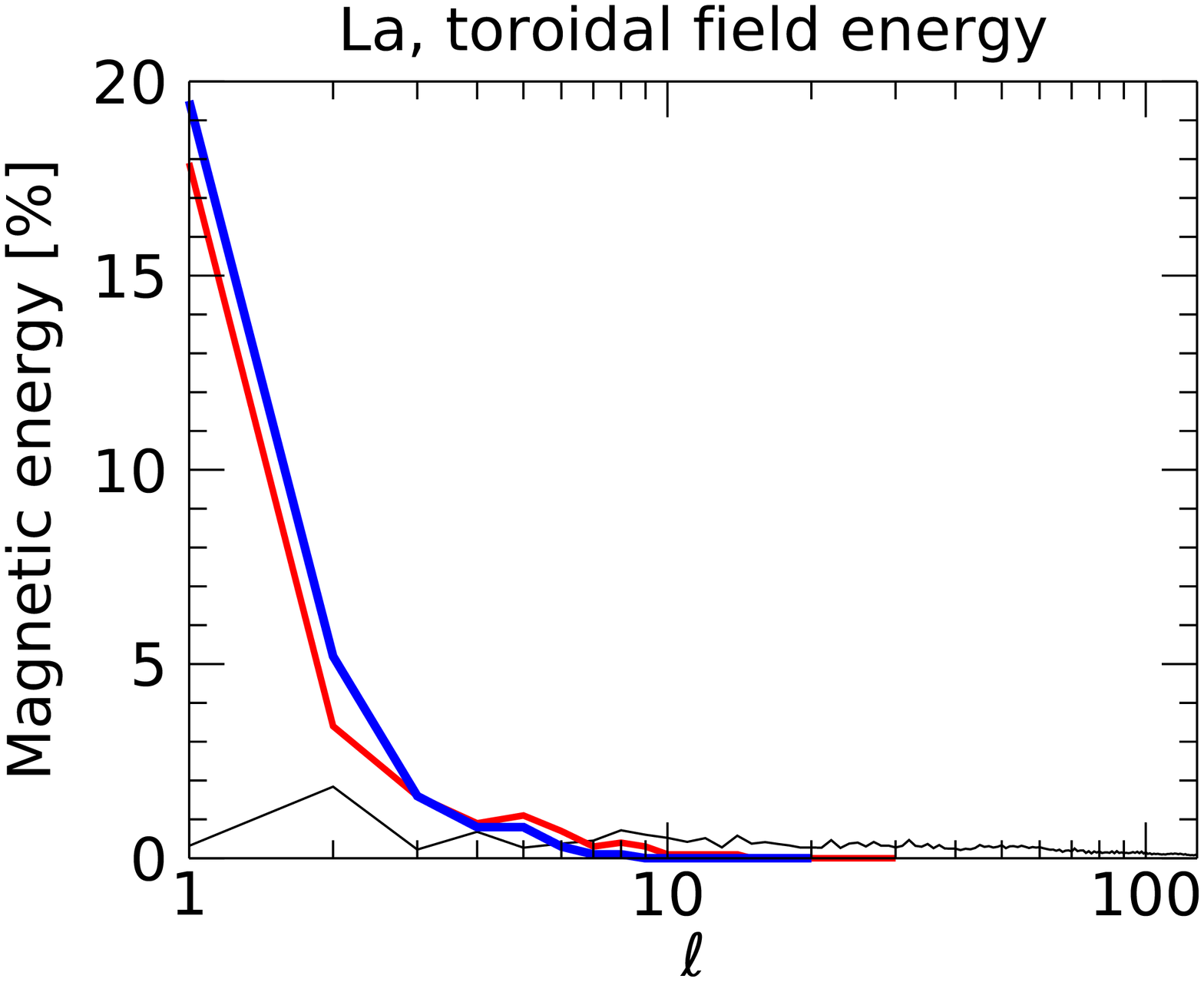}
\caption{Toroidal fraction of the total magnetic energy by $\ell$-degree. Colour markings are the same as in Fig. \ref{B2_lcomp}.}             
 \label{tor_lcomp}
    \end{figure*}

\begin{figure*}
   \centering
\includegraphics[height=4cm]{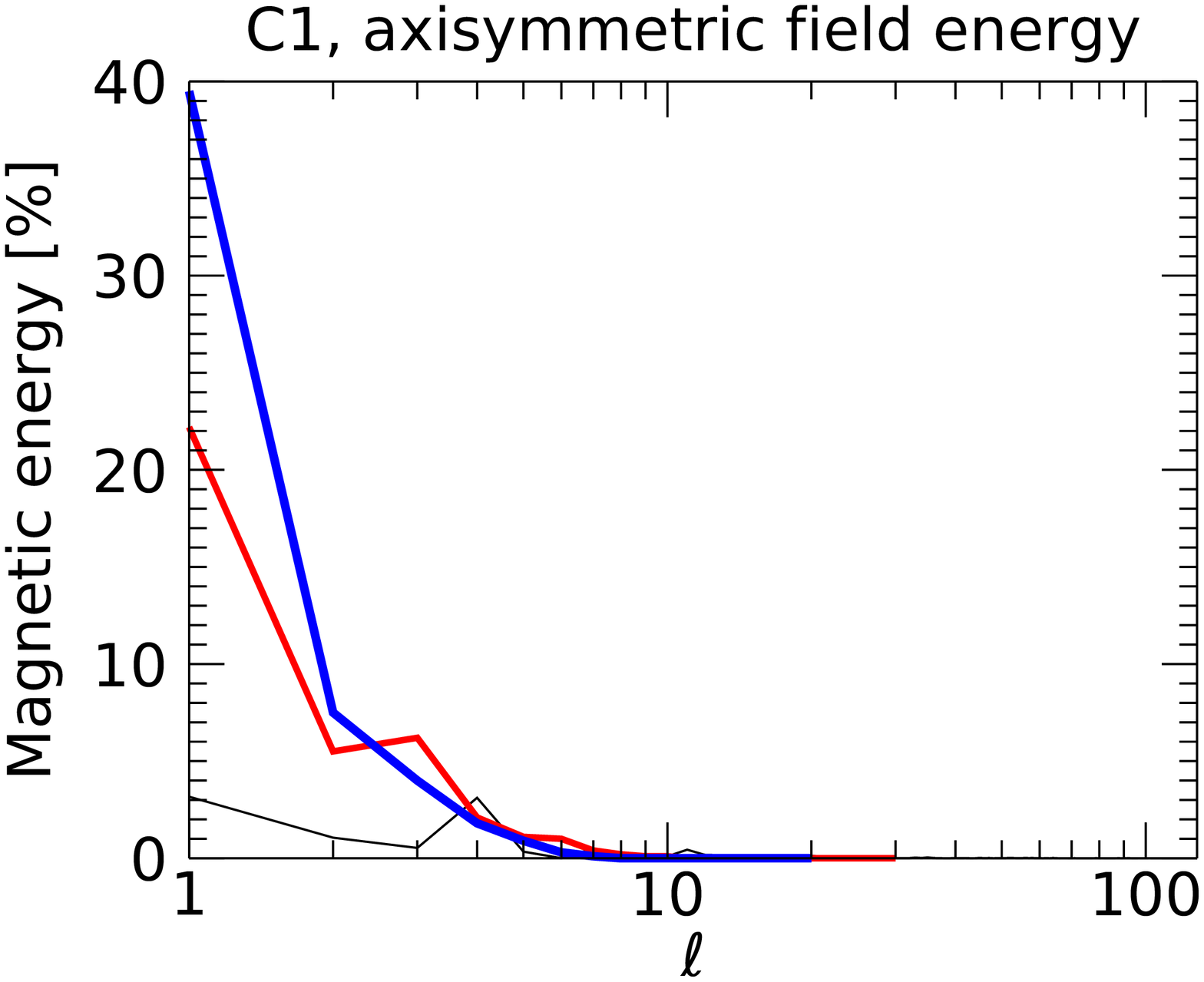}
\includegraphics[height=4cm]{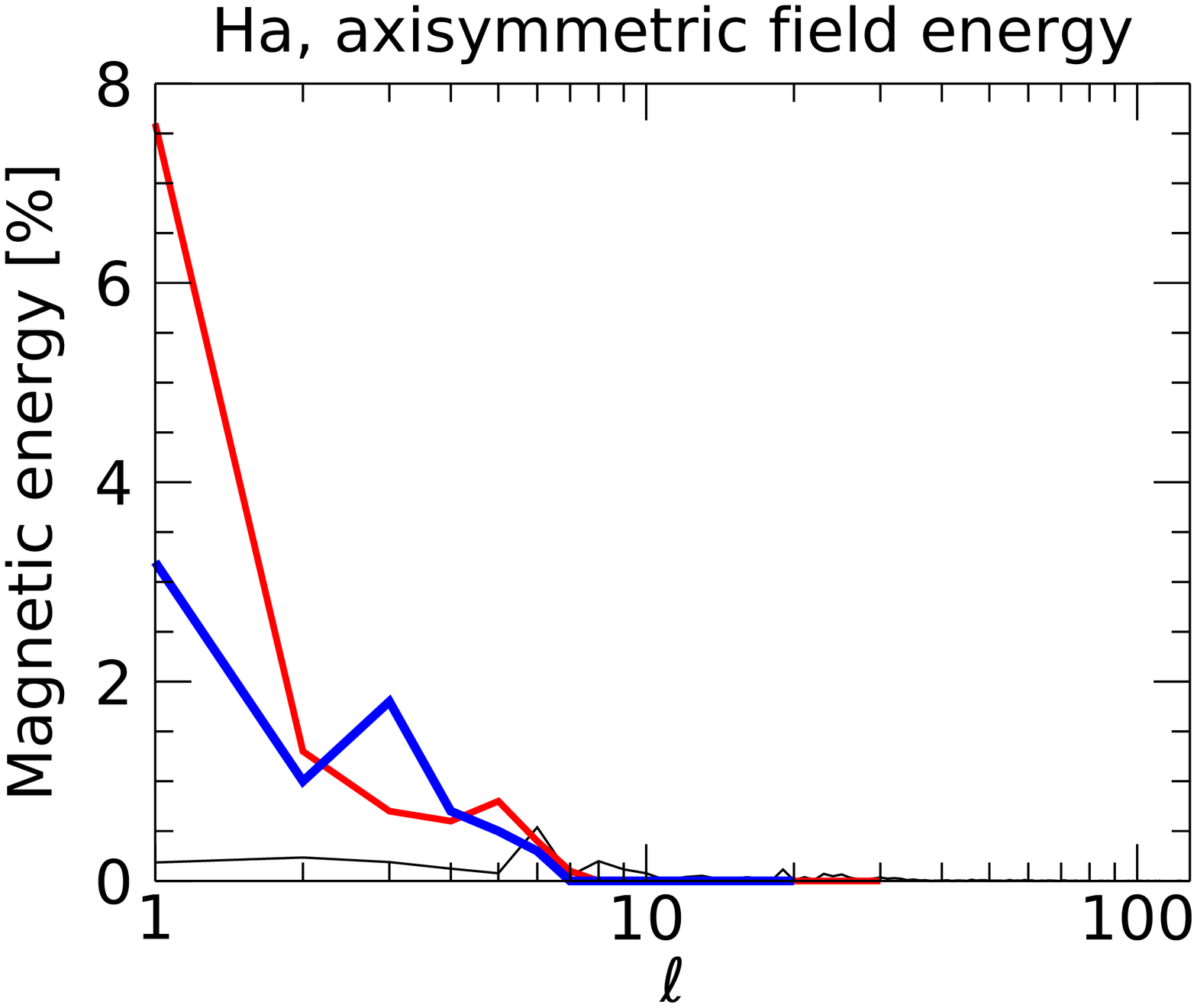}
\includegraphics[height=4cm]{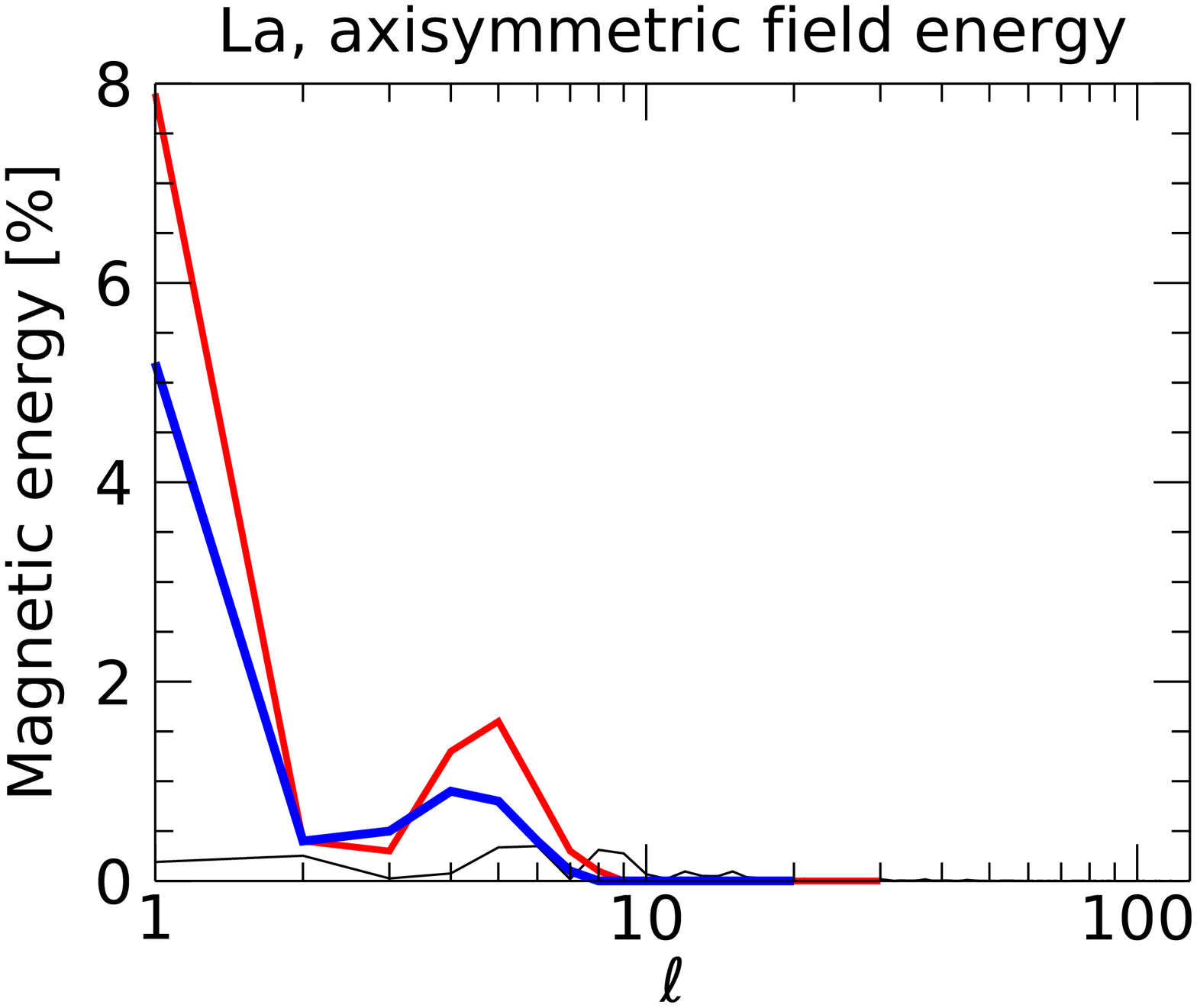}
\caption{Axisymmetric fraction of the total magnetic energy by $\ell$-degree. Colour markings are the same as in Fig. \ref{B2_lcomp}.}             
 \label{axi_lcomp}
    \end{figure*}

\begin{figure*}
   \centering
\includegraphics[height=4cm]{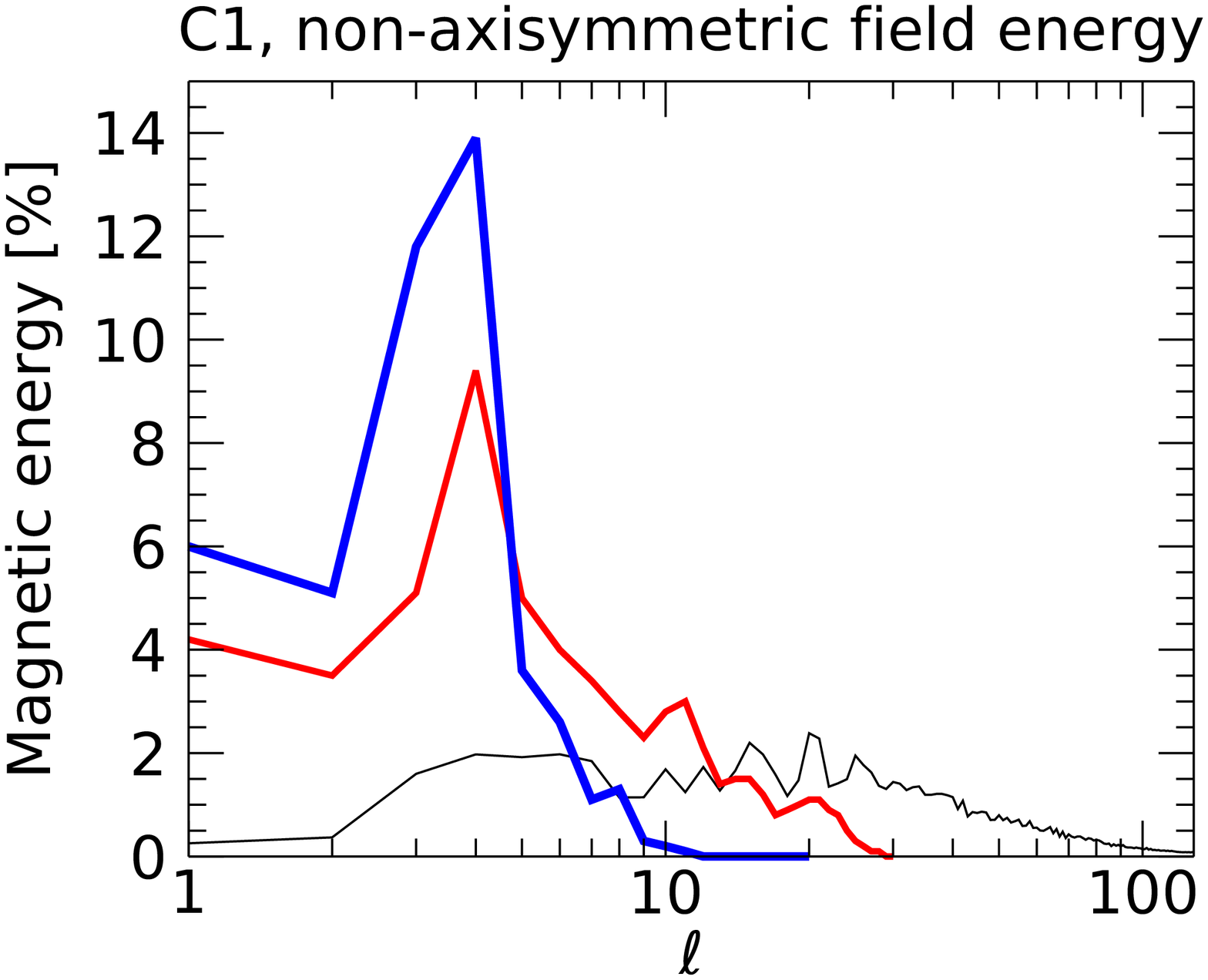}
\includegraphics[height=4cm]{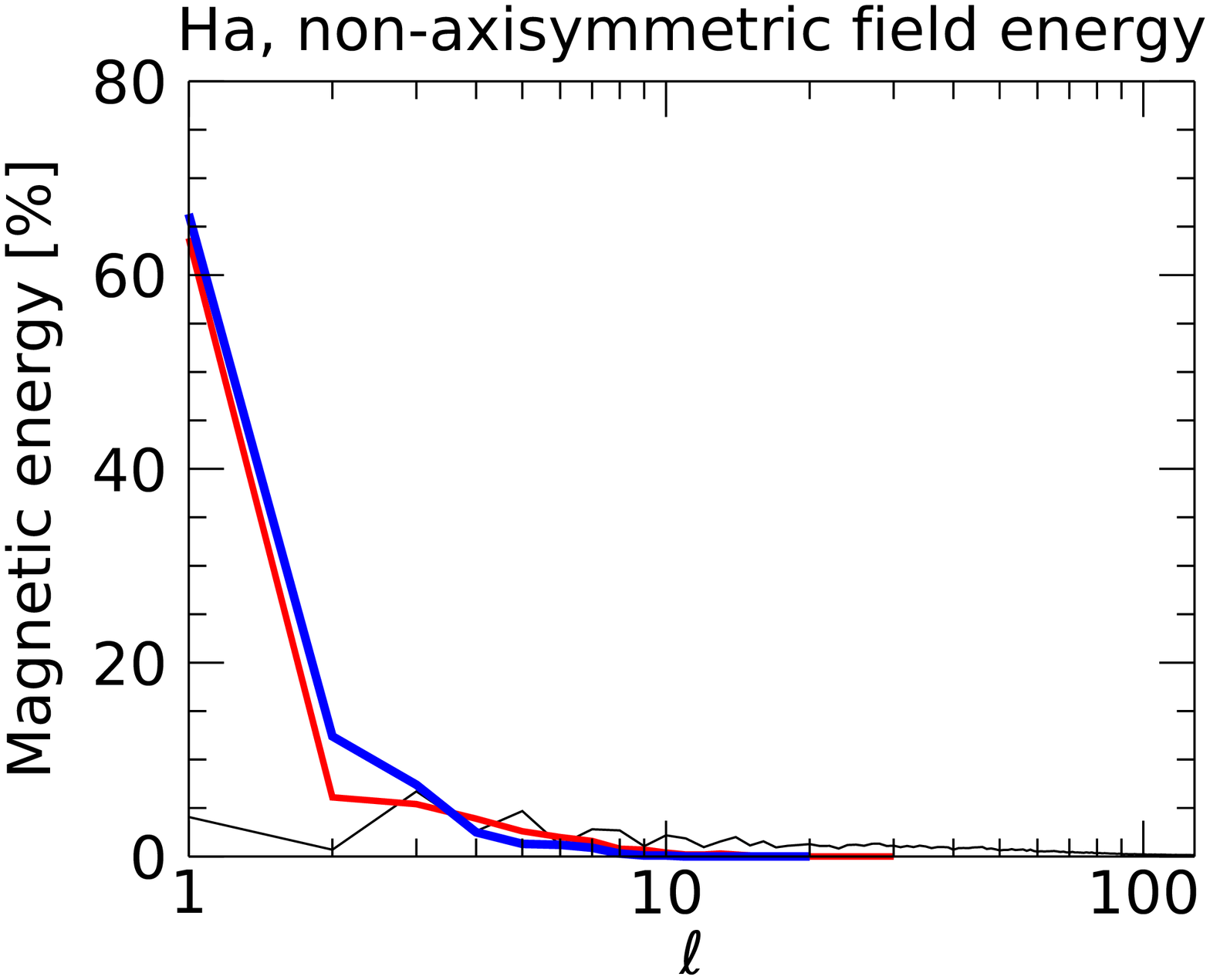}
\includegraphics[height=4cm]{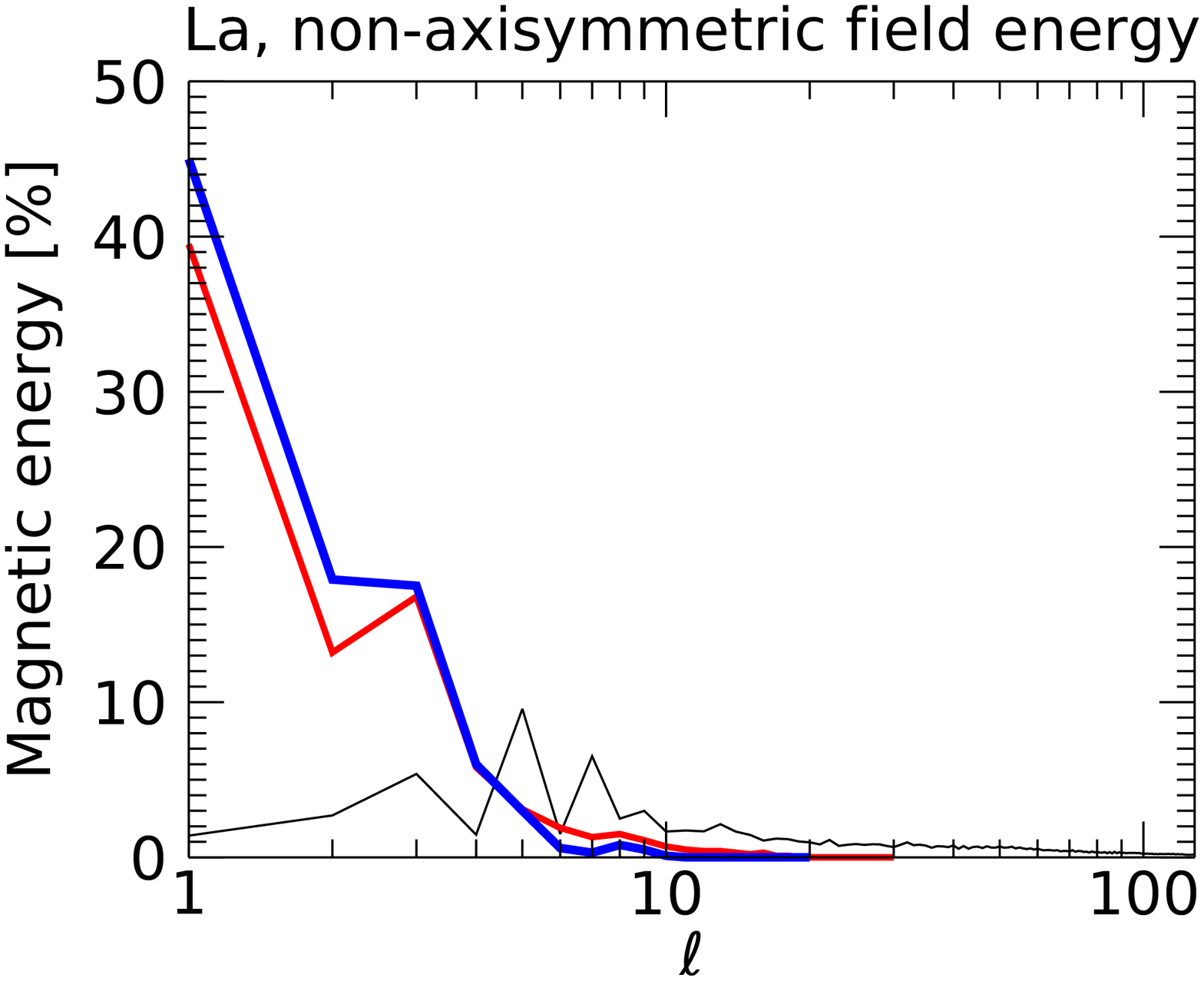}
\caption{Non-axisymmetric fraction of the total magnetic energy by $\ell$-degree. Colour markings are the same as in Fig. \ref{B2_lcomp}.}             
 \label{naxi_lcomp}
    \end{figure*}

\begin{figure*}
   \centering
\includegraphics[height=4cm]{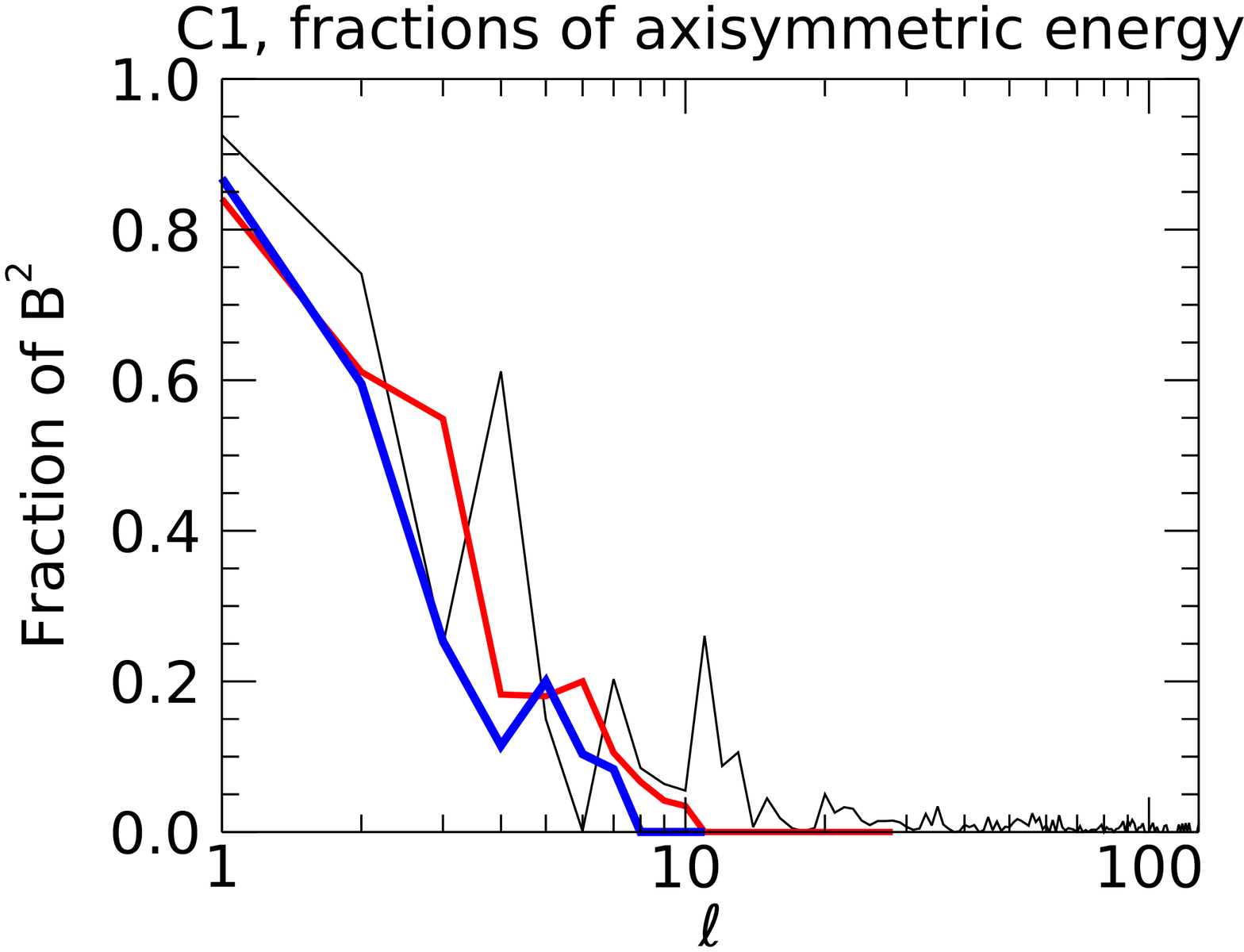}
\includegraphics[height=4cm]{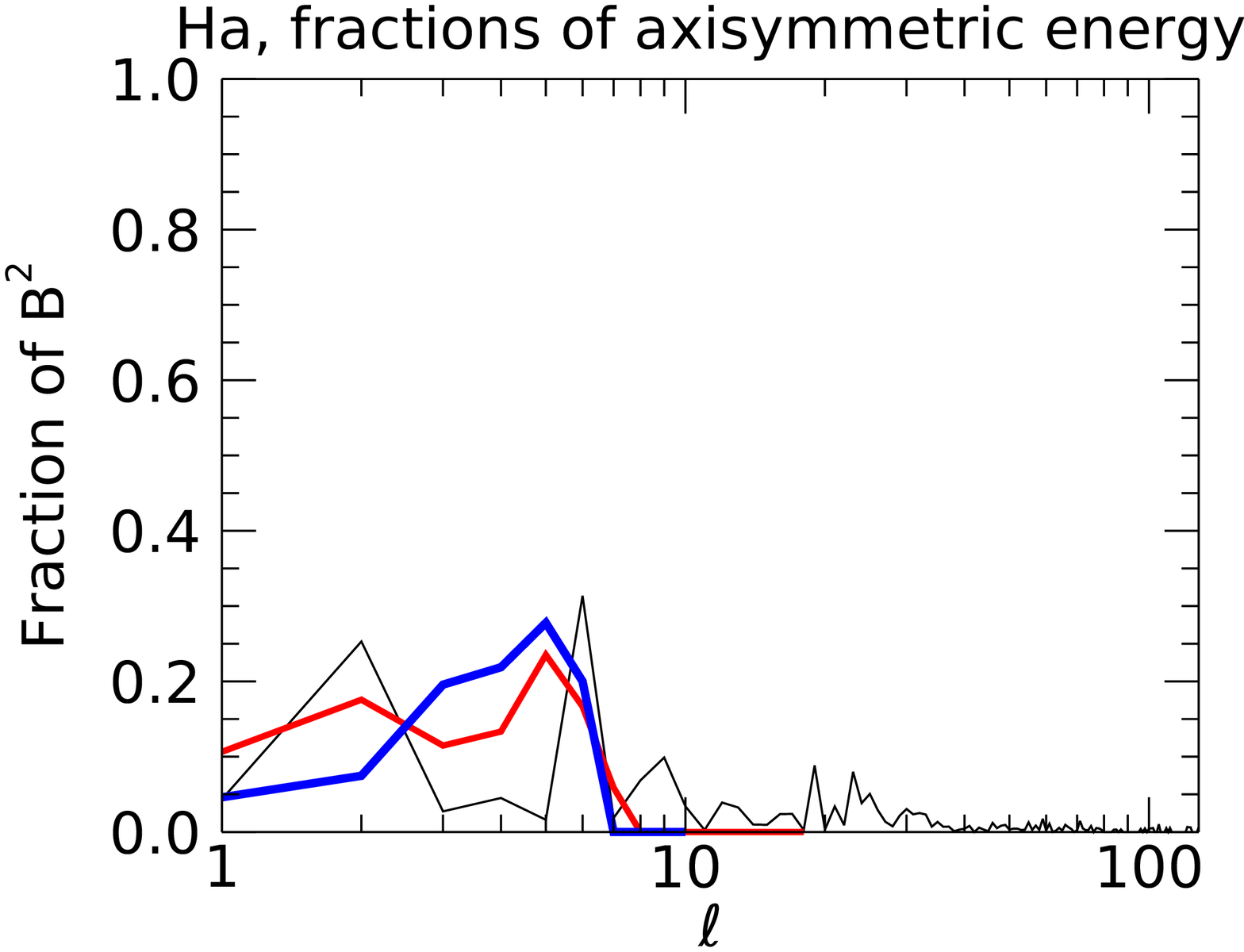}
\includegraphics[height=4cm]{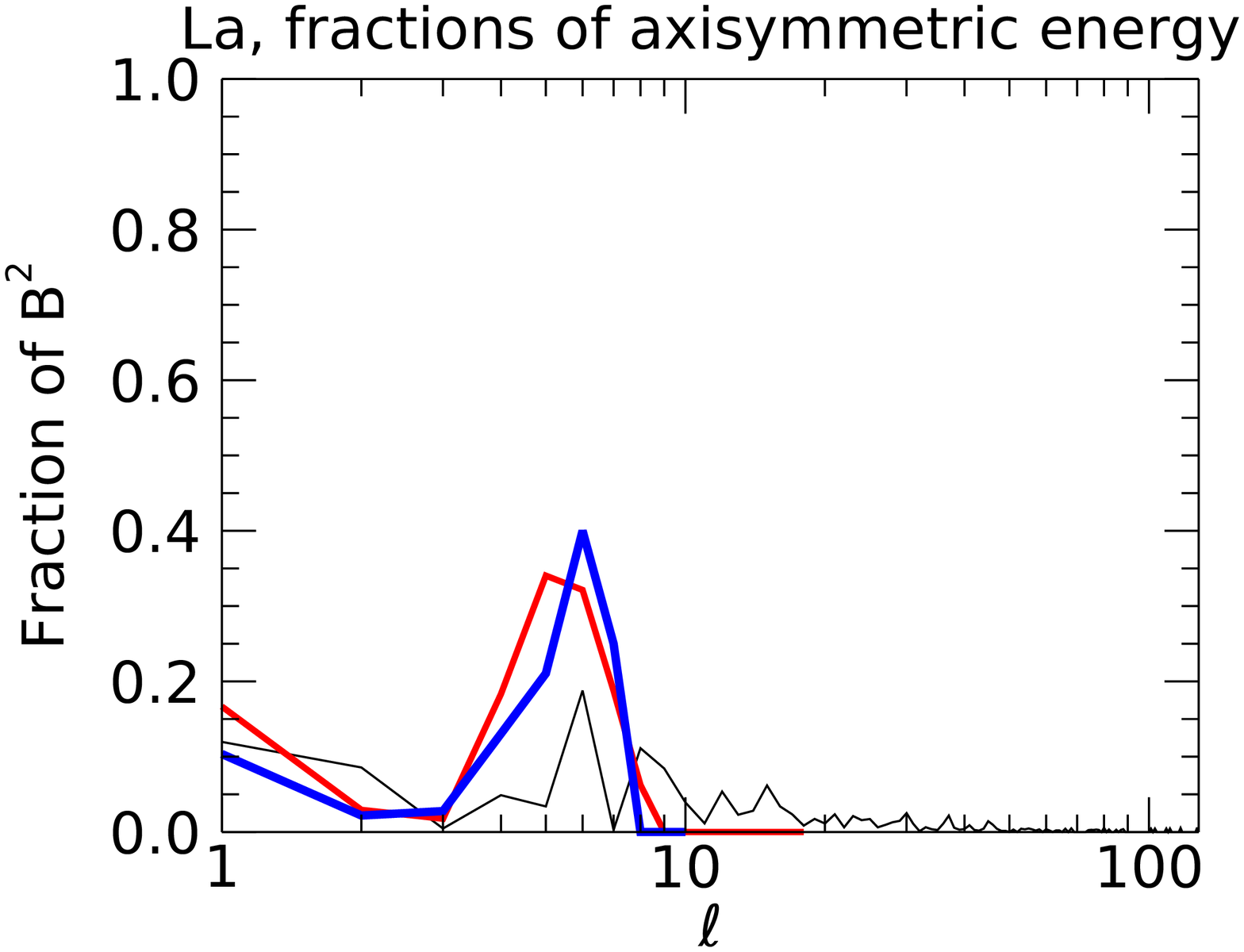}
\caption{Axisymmetric fraction of the magnetic energy for each $\ell$-degree. Colour markings are the same as in Fig. \ref{B2_lcomp}.}            
 \label{axi_l_ap}
    \end{figure*}

\begin{figure*}
   \centering
\includegraphics[height=4cm]{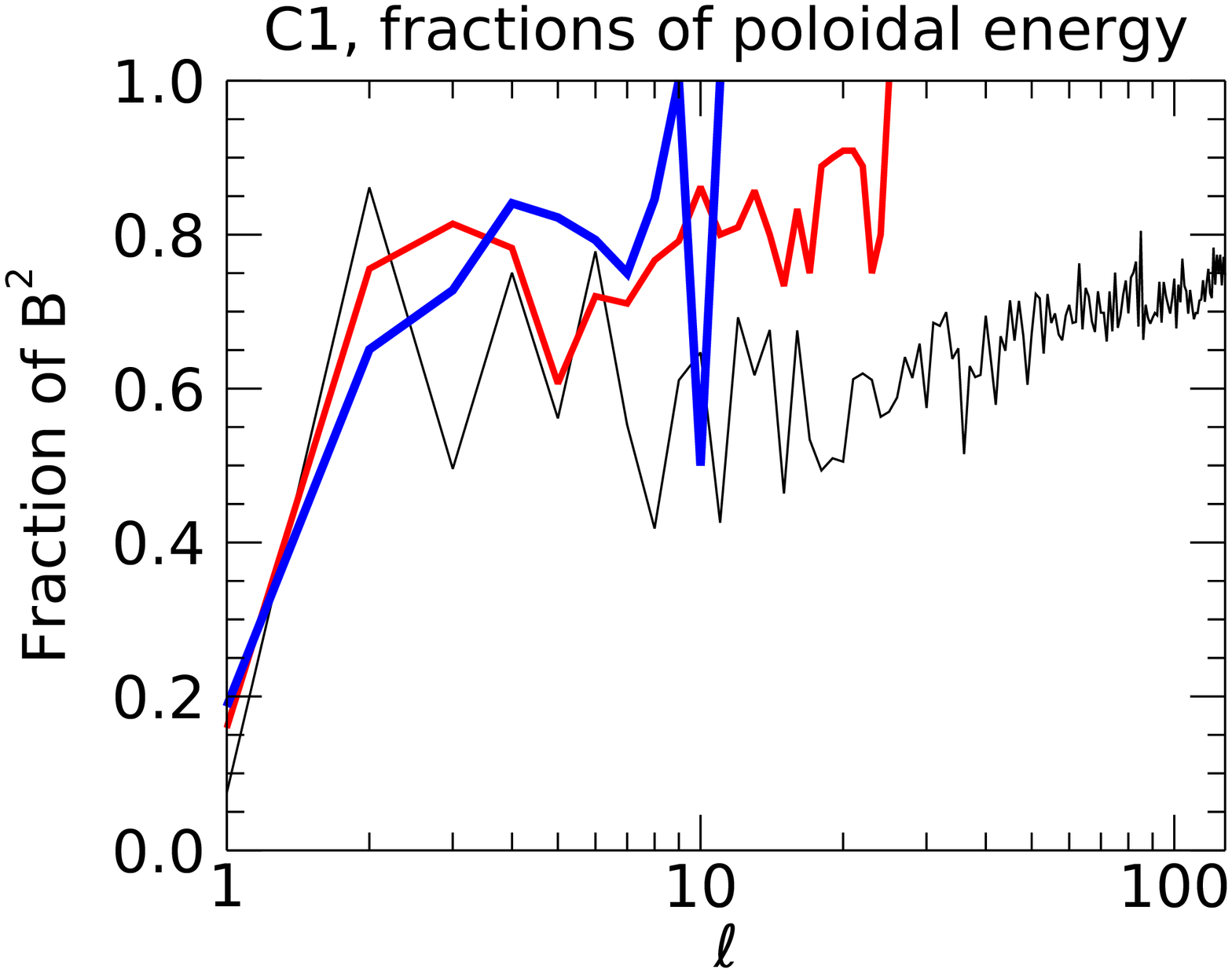}
\includegraphics[height=4cm]{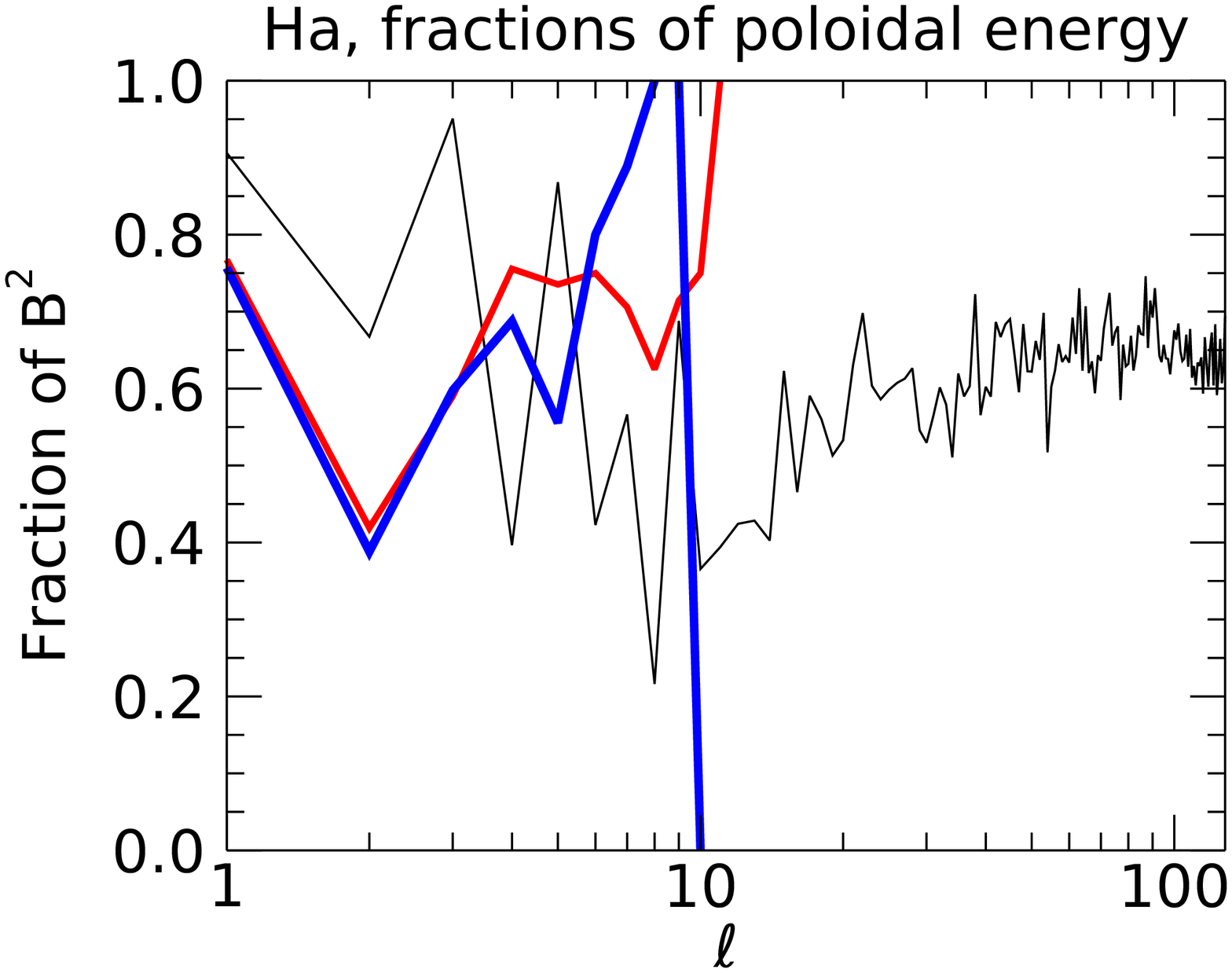}
\includegraphics[height=4cm]{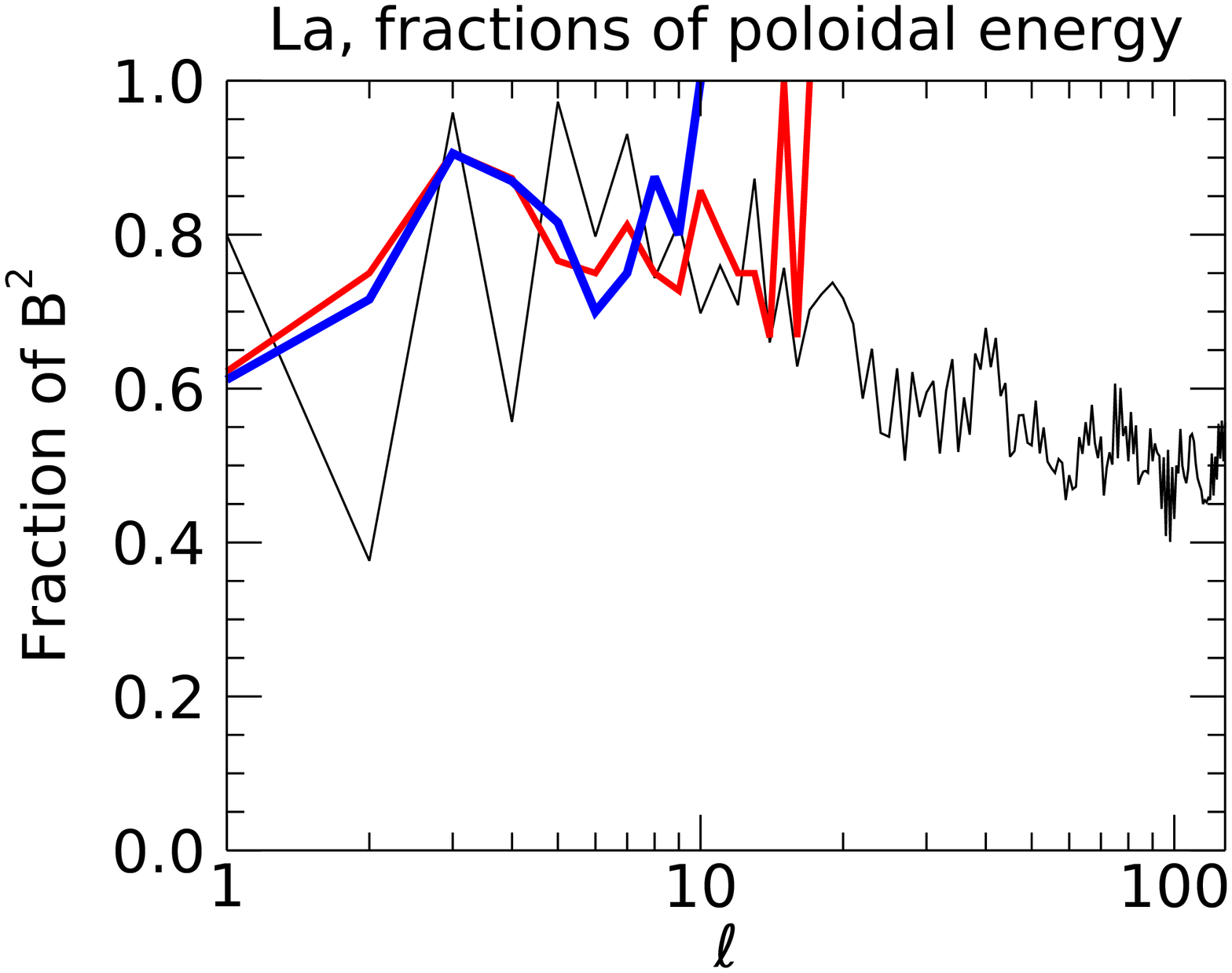}
\caption{Poloidal fraction of the magnetic energy for each $\ell$-degree. Colour markings are the same as in Fig. \ref{B2_lcomp}.}             
 \label{axi_l_pp}
    \end{figure*}

\begin{figure*}
   \centering
\includegraphics[height=4cm]{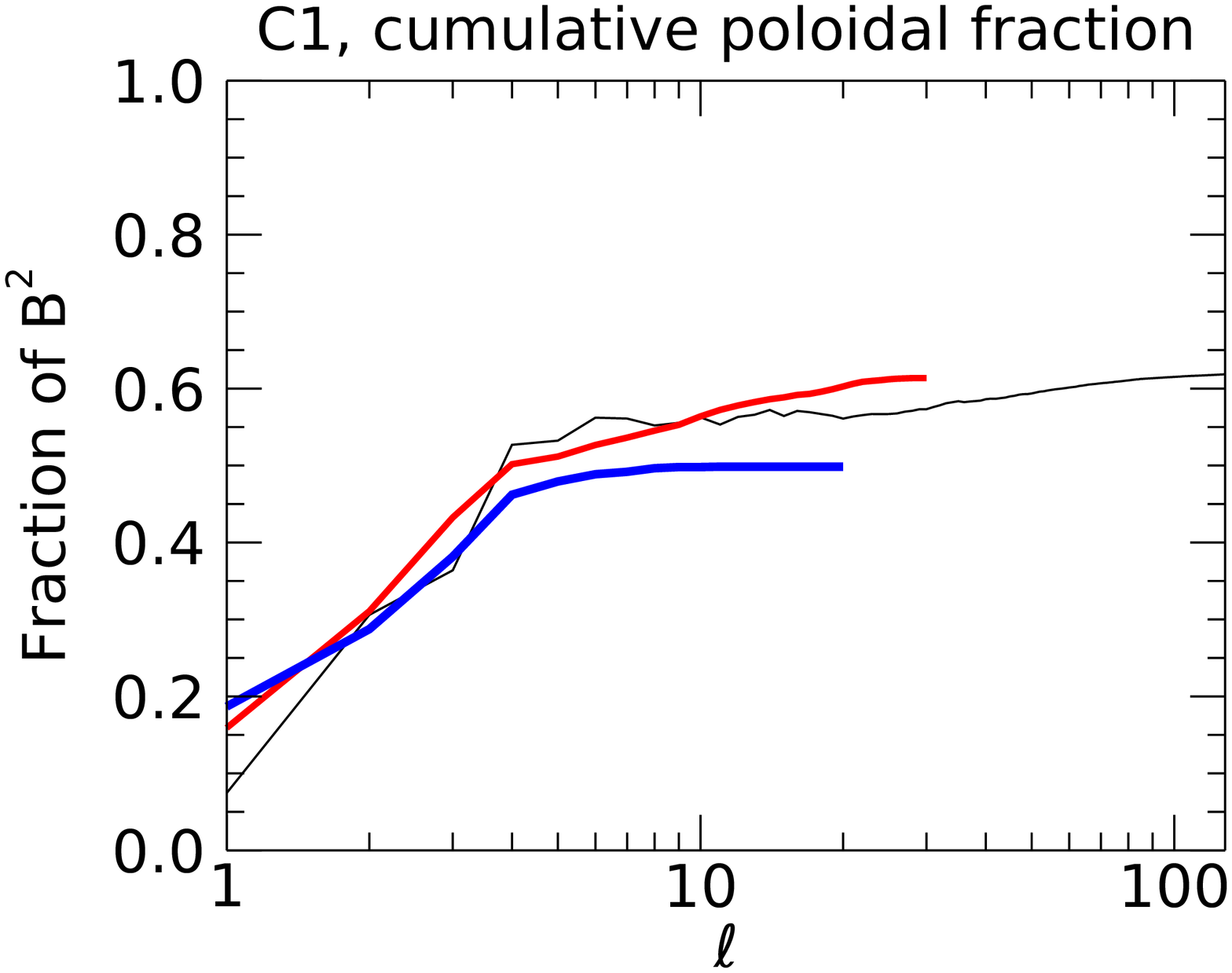}
\includegraphics[height=4cm]{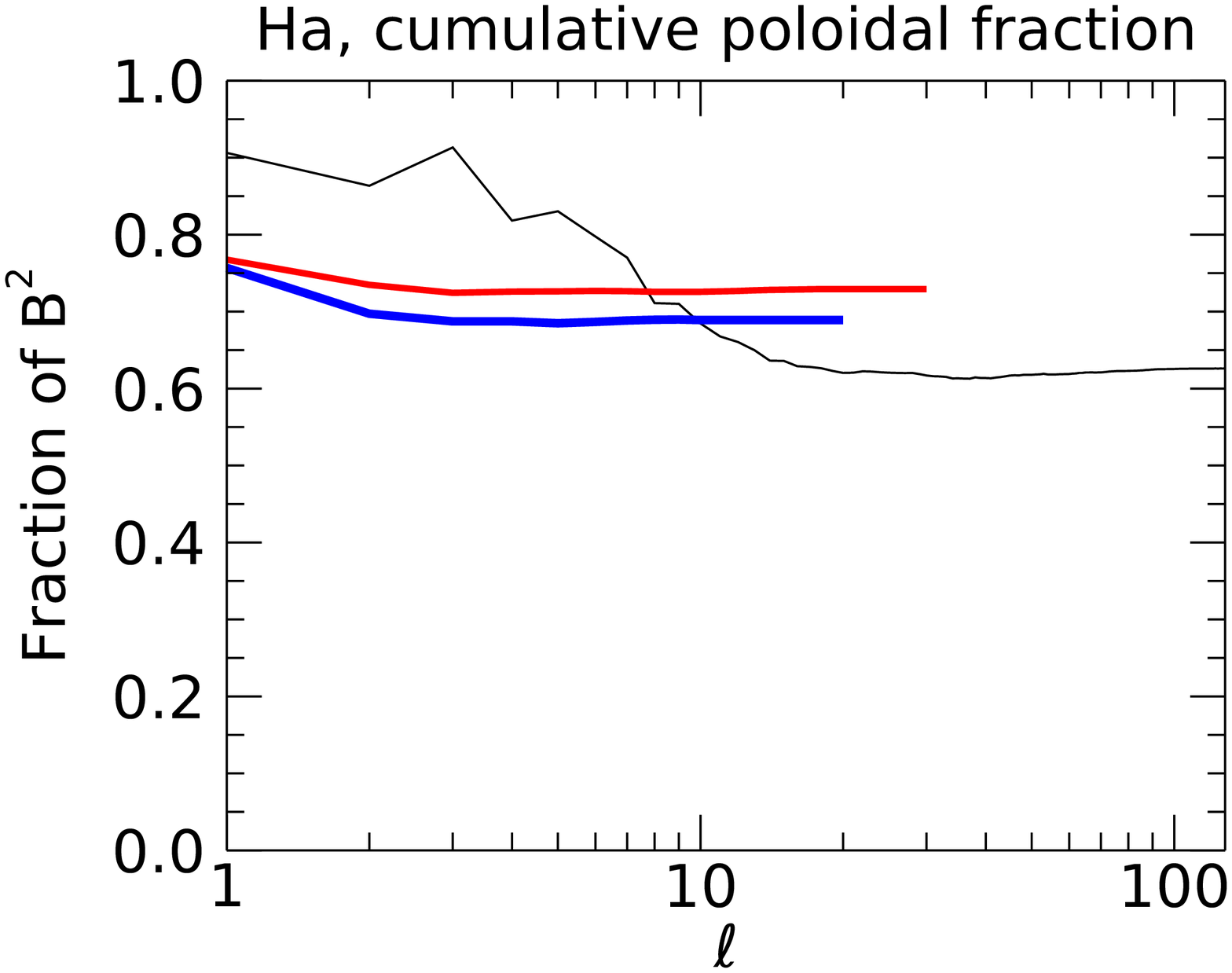}
\includegraphics[height=4cm]{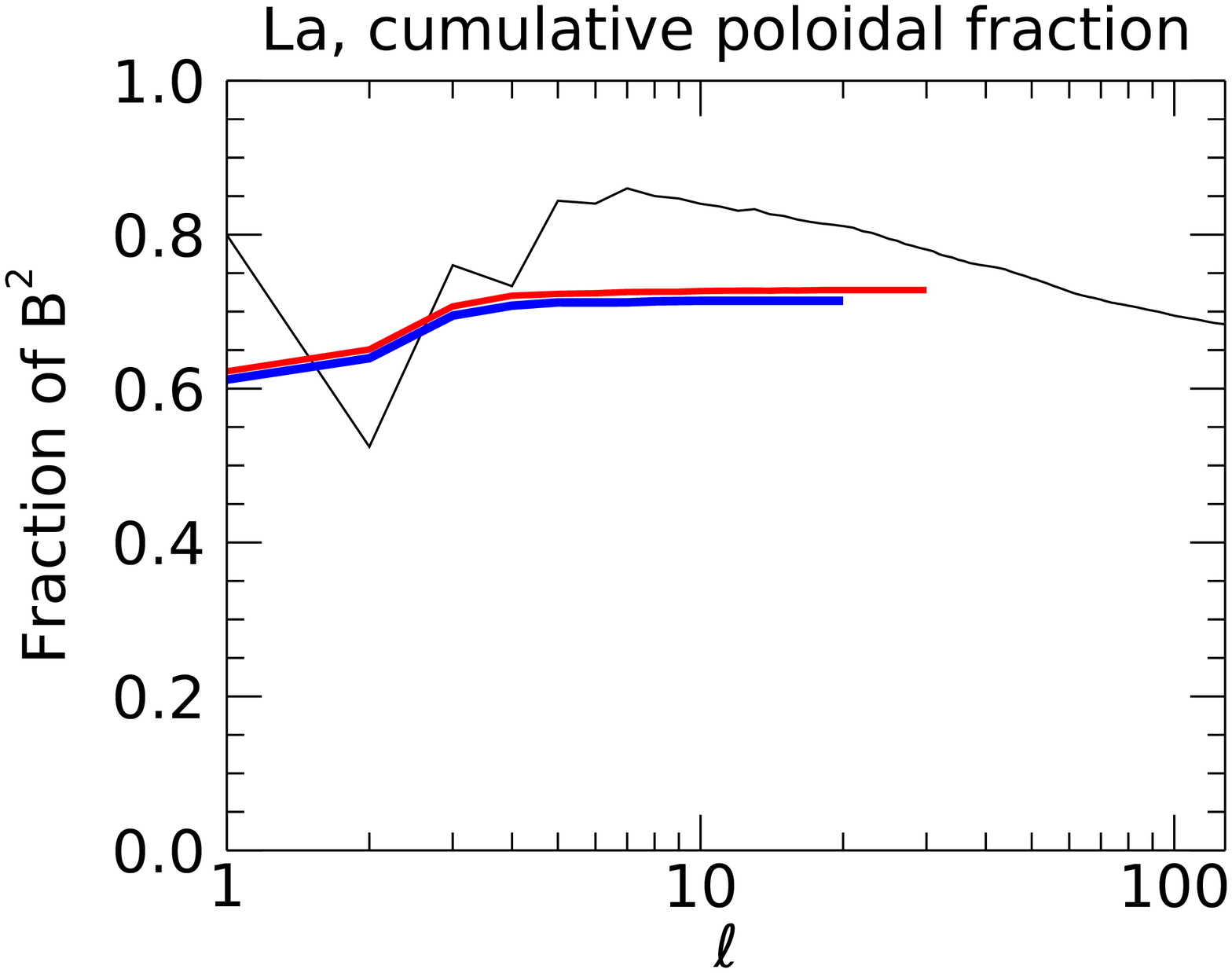}
\includegraphics[height=4cm]{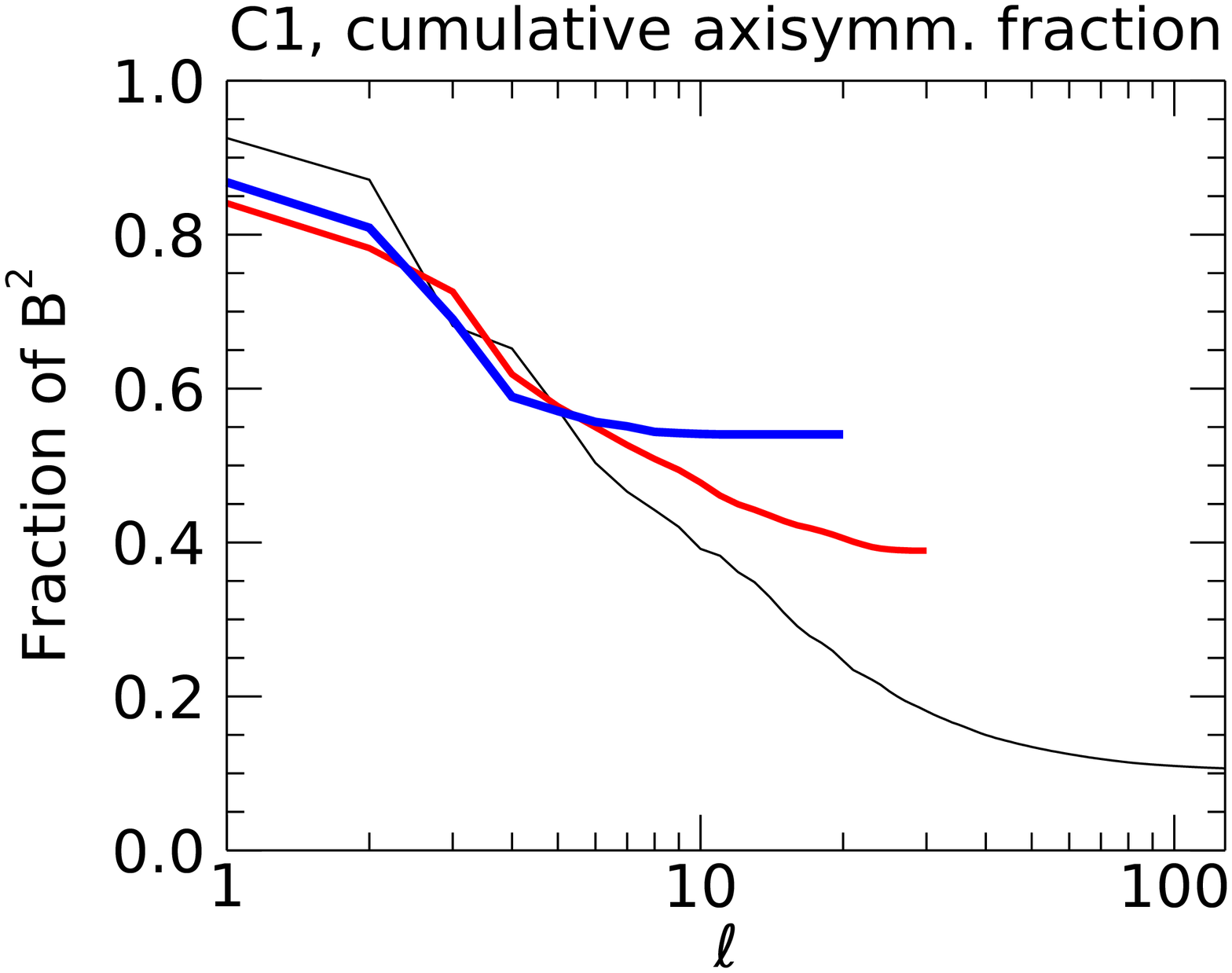}
\includegraphics[height=4cm]{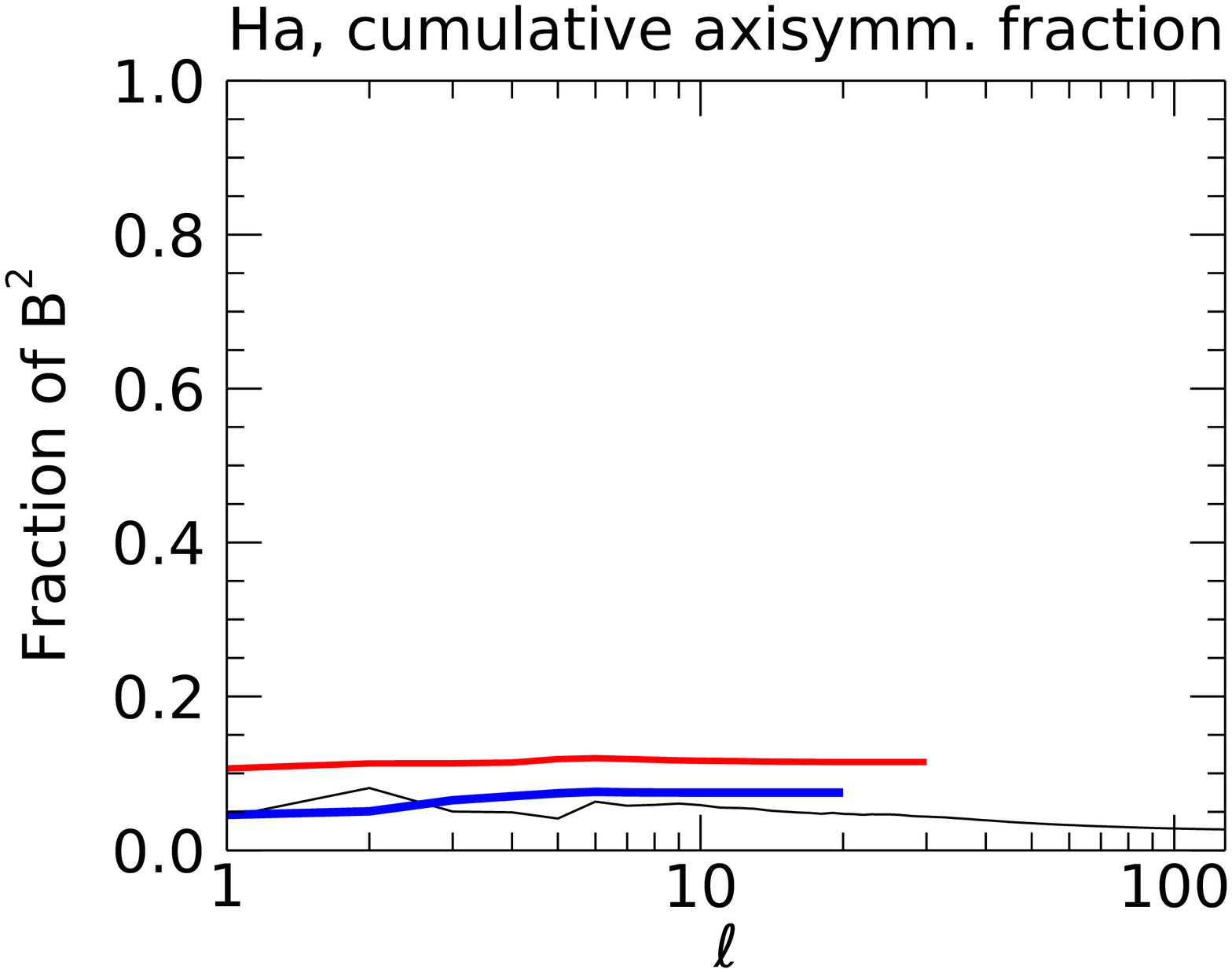}
\includegraphics[height=4cm]{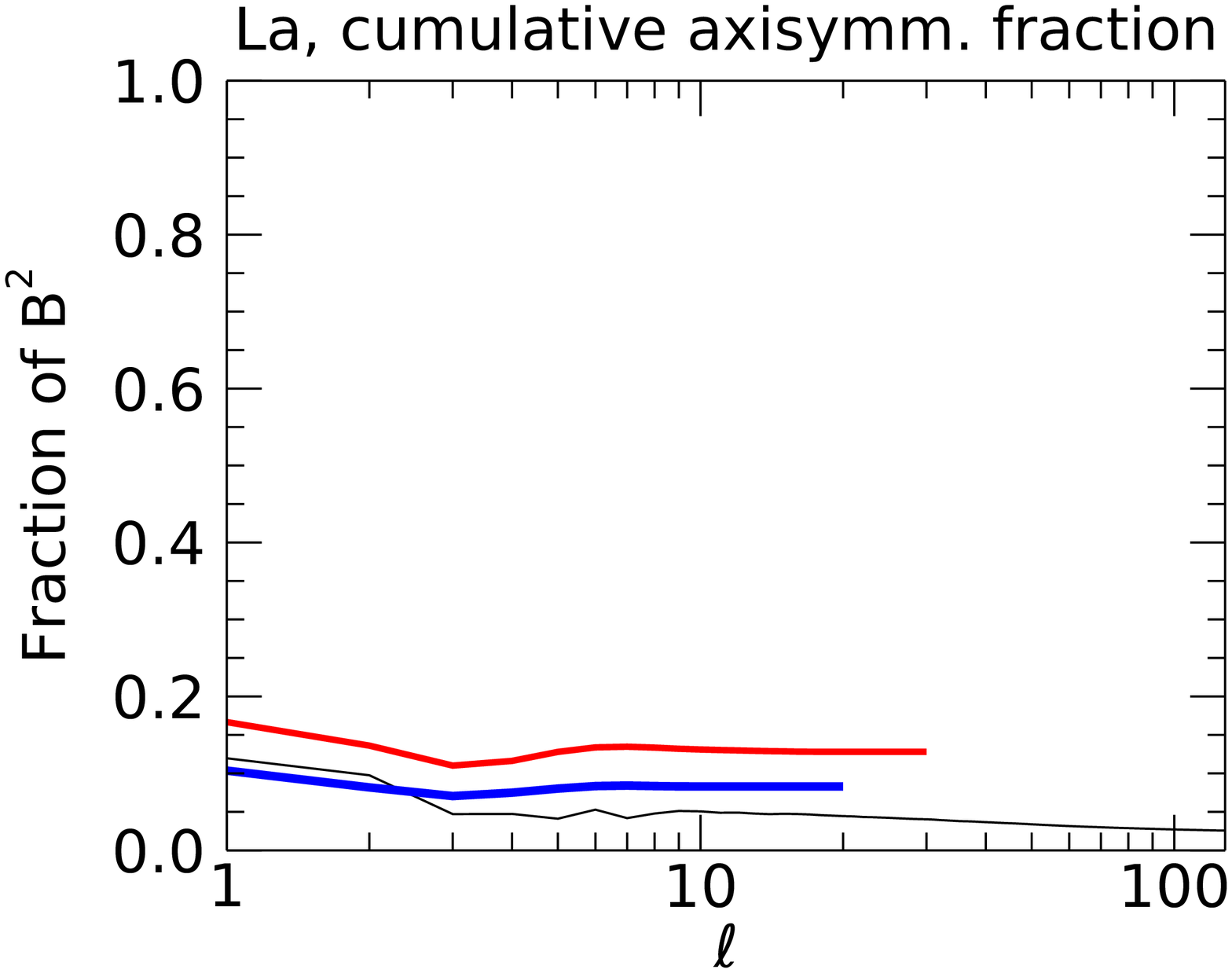}
\caption{Cumulative fractions of the magnetic energy by $\ell$-degree. Colour markings are the same as in Fig. \ref{B2_lcomp}.}             
 \label{cum_lcomp}
    \end{figure*}

\begin{figure*}
   \centering
\includegraphics[height=4cm]{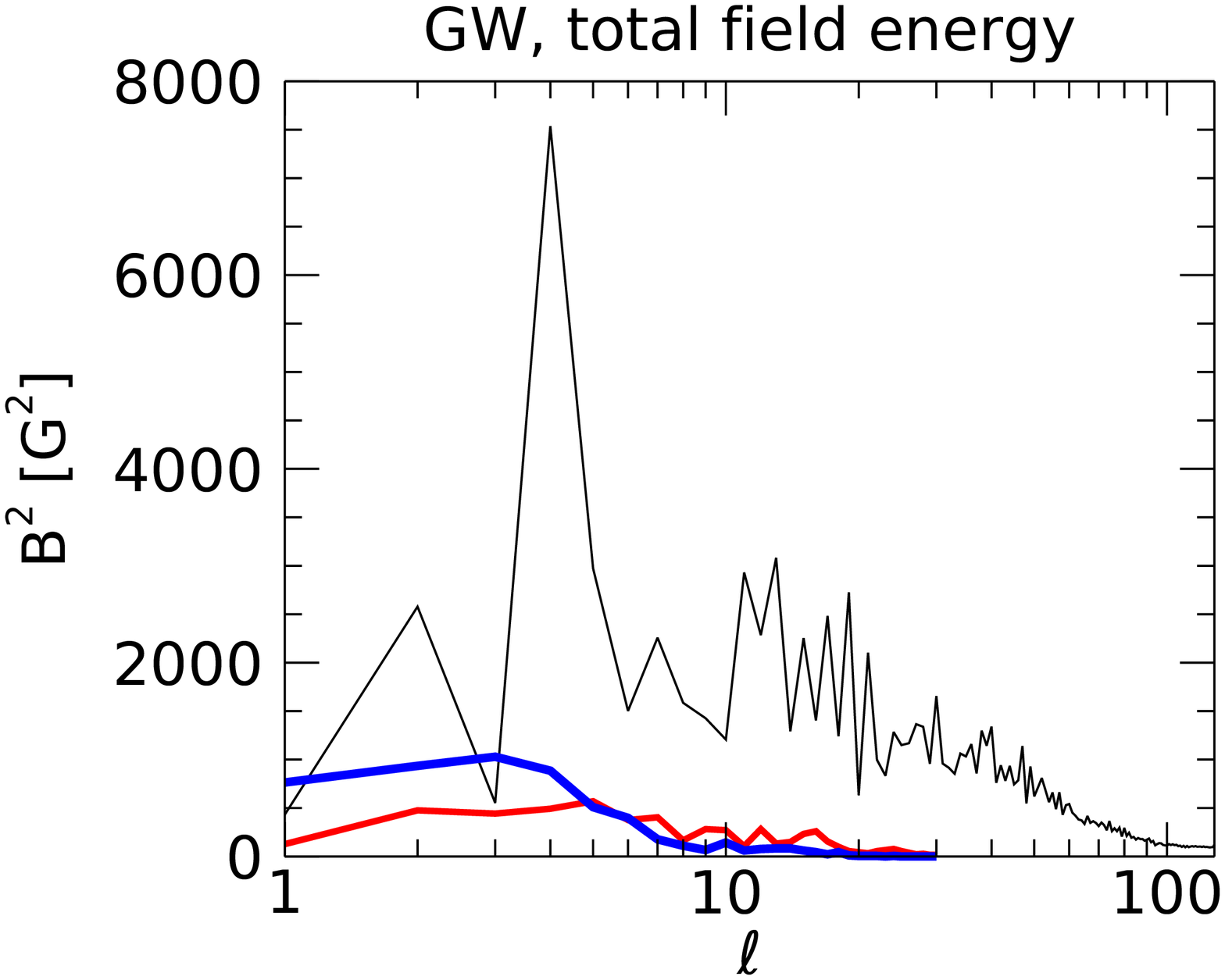}
\includegraphics[height=4cm]{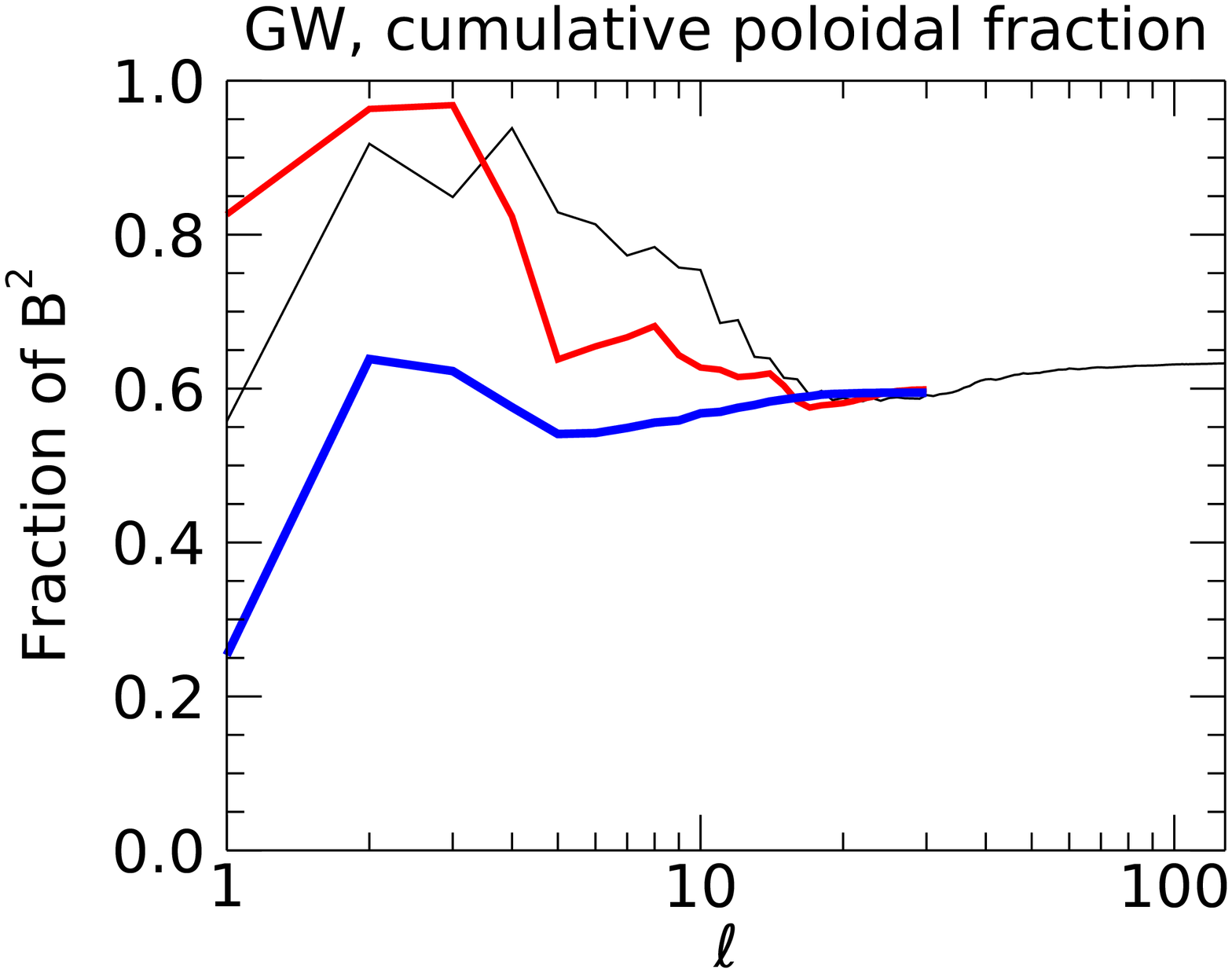}
\includegraphics[height=4cm]{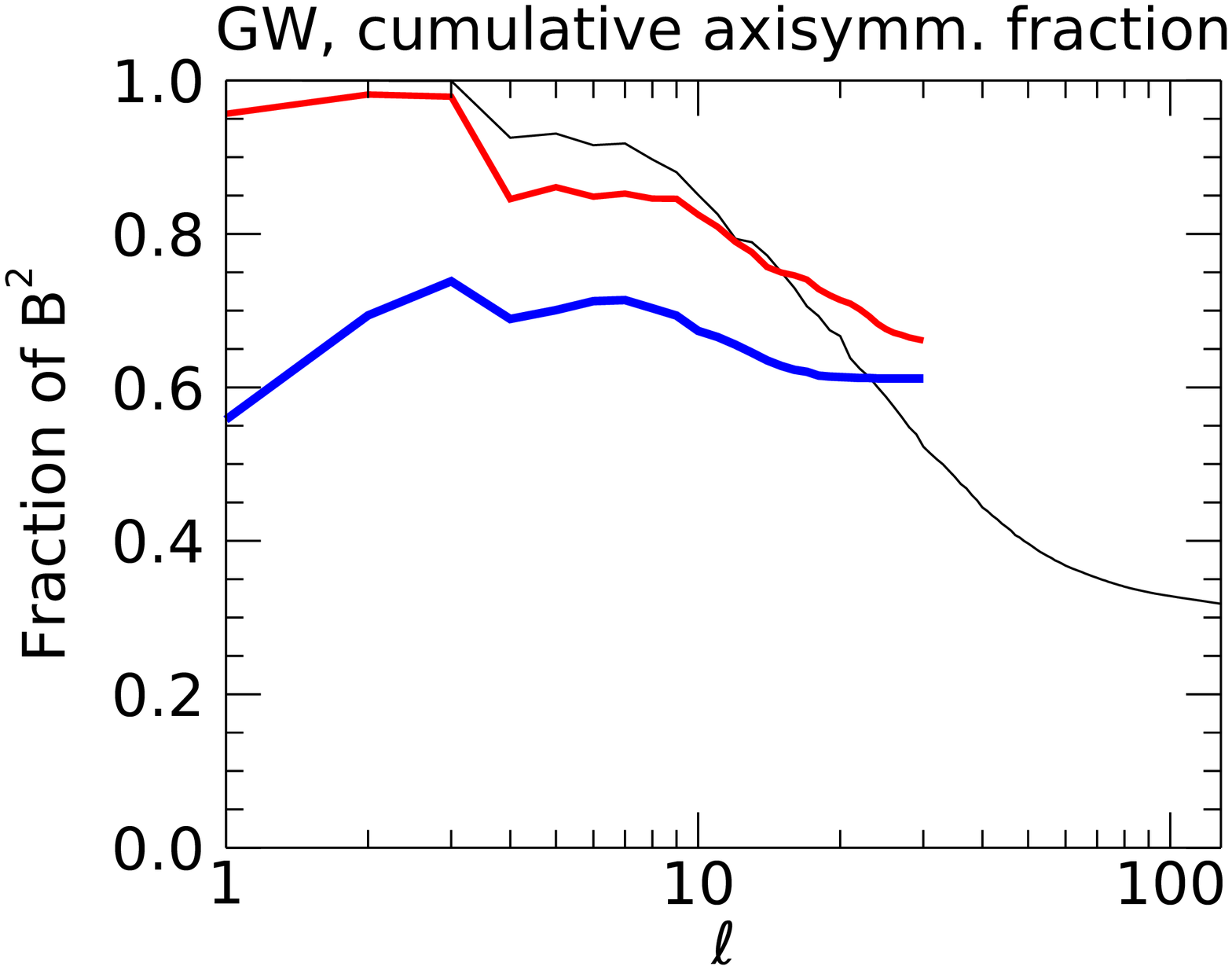}
\caption{Energy distribution and cumulative fractions of the magnetic energy by $\ell$-degree:
Original input image (black line), Test 31 (red line), and Test 32 (blue line).}             
 \label{GW_ldist}
    \end{figure*}

In order to illustrate the effect of \vsini, we plot the retrieved $B_\mathrm{RMS}$, axisymmetry $p_\mathrm{axi}$, and poloidality 
$p_\mathrm{pol}$ as functions of this. We also plot the dependence between $B_\mathrm{RMS}$, axisymmetry, and poloidality  (Figs.~\ref{C1_res} -- \ref{La_res}). Figs.~\ref{Lai_res} and \ref{Laib_res} display the dependence of the magnetic field characteristics on the inclination $i$ for the L$^a$ data.
For two representative cases (\vsini = 10 km s$^{-1}$ and  40 km s$^{-1}$) of the C1,  H$^a$, and L$^a$ MHD simulations, we compared the energy distributions as functions of $\ell$ for the preprocessed input data  and ZDI reconstruction (Figs. \ref{B2_lcomp}  -- \ref{cum_lcomp}).
For the G$^W$ data, we plot the magnetic energy distribution and cumulative fractions of the poloidal and axisymmetric field energies for Tests 31 and 32, which represent two cases of different $S/N$ and rotation phase coverage (Fig. \ref{GW_ldist}).
In these plots, a spherical decomposition up to $\ell_{\max} = 128$ of the input map was used, as this would correspond to the resolution in latitudes ($n_\theta = 128)$.
In Figs. \ref{B2_lcomp} -- \ref{naxi_lcomp}, the percentages are related to the total magnetic field energy summed over all $\ell$-degrees for the respective maps. In Figs. \ref{axi_l_ap} -- \ref{axi_l_pp},  we plot the axisymmetric and poloidal fractions for each $\ell$-degree. In the latter cases, the fractions of non-axisymmetric and toroidal field energies are naturally $1 - p_\mathrm{axi}$ and $1 - p_\mathrm{pol}$, respectively. We note that Figs. \ref{axi_l_ap} -- \ref{axi_l_pp} should be interpreted whilst keeping in mind the magnetic energy distribution with $\ell$ (Fig. \ref{B2_lcomp}) .
The cumulative fractions in Figs. \ref{cum_lcomp} -- \ref{GW_ldist} are the total fractions of poloidal and axisymmetric energies up to the $\ell$-degree of the horizontal plot axis.

\begin{figure*}
   \centering
   \vspace{0.5cm}
 \includegraphics[width=9cm]{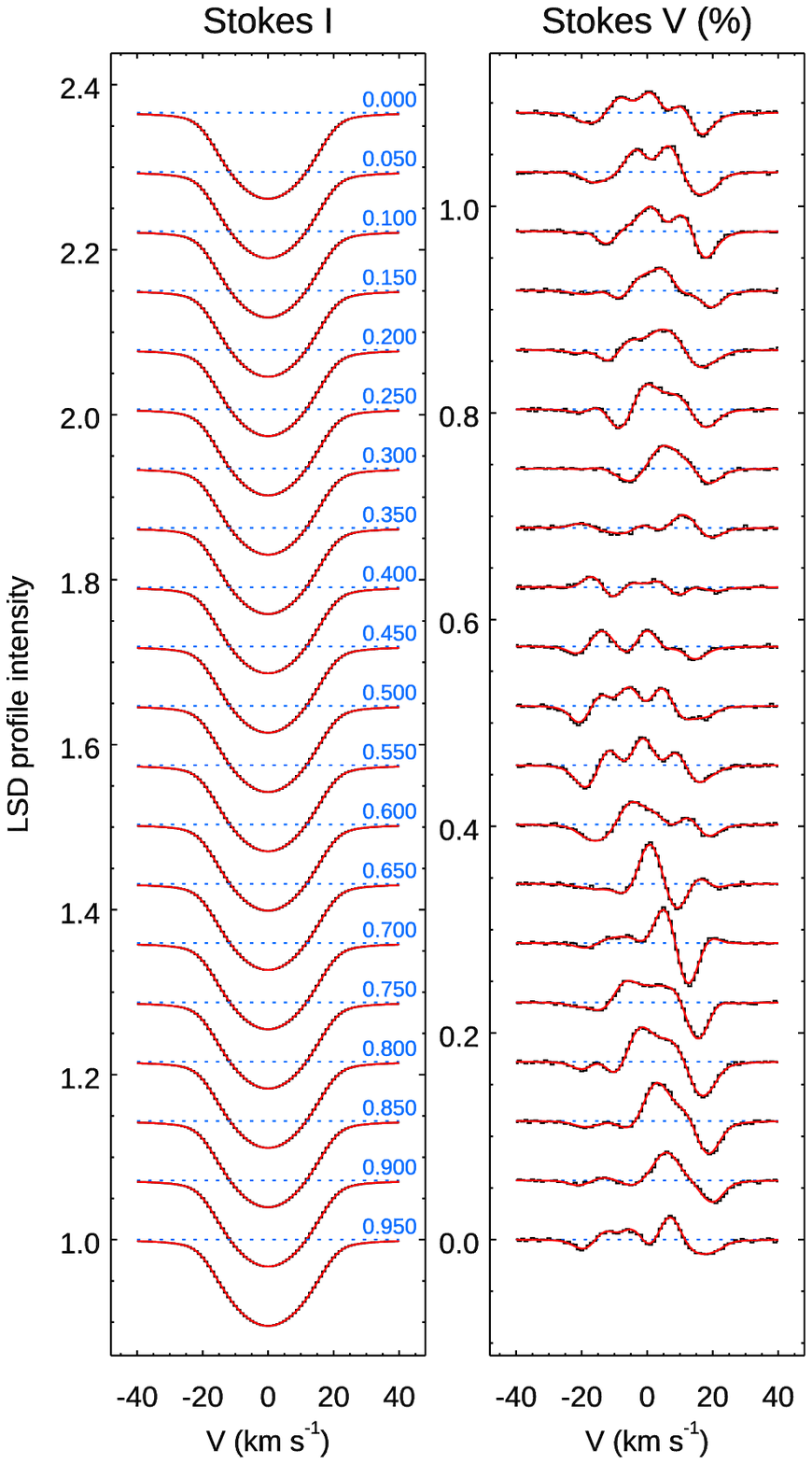} 
 \hspace{-4.5cm}
 \includegraphics[width=9cm]{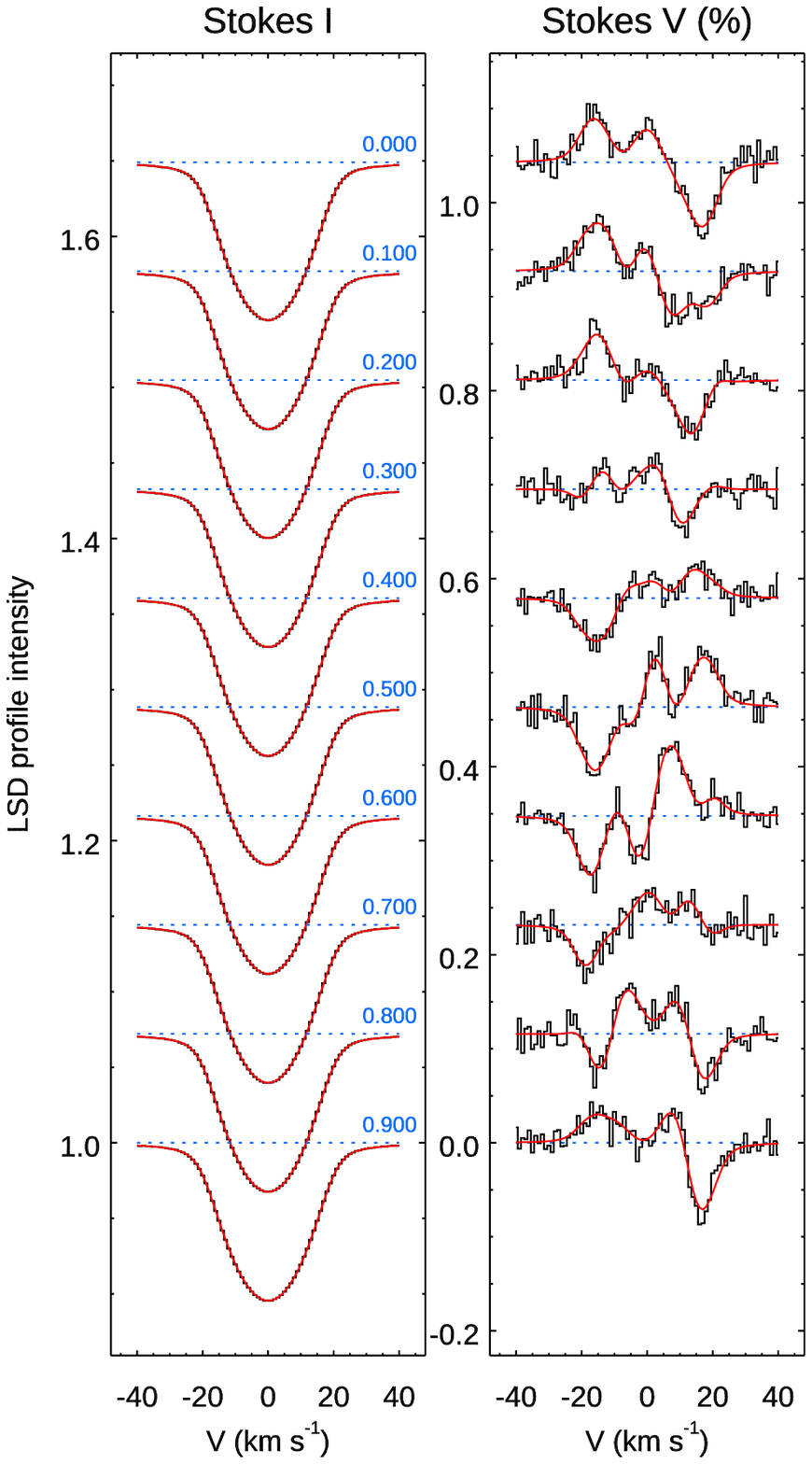}
 \hspace{-4.5cm}
 \includegraphics[width=9cm]{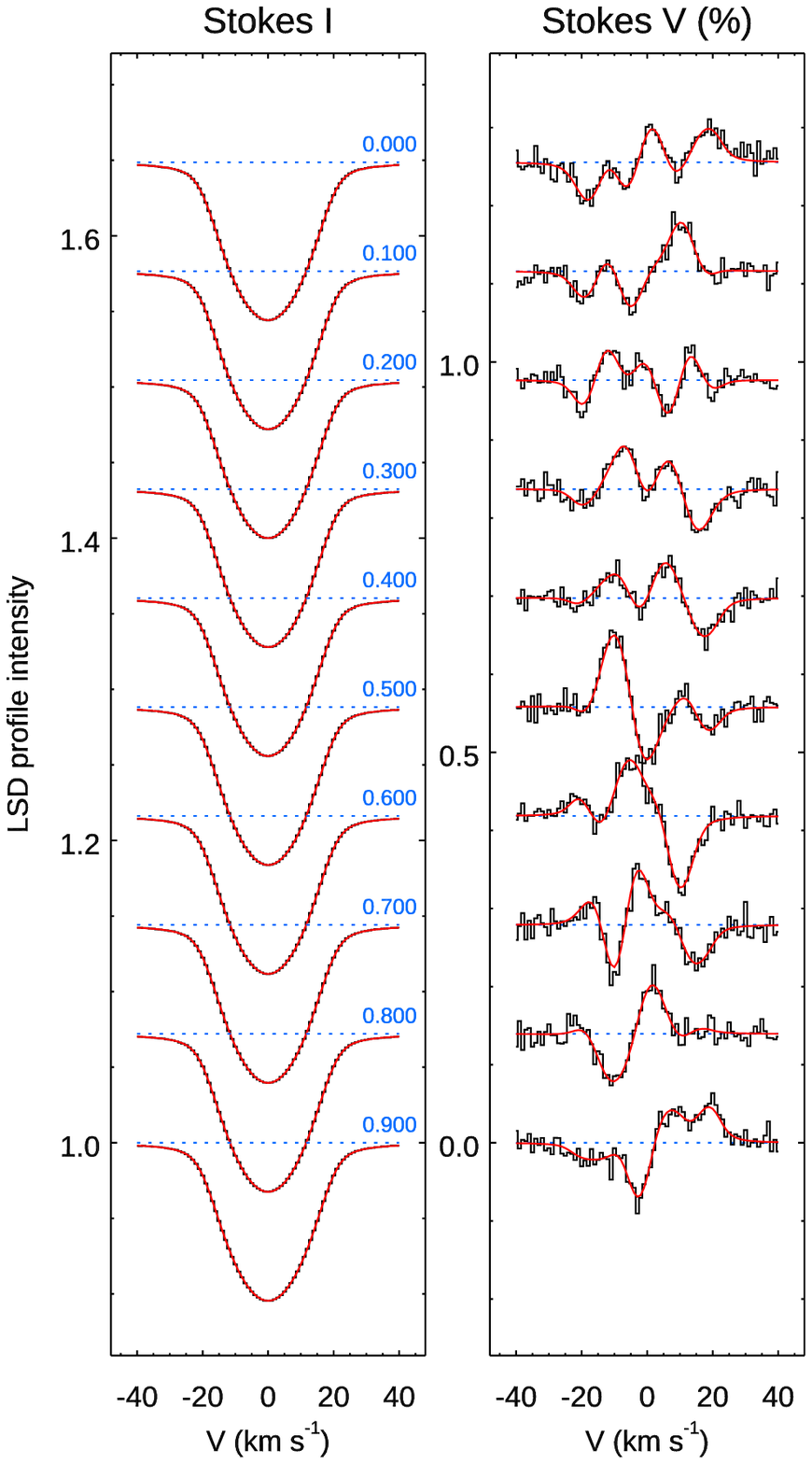}
\hspace{-4.5cm}
\vspace{0.5cm}
 \caption{Simulated (black) and calculated (red) profiles for Tests 3, 6, and 10.}             
 \label{prof}
    \end{figure*}

We only plot the modelled Stokes I and V spectra for three cases (Fig. \ref{prof}). This is mainly because all runs were continued until the reduced $\chi^2$-value corresponded to the level of the induced noise in the synthetic Stokes V data. Thus, in all cases, the spectra from the ZDI solution will fall into the noise limit of the forward calculated synthetic spectra.

In general, it is clear that the magnetic energy is always lower in the ZDI images than in the original MHD simulations. This is evident both from visually comparing the original images with the ZDI solutions, comparing the RMS values in Tables \ref{mhdpar}--\ref{zdires}, and from the magnetic energy distribution as a function of $\ell$-degree (Fig. \ref{B2_lcomp}). The percentage of the reconstructed $B_\mathrm{RMS}$ varied within 18\%--27\% for the C1 case, within 26\%--31\% for H$^a$, and
within 23\%--42\% for L$^a$. This reflects the fact that a considerable part of the magnetic field energy is `hidden' in small structures, that is, those with high $\ell$-degrees.

The characteristics of the spherical harmonics decompositions of the C1, H$^a$, and L$^a$ data reconstructed to $\ell_{\max}$ = 20 (see Table \ref{mhdpar}) can be compared with the ZDI results. Although the magnetic energy is slightly lower than in the original data, it is still much higher than in the ZDI maps. Furthermore, the fractions of axisymmetry are much smaller than in the ZDI inversions. These differences can mostly be attributed to
artifacts in the reconstructed field.  Such artifacts will always be present in ZDI images and will contribute to RMS values of the total field and its components. Therefore, the percentages above cannot as such be used as a diagnostic for the reliability of the ZDI inversion. This is demonstrated by the fact that the optimal values of the inclination ---namely $i \sim 60-70 \degr$ --- coincide with a minimum of $B_\mathrm{RMS}$ in the ZDI maps (Fig. \ref{Laib_res}). The sources of such artifacts are for example limited visibility (especially at low values of $i$) and symmetry effects (especially at $i \ge 80 \degr$).

The percentages of recovered magnetic field energy show a clear bias towards the lowest $\ell$-degrees. This tendency is naturally more accentuated in cases with lower \vsini \, and thus poorer spacial resolution.
However, as is shown in Fig. \ref{B2_lcomp}, the ZDI inversion does not fully reach the original field energies even in the lowest $\ell$-orders, except for the overestimated energy in $\ell=1$ in the cases of C1 and H$^a$. The latter excesses are clearly caused by artifacts in the reconstruction.

The axisymmetry is generally overestimated. However, our tests show that a more axisymmetric input field (C1 and G$^W$)) in general also yields a more axisymmetric reconstruction. The fraction of the poloidal field energy seems to be better reconstructed. However, we cannot confirm how well trends in the poloidality are reproduced, as all our cases had similar amounts of poloidal field energy.

By visual comparison of the original and ZDI reconstructed images, it is clear that the radial field is best reproduced in all cases. Some of the main structures of the azimuthal field are also captured in the ZDI inversions, such as the alternating tilted positive and negative field stripes in the H$^a$ image and the strong negative spot at longitude 120$\degree$ in C1. The meridional field is generally poorly reproduced. These tendencies are well-known consequences of using only Stokes V, but not Q and U \citep[see e.g.][]{Donati1997,Rosen2015}. In general, the L$^a$ simulation is best recovered. In particular, the tilt of the high-latitude stripes in $B^r$ is well reproduced even with a low \vsini\, in the case of L$^a$ compared with the similar feature of H$^a$. 

Naturally, the inversion is also dependent on the inclination $i$ and rotational velocity \vsini. Here, the best visual resemblance is achieved with the combination $i \sim 60\degree$ and \vsini \,$\ge 40$ km s$^{-1}$. With $i \ge 80\degree$, strong artifacts due to equatorial symmetry will appear in the 
ZDI maps (Fig. \ref{resmapLai}; in the three bottom panels). 
With $i<40\degree$, increasing parts of the stellar surface will be invisible. Indeed, our calculations demonstrate the difficulties in reliably recovering the part of the stellar hemisphere below the equator with any value of $i$. In our cases, the fact that the original data are quite symmetric with respect to the equator will pose extra problems in this respect. In the case of less equatorial symmetry and strong magnetic field structures on the southern hemisphere, these would be better resolved.

The dependence on $i$ explored for the L$^a$ case shows interesting tendencies (Figs. \ref{Lai_res} -- \ref{Laib_res}). There is clear minimum of $B_\mathrm{RMS}$ at $i=70 \degree$ and a maximum of $p_\mathrm{axi}$ at $i=80 \degree$.  The inclination will especially influence the
reconstruction of the meridional and azimuthal field components.
This is clearly illustrated as an increasing $B^\phi_\mathrm{RMS}$ and decreasing $B^\theta_\mathrm{RMS}$ with increasing $i$ (Fig. \ref{Laib_res}). A natural explanation for this is how these components contribute to the `observed' longitudinal field with different values of $i$.

Figs. \ref{C1_res}--\ref{La_res} illustrate that there are clear dependencies between the reconstructed $B_\mathrm{rms}$, $p_\mathrm{pol}$, and $p_\mathrm{axi}$. In particular, there are correlations between {\vsini} and $p_\mathrm{axi}$ in all cases.
This was seen in all ZDI reconstructions, and also in the reconstructions used for testing different values of $S/N$, $\ell_{\max}$, and $n_{\phi}$  not reported here.
For the C1 case, all three characteristics ($B_\mathrm{RMS}$, $p_\mathrm{pol}$, and $p_\mathrm{axi}$) of the surface magnetic field are clearly correlated (Fig. \ref{C1_res}). This implies that the spurious correlations are stronger as the Stokes V signal is decreased and the ZDI reconstruction becomes less reliable.

Figs. \ref{cum_lcomp} and \ref{GW_ldist} show the cumulative fractions of the poloidal and axisymmetric magnetic energies. These plots show how the summed fractions change with increasing $\ell$ degree.
The plots
reveal clear differences between the MHD simulations. 
In the cases of 
H$^a$ and L$^a$, the fractions of poloidal and non-axisymmetric field energies are relatively constant, while in the C1 case these fractions increase strongly with larger degrees of $\ell$. We also note that the cumulative functions of the fractions are similar both in the original input data and the resulting ZDI solutions. These are also similar for the cases of different \vsini~values. Furthermore, the cumulative functions for the C1 case resemble those calculated for the Sun by \cite{Vidotto2016}. Fig. \ref{GW_ldist} clearly shows why axisymmetry is higher in the ZDI reconstructions than in the original G$^W$ data: the percentage of axisymmetry in the original data decreases with $\ell$ from 100\% in the lowest angular degrees to a final value of $\sim$ 30 \% with all angular degrees included. This figure also demonstrates, that the reconstruction is considerably closer to the input data when using
higher $S/N$ and denser phase coverage.

In the C1 case, the higher \vsini~values yield more reliable images in terms of $p_\mathrm{pol}$ and $p_\mathrm{axi}$ (Fig. \ref{C1_res}). This tendency is not as clear for H$^a$ and L$^a$. In these cases, a smaller fraction of the magnetic energy is contained in the higher $\ell$-degrees (Fig. \ref{B2_lcomp}). Thus, a lower {\vsini} will provide a more sufficient resolution. Furthermore, for the H$^a$ and L$^a$ data, the cumulative fractions of poloidal and axisymmetric magnetic field energy are more constant with $\ell$ (Fig. \ref{cum_lcomp}).

The use of different exponents $n$ in the regularisation (Eq. \ref{reg}) did not significantly influence the results. This is clear from comparing the indicators ($B_\mathrm{RMS}$, $B^r_\mathrm{RMS}$, $B^\theta_\mathrm{RMS}$, $B^\phi_\mathrm{RMS}$, $p_\mathrm{pol}$ and  $p_\mathrm{axi}$) of Tests  13--22 ($n=2$) with those of Tests 23--30 ($n=1$) in Table \ref{zdires}. Naturally, using the lower $n$-value led to a marginal increase in the magnetic field energy of higher $\ell$-degrees. However, the difference in the distribution of magnetic energy, for example, was insignificant. 

\section{Conclusions}

We tested the ZDI inversion method by calculating synthetic Stokes IV LSD profiles for a set of MHD simulations from \cite{Viviani2018}. While all these simulations are based on a solar-type star, none of them represent the Sun. In particular, they all have a different rotational profile and stronger magnetic activity than the Sun. 

In general, the ZDI is capable of reproducing the main structures of the field. In particular, structures that represent active longitudes are well recovered.
We demonstrate that the reconstructions are dependent on stellar parameters such as the inclination $i$ of the rotational axis. Furthermore, we show that the axisymmetry of the surface field is in general overestimated in the ZDI images.
But the general trend is correctly reproduced, with more axisymmetric data
yielding more axisymmetric ZDI maps.
The total field is in turn underestimated due to insufficient resolution. The energy distribution between the poloidal and toroidal field seems to be better reproduced.

Unsurprisingly, our study confirms the previously reported problem of unequal recovery of the field components \citep[e.g.][]{Donati1997}. Using only Stokes V, the $B^r$ components will be much more reliably mapped than $B^\phi$ or $B^\theta$. The recovery of the latter components will also depend on the rotational axis inclination $i$.

Inclination is a particularly problematic parameter in ZDI, as it is usually hard to determine accurately from observations. Furthermore, the range of optimal values of $i$ is narrow. With $i < 40 \degree$, a large part of the star will be hidden, hampering conclusions about the magnetic field topology. With $i > 80 \degree$, there will be problems with equatorial symmetry. We also see a clear dependence between the fractions of axisymmetric versus non-axisymmetric field energy and rotation velocity. The dependence is naturally caused by the increasing achievable surface resolution in ZDI with growing \vsini, which will allow better reconstruction of higher $\ell$-degrees. This is problematic, because faster rotation means higher magnetic activity and the spurious correlation will contribute to measured differences in the magnetic field axisymmetry versus Rossby number. There is also a similar effect on the fraction of poloidal versus toroidal field energy, but our results indicate that this
may only be a problem for cases with weaker magnetic fields.

The retrieved field distribution in terms of the angular degree $\ell$ can generally not be trusted. There is a clear transition from higher degrees in the original field to lower degrees in the reconstruction. This can partly be understood as a blurring effect in the ZDI inversion. Furthermore, we note that limiting the angular degrees to very small values may be problematic. This is highly evident for the C1 data, which represent a less active star. \cite{Lehmann2019} recommended that $\ell_{\max} = 5$ would be sufficient for less active slow rotators. This is true in the sense that with \vsini\,$\sim$ 10 km s$^{-1}$, the spacial resolution will usually not be enough to resolve details of $\ell > 5$. However, our results indicate that a large fraction of the magnetic energy is in the higher $\ell$-degrees for less active stars. The relation between the optimal $\ell_{\max}$ and {\vsini} is not trivial, and using an overly low
$\ell_{\max}$ value may introduce unnecessary smearing of the ZDI map. Therefore, we would recommend using a higher $\ell_{\max} \sim 10$--20, even for slow rotators.

Although the magnetic field energy in specific $\ell$ degrees cannot be trusted, the cumulative distribution of the fractions of poloidal versus toroidal and axisymmetric versus non-axisymmetric field energy are surprisingly well reproduced. Again, the C1 data differed from the two other cases in that the cumulative distribution changed significantly as a function of $\ell$. However, this change was similar in both the original data and the ZDI reconstructions.

The above-mentioned discrepancies can be seen as a critique against how ZDI results are sometimes reported. It is somewhat problematic to compare the magnetic topology of different stars without accounting for systematic biases related to stellar parameters. Here, the impact of the inclination has usually been neglected in ZDI studies. 

Our study also emphasises the importance of high-quality observations. With a weaker field (C1 in our study), there is a need for extremely high $S/N$ and dense phase coverage, in addition to high spectral resolution. Perhaps a more surprising result is that the quality of the reconstruction is not strongly dependent on rotation velocity; even low \vsini~values ($\le 10$ km s$^{-1}$) allow for reasonable reconstruction. However, the \vsini~values will be more important if temperature mapping using Stokes I is included.

\begin{acknowledgements}
  This work received funding from the Research Council of Finland (project SOLSTICE,
  decision No. 324161).
O.K. acknowledges support by the Swedish Research Council (grant agreement no. 2019-03548), the Swedish National Space Agency, and the Royal Swedish Academy of Sciences.
  This project has received funding from the European Research Council (ERC)
under the European Union's Horizon 2020 research and innovation
program (Project UniSDyn, grant agreement n:o 818665).
This work was done in collaboration with the COFFIES DRIVE Science Center. We thank the anonymous referee for detailed comments, which helped to improve the paper.
\end{acknowledgements}

\bibliographystyle{aa}
\bibliography{hackman}
  
\end{document}